\documentclass[reprint,superscriptaddress,amsmath,amssymb,aps,pre,nofootinbib]{revtex4-2}
\usepackage{graphicx}
\usepackage{dcolumn}
\usepackage{bm}
\usepackage{color}
\usepackage{wasysym}
\usepackage{MnSymbol}
\usepackage{comment}
\usepackage{times}
\usepackage{siunitx}
\DeclareSIUnit{\dalton}{Da}
\DeclareSIUnit\angstrom{\text{Å}}
\usepackage[unicode=true,pdfusetitle,bookmarks=false,colorlinks=true,citecolor=blue,urlcolor=blue,linkcolor=blue]{hyperref}
\begin{document}

\title{Tunable Nanostructures from Inverse Surfactants}

\author{Nivedina A. Sarma}
\affiliation{Department of Materials Science and Engineering, University of California, Berkeley, California 94720, USA}

\author{Alexandra Grigoropoulos}
\affiliation{Department of Materials Science and Engineering, University of California, Berkeley, California 94720, USA}

\author{Mustafa Arslan}
\affiliation{Department of Bioengineering, University of California, Berkeley, California 94720, USA}
\affiliation{Department of Chemistry, Faculty of Science and Letters, Kirklareli University, Kirklareli 39100, Türkiye}

\author{Erika E. Salzman}
\affiliation{Department of Materials Science and Engineering, University of California, Berkeley, California 94720, USA}

\author{Panagiotis Christakopoulos}
\affiliation{Center for Nanophase Materials Sciences, Oak Ridge National Laboratory, Oak Ridge, Tennessee 37831, USA}

\author{Honghai Zhang}
\affiliation{Center for Nanophase Materials Sciences, Oak Ridge National Laboratory, Oak Ridge, Tennessee 37831, USA}

\author{Kelsey G. DeFrates}
\affiliation{Department of Bioengineering, University of California, Berkeley, California 94720, USA}

\author{Joakim Engstr\"om}
\affiliation{Department of Materials Science and Engineering, University of California, Berkeley, California 94720, USA}
\affiliation{Department of Bioengineering, University of California, Berkeley, California 94720, USA}

\author{Peter V. Bonnesen}
\affiliation{Center for Nanophase Materials Sciences, Oak Ridge National Laboratory, Oak Ridge, Tennessee 37831, USA}

\author{Sai Venkatesh Pingali}
\affiliation{Neutron Scattering Division, Oak Ridge National Laboratory, Oak Ridge, Tennessee 37831, USA}

\author{Ting Xu}
\affiliation{Department of Materials Science and Engineering, University of California, Berkeley, California 94720, USA}
\affiliation{Department of Chemistry, University of California, Berkeley, California 94720, USA}
\affiliation{Materials Sciences Division, Lawrence Berkeley National Laboratory, Berkeley, California, 94720, USA}
\affiliation{Kavli Energy NanoScience Institute, Berkeley, California, 94720, USA}

\author{Phillip B. Messersmith}
\affiliation{Department of Bioengineering, University of California, Berkeley, California 94720, USA}
\affiliation{Department of Materials Science and Engineering, University of California, Berkeley, California 94720, USA}
\affiliation{Materials Sciences Division, Lawrence Berkeley National Laboratory, Berkeley, California, 94720, USA}

\author{Ahmad K. Omar}
\email{aomar@berkeley.edu}
\affiliation{Department of Materials Science and Engineering, University of California, Berkeley, California 94720, USA}
\affiliation{Materials Sciences Division, Lawrence Berkeley National Laboratory, Berkeley, California, 94720, USA}

\begin{abstract}
Hierarchical materials in the natural world are often made through the self-assembly of amphiphilic molecules.
Achieving similar structural complexity in synthetic materials requires understanding how various molecular parameters affect assembly behavior.
In recent years, inverse surfactants---molecules with hydrophobic head groups and hydrophilic macromolecular tails---have been shown to self-assemble into supramolecular assemblies in aqueous solutions that show promise for a number of applications, including drug delivery.
Here, we build an understanding of the morphological phase diagram of inverse surfactants using insights from scattering experiments, computer simulations, and statistical mechanics.
The scattering and simulation results reveal that changing the head-group size is an important molecular knob in controlling morphological transitions. 
The molecular size ratio of the hydrophobic group to the hydrophilic emerges as a crucial dimensionless quantity in our theory and plays a determining role in setting the micelle structure and the transition from mesoscale to macroscale aggregates.
Our minimal theory is able to qualitatively explain the key features of the morphological phase diagram, including the prevalence of fiber-like structures in comparison to spherical and planar micelles.
Together, these findings provide a more complete picture for the molecular dependencies of assemblies of inverse surfactants, which we hope may aid in the de novo design of supramolecular structures. 
\end{abstract}
\maketitle

\section{Introduction}

Self-assembly, the process by which constituent building blocks spontaneously form ordered structures, is a powerful strategy for creating materials with well-defined order and emergent properties~\cite{Whitesides2002BeyondComponentsb, Whitesides2002Self-assemblyScalesb, Philp1996Self-AssemblySystems, Lehn2002TowardMatter, Hagan2021EquilibriumAssembly, Qiu2021PushingStorage, Santos2021MacroscopicSuperlattices, Vargo2023FunctionalGrowth}.
There is perhaps no self-assembling system as widely studied as surfactants.
By virtue of containing both solvophobic and solvophilic domains, these unique molecules can induce immiscible liquids---such as oil and water---to undergo emulsification through their affinity to interfaces.
In the bulk of a liquid, the amphiphilic character of surfactants enables them to assemble into a variety of structures.
As a result of this ``surface active'' nature, surfactants have been implicated as essential to the origins of life~\cite{Szostak2001SynthesizingLife, Luisi2016TheBiologyb, Deamer2002TheMembranes, Pohorille2009Self-assemblyMembranes} and have underpinned numerous technologies since antiquity~\cite{Levey1954TheChemistry, Konkol2015AnAntiquity, Lucas2012AncientIndustries}.
Today, materials made through surfactant self assembly have many applications in drug delivery~\cite{Rosler2012AdvancedCopolymers, Stevens2021Self-assemblyDelivery,Dong2012Long-CirculatingConjugates}, energy conversion~\cite{Walcarius2021ElectroinducedFilms, Miyake2012SynthesisCrystals}, and the hierarchical synthesis of macroscopic structures~\cite{Ikkala2004HierarchicalMaterials, Faul2014IonicMaterials, Lu2020Self-AssemblyStructure, Hartgerink2001Self-assemblyNanofibers,Cui2010Self-assemblyBiomaterials,Hendricks2017SupramolecularAmphiphiles}, in addition to being widely used in consumer products such as shampoos and detergents. 
The success of these applications depends on being able to precisely tune the properties, such as size and shape, of the self-assembled materials for optimal performance. 
It is therefore crucial to have predictive control over the assembled morphologies in order to realize the promise of self-assembly through the creation of  bespoke materials.

Phospholipids, which have a hydrophilic phosphate ``head'' group and a hydrophobic lipid ``tail'' are among the most well-known surfactants due to their central role in biological membranes~\cite{Phillips2012PhysicalCell}.
Many commonly studied surfactants have a similar structure.
Recently, we reported on the self-assembly of ``inverse surfactants''---molecules with \textit{rigid} hydrophobic groups that are shielded from solvent interactions by a corona of \textit{flexible} hydrophilic chains---for wound healing applications~\cite{Cheng2019SupramolecularRegenerationc, DeFrates2022TheProdrugc}.
In aqueous solutions, these molecules have been routinely created through the conjugation of hydrophilic polymers to hydrophobic groups~\cite{Otsuka2003PEGylatedApplications, Krishnadas2003StericallyDrugs, Vukovic2011StructureMedia, Kastantin2010ThermodynamicAlbumin}.
We observed the formation of a variety of structures including spheres, fibers, and planar sheets that feature hydrophobic cores consisting of the head groups and hydrophilic coronas made from the polymeric tails~\cite{DeFrates2022TheProdrugc}.
Intriguingly, it was found that increasing the size of the hydrophobic group of the surfactant can lead to the growth of \textit{macroscopically large fibers}.
These observations eluded the description of the standard heuristic tools used to map molecular surfactant geometry to the aggregate morphology~\cite{Israelachvili2011IntermolecularForces}. 

The history of predicting surfactant self-assembly spans nearly seventy-five years.
In one of the earliest works, Phillips and co-workers~\cite{Williams1955TheC, Phillips1955TheFormation} attempted to locate the critical micelle concentration---the concentration above which appreciable surfactant assembly occurs.  
This work and others made it clear that assembly begins at extraordinarily low concentrations.
Since then, many models have been developed to understand the size and shape of an aggregate depending on physical properties of the building blocks~\cite{Israelachvili2011IntermolecularForces, Israelachvili1975AMembranes, Israelachvili1975TheoryBilayers, Israelachvili1977TheoryVesicles, Daoud1976Temperature-concentrationSolutionsd, Daoud1982StarDependence, Bug1987TheoryAggregates, Wang1988SizePacking, Zhulina2005DiblockSolution, Pryamtisyn2009ModelingNanoparticlesc, Hurter1993MolecularTheory, Chang2006DiblockTension, Matsen2012Self-ConsistentCopolymer, Mysona2019SimulationEnergies, Xi2019ThermodynamicsTheory, Matsen2020FieldFTS, Duan2025QuantifyingTheory, Woo1996AssemblyMixtures, Deem1994Charge-frustratedPhases, Maibaum2004MicelleEffectb}.
An important quantity that emerged in the field is Israelachvili's packing parameter, which emphasizes that molecular geometry can encoded aggregate structure~\cite{Israelachvili2011IntermolecularForces, Israelachvili1975AMembranes, Israelachvili1975TheoryBilayers, Israelachvili1977TheoryVesicles}.
This simple geometric parameter posits that the solvophilic area/solvophobic volume ratio for a molecular surfactant maps to the surface area/volume of the aggregates it forms.
While Israelachivili's geometric perspective has been a useful heuristic, thermodynamic theories are needed to describe the competition between different morphologies, the transition from mesoscale to macroscale aggregates, and the assembly behavior of surfactants with more complicated geometries. 
Numerous theoretical approaches have been used to deepen our understanding of amphiphile assembly including scaling theory~\cite{Daoud1976Temperature-concentrationSolutionsd, Daoud1982StarDependence, Bug1987TheoryAggregates, Wang1988SizePacking, Zhulina2005DiblockSolution, Pryamtisyn2009ModelingNanoparticlesc}, self-consistent field theory~\cite{Hurter1993MolecularTheory, Chang2006DiblockTension, Matsen2012Self-ConsistentCopolymer, Mysona2019SimulationEnergies, Xi2019ThermodynamicsTheory, Matsen2020FieldFTS, Duan2025QuantifyingTheory} and a host of other techniques~\cite{Woo1996AssemblyMixtures, Deem1994Charge-frustratedPhases, Maibaum2004MicelleEffectb}.

Our aim in this work is to understand the morphological transitions, including the transition from mesoscale to macroscale aggregates, of inverse surfactants as a function of tunable molecular parameters using the complementary tools of neutron scattering, coarse-grained molecular dynamics, and mean-field thermodynamics.
Each method provides a window into the transition that together allows us to form a more complete understanding. 
We begin in Section~\ref{sec:ModelSystem} by introducing our experimental and coarse-grained model system for amphiphile self-assembly that features a hydrophilic polymer tethered to various rigid hydrophobic molecules.
We subsequently present our experimental and computational findings for the morphologies formed by these inverse surfactants.
These results motivate our thermodynamic picture, presented in Section~\ref{sec:Theory}, which we use to construct our morphological phase diagrams and identify the origins of the transition between mesoscale and macroscale structures.
Finally, the implications of our perspective and our outlook are provided in Section~\ref{sec:conclusions}.

\section{Experimental and Computational Model System and Results}
\label{sec:ModelSystem}

\begin{figure}
    \centering
    \includegraphics[width=1\linewidth]{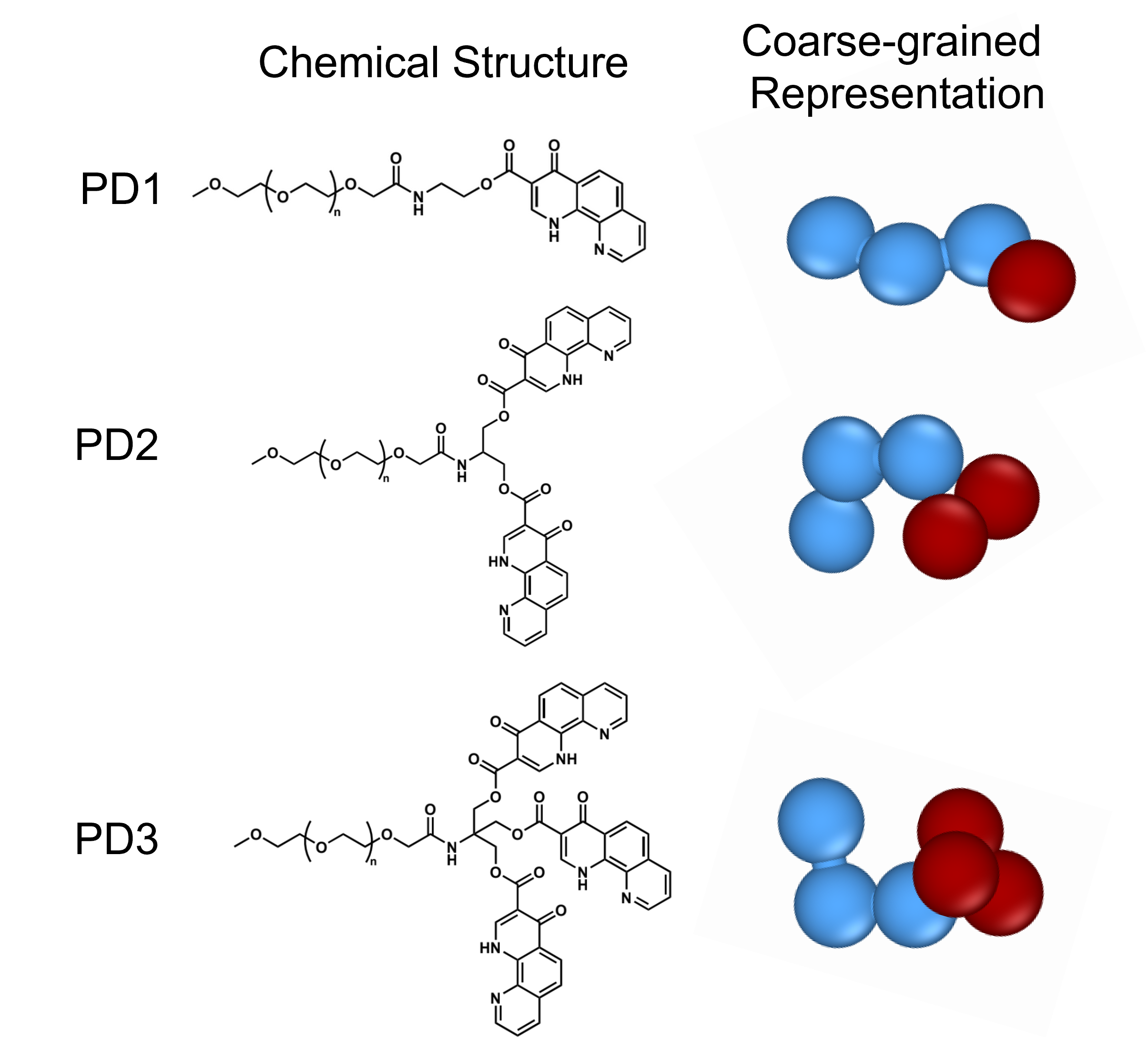}
    \caption{Chemical structures and coarse-grained representations of the inverse surfactants explored in this work.
    The coarse-grained surfactants represent each $250$ Da DPCA molecule as a single attractive bead (red) and the $1140$ Da PEG chain as three linearly connected repulsive beads (blue).}
    \label{fig:model_system}
\end{figure}

To build our understanding of inverse surfactant self-assembly, we conduct experiments and computer simulations on the three surfactants shown in Fig.~\ref{fig:model_system}, in which a polymer chain is conjugated to either $1$, $2$, or $3$ small hydrophobic molecules through a hydrolyzable ester bond (Fig.~\ref{fig:model_system}).
The polymer is a hydrophilic \SI{1140}{\dalton} poly(ethylene glycol) (PEG) chain and the hydrophobic groups are 1,4-dihydrophenonthrolin-4-one-3-carboxylic acid (DPCA).
We refer to the molecules with $1$, $2$, and $3$ rigid DPCA groups conjugated to the chain as PD1, PD2, and PD3, respectively.
These molecules were previously examined by some of us~\cite{Cheng2019SupramolecularRegenerationc, DeFrates2022TheProdrugc} for their utility as supramolecular prodrugs, as DPCA is a small-molecule therapeutic that enables regenerative wound healing in mammalian systems~\cite{Zhang2015Drug-inducedMice, DeFrates2021UnlockingSignalingb, DeFrates2024AColitis}.
In these past studies, the PEG chains were shorter (\SI{750}{\dalton}) than those used in the present study. 
In aqueous environments, the polymer surfactants self-assemble to shield the rigid core with a polymer corona, allowing the resulting structures to possibly be used as drug delivery vectors~\cite{Ahmad2014PolymericVehicles, Aiertza2012Single-chainNanoparticles, Lawrence1994SurfactantDelivery}.
The size and morphologies of the aggregates formed by these molecules is thus of particular importance for their successful application in wound healing contexts, motivating the present study.

Our previous work, which characterized the aggregates using light scattering, \textsc{X}-ray scattering, and a combination of dry and wet-state microscopy techniques, has shown that the molecules with $1$ or $2$ head groups on the chain aggregate into nanoscale spheres while those with $3$ assemble into macroscopic fibers~\cite{Cheng2019SupramolecularRegenerationc, DeFrates2022TheProdrugc}.
Preliminary computer simulations confirmed that fibers are increasingly prevalent with increasing head group size, however, suggested these simulations also suggest that \textit{all of these molecules} can assemble into a diverse range of structures~\cite{DeFrates2022TheProdrugc}.
This highlights the need for a deeper understanding of the origins of the resulting structures and their dependence on tunable molecular parameters.

Many high resolution characterization techniques that could be used to probe mesoscale structures are often unable to probe systems in the in situ wet state. 
Rather, the system might need to be rapidly dried or frozen (to preserve the original structures) in the case of transmission electron microscopy (TEM).  
We therefore turn to small-angle neutron scattering (SANS) as an in situ technique to complement previous characterization of the mesoscale self-assembly of these inverse surfactants. 
SANS allows us to probe the size and morphology of the aggregate core, which would reveal if PD1 and PD2 can in fact also make structures beyond the spherical micelles observed with light scattering and TEM.
The following sections will discuss the results of SANS experiments and coarse-grained molecular dynamics simulations that we performed to understand the role of head-group structure on assembly.

\begin{figure*}
	\centering
	\includegraphics[width=1\textwidth]{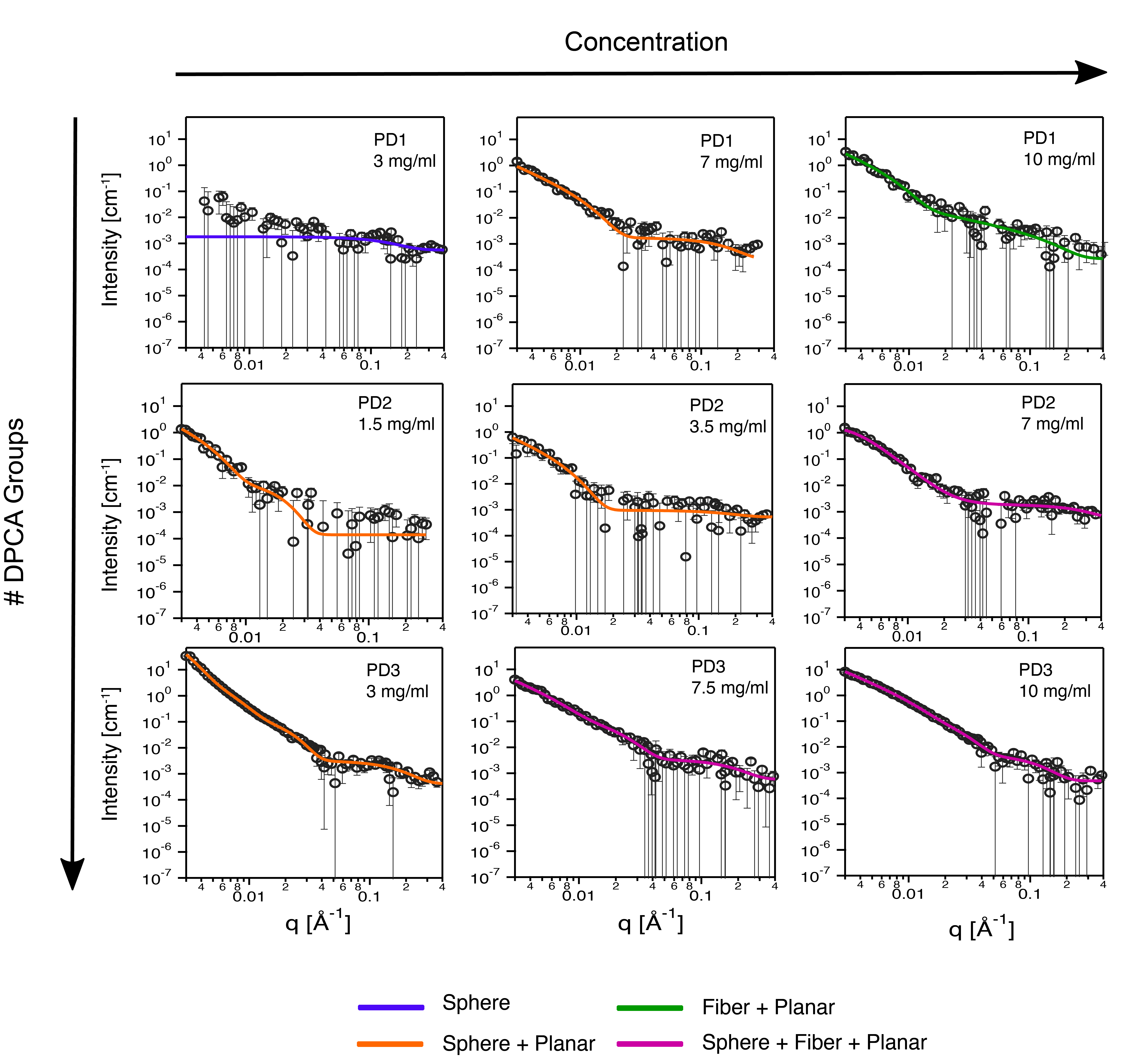}
	\caption{SANS profiles at room temperature ($25^\circ$C).
    The polymer chain is contrast-matched to the solvent so that scattering signal only comes from the hydrophobic aggregate core.
    The solid lines represent fits that are detailed in the SI. Error bars represent uncertainty after azimuthal detector averaging.} 
	\label{fig:sans}
\end{figure*}

\subsection{Scattering Experiments}
\label{sec:SANS}

We use SANS to understand how the different head group structures affect the aggregates formed by the three polymer surfactants shown in Fig.~\ref{fig:model_system}.
To isolate the scattering signal from the aggregate cores, we contrast-matched the polymer chain to the solvent through isotopic substitution of hydrogen with deuterium.
Detailed procedures for the synthesis and characterization of both the deuterated PEG chains and the prodrug conjugates can be found in the Supporting Information (SI).
Briefly, DPCA was synthesized and coupled to the deuterated PEG chains according to methods outlined in our previous works~\cite{Cheng2019SupramolecularRegenerationc, DeFrates2022TheProdrugc}.
The reaction mixtures were purified through precipitation into cold diethyl ether and separated by centrifugation.
Additional purification was conducted as necessary, as described in the SI.
The final products were characterized using both nuclear magnetic resonance (NMR) spectroscopy and mass spectrometry (MS) to ensure proper coupling and that the molecular structures, aside from the isotopic substitution on the polymer chains, matched those of non-deuterated PEG-DPCA structures, allowing us to expect similar self-assembly behavior.
Lyophilized powders of deuterated PEG-DPCA conjugates were dissolved into \SI{100}{\%} heavy water at \SI{50}{\degreeCelsius} for $30$ minutes.
The solutions were then moved to a room temperature (\SI{25}{\degreeCelsius}) environment and allowed to cool for $1$ hour prior to characterizing the resulting aggregates, following previous methods for the non-deuterated analogs~\cite{Cheng2019SupramolecularRegenerationc, DeFrates2022TheProdrugc}. 
We measured the neutron scattering intensity over the range ${0.003~\mathrm{\AA}^{-1} < q < 0.85~\mathrm{\AA}^{-1}}$ at various concentrations near and above the experimentally-determined critical micelle concentrations (CMCs) of \SI{2\pm 0.5}{\milli\gram\per\milli\litre}, \SI{1\pm 0.2}{\milli\gram\per\milli\litre}, and \SI{3.2\pm 0.7}{\milli\gram\per\milli\litre} for PD1, PD2, and PD3, respectively.
These CMCs were determined from fluorescence measurements as detailed in Ref.~\cite{DeFrates2022TheProdrugc} for similar molecules that differ only in the PEG length, as previously discussed. 
The reduced SANS scattering one-dimensional data was analyzed to identify changes in the aggregate shape and size. 

Figure~\ref{fig:sans} displays the SANS data obtained for three concentrations for each of the three molecules considered. 
To fit the SANS data, we assume that our molecules can form three possible aggregates: globular (or spherical) micelles, elongated (or fiber-like) micelles, and planar sheets.
Dimensions exceeding approximately \SI{200}{\nano\meter} are beyond the resolution of our beamline. 
We anticipate that the radius of the spheres, the cross-sectional diameter of the fibers, and the thickness of the planar sheets will be within our resolution but that the longer dimensions of the fibers and sheets likely exceed our size resolution.
We therefore cannot determine the relative concentration of these different populations but can qualitatively identify whether combinations of aggregate morphologies are present and the approximate dimensions of the morphologies within the length scales accessible by the measured SANS data.
While we of course cannot determine \textit{unique} fits, we attempt to fit the data with as few parameters as possible to understand which morphologies may dominate or coexist in a given sample. 
Here, we will emphasize the general features and trends of our SANS profiles, while the precise fitting methods~\cite{Heller2022,Kline2006ReductionPro} and details can be found in the SI.

We find that varying the number of head groups conjugated to the chain affects both the slope and intensity of the scattering profiles.
For PD1 at \SI{3}{\milli\gram\per\milli\litre}, the relatively weak intensity precludes a high-confidence fit to the data, but might itself indicate the presence of small aggregates. 
Assuming a spherical aggregate results in an estimated core radius on the order of a nanometer, which may correspond to small aggregates such as dimers or trimers (see SI for details). 
As we increase the concentration of PD1, we observe a monotonic increase in low-$q$ scattering intensity (and a reduced relative uncertainty) with concentration, which is consistent with stepwise aggregation.
At \SI{7}{\milli\gram\per\milli\litre}, the low-$q$ region has a slope of $-2.85$, while at \SI{10}{\milli\gram\per\milli\litre} the slope decreases to $-3$ for $q < 0.01~\mathrm{\AA}^{-1}$.
For both of these concentrations, these low-$q$ slopes are suggestive of the existence of non-spherical morphologies in addition to larger spherical aggregates than those found at the lowest concentration. 
The data can be well described by fits consistent with this physical picture.

For PD2, we performed SANS measurements on lower concentrations than for PD1 or PD3 due to our previous finding of a lower CMC for the PD2 system.
At the lowest concentration of PD2 (\SI{1.5}{\milli\gram\per\milli\litre}) we find a steeper low-$q$ slope that may again be indicative of planar aggregates~\cite{Beaucage2012CombinedTheory, Rai2016QuantificationConditionsb, Beaucage2004ParticleFunctionsb, Kammler2005MonitoringScattering}.
Interestingly, the magnitude and steepness of the slope of low-$q$ scattering intensity \textit{decrease} when the PD2 concentration is increased to \SI{2}{\milli\gram\per\milli\litre}.
This scattering signal is reminiscent of the PD1 system at \SI{3}{\milli\gram\per\milli\litre} which may be consistent with small spherical aggregates.
Increasing the concentration further to \SI{3.5}{\milli\gram\per\milli\litre}, we see a return to a higher intensity and steeper slope and the low-$q$ limit

The PD3 system also shows a steep slope with high intensity at the lowest concentration examined (\SI{3}{\milli\gram\per\milli\litre}) at the lowest $q$.
This data can again be well-fit by a model consisting of spherical and planar structures.
For both of the higher concentrations, \SI{7.5}{\milli\gram\per\milli\litre} and \SI{10}{\milli\gram\per\milli\litre}, the entire low-$q$ region follows a $-2$ slope.
A model fit consisting of coexisting spheres, fibers, and planar structures is able to fit this data well at these concentrations.
Unlike the molecules with $1$ or $2$ head groups, those with $3$ have a high signal-to-noise ratio.
We attribute this to the higher atomic density in the core.
The lowest-$q$ power law slopes found using SANS were validated for samples without deuteration using small angle \textsc{X}-ray scattering experiments (which detects the PEG corona of our aggregates in addition to the DPCA core), as detailed in the SI.

The scattering data reveals that the formation of larger aggregates takes place at lower concentrations for the systems with bulkier hydrophobic groups.
While the molecules with just $1$ hydrophobic group display gradual growth and transitions from spherical to planar morphologies, those with more hydrophobic groups form very large structures immediately above their critical micelle concentrations.
Altogether, our current experimental picture of the PEG-DPCA system shows that it forms a variety of structures, each of which is best-studied with different techniques.
Structured illumination microscopy (SIM) has shown that the PD3 system forms large flexible fibers~\cite{DeFrates2022TheProdrugc}.
SANS, however, shows that each of the PEG-DPCA systems is able to form mesoscale structures beyond spherical micelles.
With this experimental evidence adding to our growing understanding, we now turn to computation and theory to further add to our perspective. 
In particular, we will use computer simulations to probe the structure and dependencies of the mesoscale structures in greater detail while using theory to build a deeper understanding of the mesoscale-macroscale transition.

\subsection{Molecular Dynamics Simulations}
\label{sec:MD}

We now have experimental evidence that there exists a diverse range of possible mesoscale structures formed by not only PD3, but also by PD1 and PD2.
Coarse-grained molecular dynamics simulations can allow us to probe these structures in detail. 
Even for our coarse-grained system, the accessible time and length scales still place limitations on the aggregate sizes we can observe in practice. 
Nevertheless, much like SANS, these simulations may provide a complimentary tool to understand the morphologies of these systems.

Our coarse-grained representation for each of the three molecules is shown in Fig.~\ref{fig:model_system}. 
The details of this model are fully described in some of our previous work~\cite{DeFrates2022TheProdrugc}.
Briefly, each surfactant features a rigid head group comprised of either $1$, $2$, or $3$ attractive beads, with the relative positions of these beads fixed through a rigid body constraint~\cite{Nguyen2011RigidUnits, Nguyen2014EmergentRotation}.
The head group is tethered to a linear chain of connected beads representing the polymer~\cite{Kremer1998DynamicsSimulation}.
All interactions are pairwise and include volume exclusion.
The attractive intermolecular interactions between head group beads is modeled through a Lennard--Jones 6--12 potential with attraction strength $\epsilon_\text{LJ}$.
The equilibrium solvent is represented implicitly using Brownian dynamics simulations that were run using the graphics processing unit (GPU)-enabled open-source \texttt{HOOMD-blue} simulation package~\cite{Anderson2020HOOMD-blue:Simulations}.

\begin{figure}
	\centering
	\includegraphics[width=.49\textwidth]{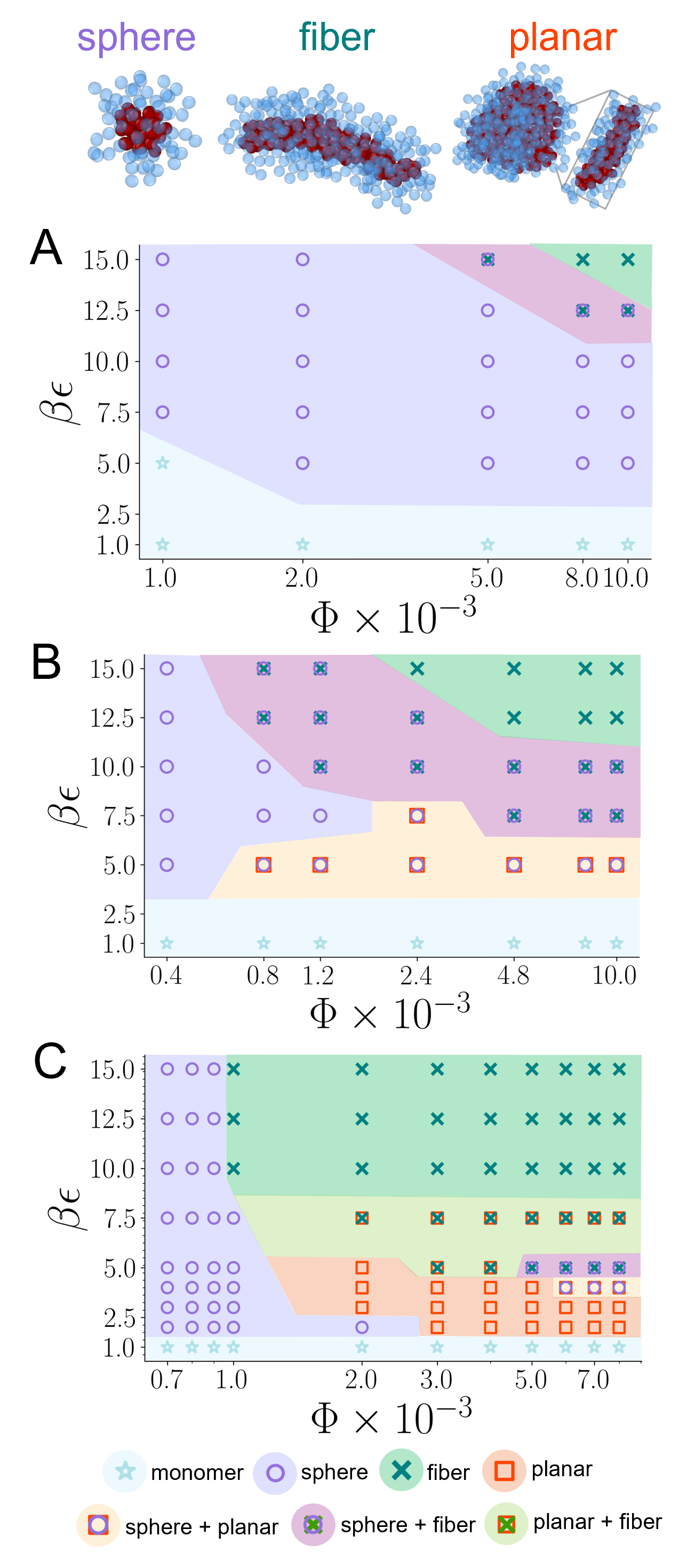}
	\caption{Morphological phase diagrams from coarse-grained molecular dynamics simulations of (A) PD1, (B) PD2, and (C) PD3.
    Equilibrated simulations display spheres, fibers, and planar aggregates, as shown in the representative snapshots at the top of the figure, as well as regions where multiple morphologies coexist.
    Phase boundaries used to shade the regions are drawn to guide the eye.}
	\label{fig:simulation phase diagrams}
\end{figure}

We define an effective attraction to facilitate comparison between  head groups of roughly equal attraction strengths to allow us to isolate the role of the head group geometry on aggregate morphology.
This \textit{effective head group attraction strength} is defined as $\epsilon = \epsilon_{\text{LJ}}$, $\epsilon = 2 \epsilon_\text{LJ}$, and $\epsilon = 3 \epsilon_{LJ}$ for the PD1, PD2, and PD3 systems, respectively.
The relative strength of this enthalpic head group attraction to the thermal energy is then $\beta\epsilon$, where $\beta \equiv 1/k_BT$ is the inverse thermal energy.
The two key parameters in the simulations are thus the volume fraction (or concentration) of surfactants, $\Phi$, and the dimensionless effective attraction, $\beta\epsilon$.
We explored assembly behavior for $0.002 < \Phi < 0.008$ and $1 < \beta \epsilon < 15$ to match the conditions under which aggregated structures were observed experimentally \cite{Cheng2019SupramolecularRegenerationc, DeFrates2022TheProdrugc}. 
(Mass concentrations in Fig.~\ref{fig:sans} were converted to volume fractions using the approximate mass densities of pure PD1, PD2, and PD3 reported in Ref.~\cite{DeFrates2022TheProdrugc}, which also provides the melting curves of pure DPCA used to estimate the relevant attraction-energy range.)
The shape of each aggregate in our equilibrated simulations was then analyzed to quantitatively determine the aggregate morphology (see SI for details). 
The morphological phase diagrams as a function of $\beta\epsilon$ and $\Phi$ are shown for each molecule considered in Fig.~\ref{fig:simulation phase diagrams}. 

Across all systems, isolated molecules make up the largest morphological state at low $\beta \epsilon$ for all volume fractions $\Phi$.
Increasing $\beta \epsilon$ introduces self-assembled mesoscale structures: for \textit{every system} we are able to observe surfactant assembly into spherical and cylindrical (or fiber-like) aggregates.
In these finite-size simulations, which contain $1000$ total molecules, the spherical aggregates typically contain $9-12$ molecules while the cylindrical aggregates can feature several dozen molecules~\cite{DeFrates2022TheProdrugc}.
For the smallest head group, increasing the attraction strength leads the monomers (isolated molecules) to form spherical aggregates (Fig.~\ref{fig:simulation phase diagrams}A).
As $\beta \epsilon$ increases further, the system transitions into coexistence between spheres and fibers, and eventually into a fiber-dominated regime at the highest attraction strengths and concentrations.
The transition to fiber morphologies shifts to lower $\Phi$ and $\beta \epsilon$ as the effective head group size increases (Fig.~\ref{fig:simulation phase diagrams}B, C), suggesting that bulkier head groups promote fiber formation.
In the systems with $2$ and $3$ head groups, we also see planar aggregates dominating the high $\Phi$ and weak attraction $\beta\epsilon$ regions of the phase diagrams.
Increasing the attraction strength further results in fibers and spheres dominating over planar structures.

The phase diagrams from our coarse-grained simulations align with the experimental trend that bulkier head groups and higher attraction strengths promote fiber assembly~\cite{DeFrates2022TheProdrugc}.
However, the simulations also reveal states that have not been observed experimentally within the detection limit of the techniques that were previously used~\cite{Cheng2019SupramolecularRegenerationc, DeFrates2022TheProdrugc}.
Such states include a planar morphology, all of the systems forming fibers (not only the PD3 system~\cite{DeFrates2022TheProdrugc}), and systems featuring coexistence between spheres, planar structures, and fibers.
The presence of planar morphologies also qualitatively agrees with the observation of two-dimensional aggregate cores in our SANS experiments.

Both the simulations and SANS experiments suggest that each of the examined inverse surfactants displays a mosaic of mesoscale morphologies with larger head groups biasing the system towards fiber and planar morphologies.  
However, due to finite-size and kinetic limitations, it is challenging to determine whether the simulated aggregates would grow further in size with increasing simulation size and time. 
The nature of the mesoscale-macroscale transition of these aggregates also remains an open question, which we will now explore theoretically.

\section{Theory of Inverse Surfactant Assembly}
\label{sec:Theory}

Having probed the effects of inverse surfactant geometry on aggregation through both experiments and computer simulations, we now discuss the thermodynamic picture for a general self-assembling system.
We will then apply that framework to understand the morphological phase diagram for an inverse surfactant that can assemble into a variety of morphological phases.
We will first briefly present the basics of classical aggregation theory~\cite{Hagan2021EquilibriumAssembly}, the mean-field framework we will use to describe the morphological phase diagram of inverse surfactants, and provide modest extensions needed for our purposes. 
As the origins of the mesoscale-macroscale transition for inverse polymer surfactants is of particular interest and is difficult to probe computationally, we develop general conditions on the free energy of formation needed for such a transition. 
The developed mean-field framework can then be applied to a variety of systems to understand both self-limited growth and the nature of the transition to macroscopic aggregates. 
We subsequently apply this framework to inverse surfactants by developing a simple but informative model for the free energy of formation of spheres, fibers, and planar structures. 

\subsection{Classical Aggregation Theory}

We work in the canonical ensemble with temperature $T$, system volume $V$, and $n$ surfactant molecules of molecular volume $v_0$. 
Surfactants can assemble into aggregates of various sizes and morphologies represented by the set ${\mathcal{M} = \{\text{monomer}, \text{sphere}, \text{fiber}, \text{planar\}}}$ where we have included individual monomers as well for compactness.
By definition, some morphologies may be limited in the range of aggregate sizes they can exhibit (e.g.,~a monomer can only have $p=1$). 
We therefore define a morphology-dependent set of sizes ${\mathcal{P}_m = \{ p \in \mathbb{Z} \mid p_m^{\rm min} \leq p \leq p_m^{\rm max} \}}$.

Each morphology will have a distinct internal partition function, $Q_{pm}$, that depends on the number of surfactants within the aggregate, $p$.  
While surfactants within the same morphological cluster clearly interact (captured in $Q_{pm}$), we neglect interactions between aggregates as is standard in classical aggregation theory~\cite{Hagan2021EquilibriumAssembly}. 
The canonical partition function of these aggregates then takes the form:
\begin{equation} \label{eq: partition function}
     Z = \sum_{c_1 = 1}^{n}...\sum_{c_{n}=1}^{n} \left(\prod_{m\in \mathcal{M}}\prod_{p \in P_m} \dfrac{1}{n_{pm}!}\left(\dfrac{Q_{pm} V}{v_\text{ref}}\right)^{n_{pm}}\right) \delta_{nn'},
\end{equation}
where $n_{pm}$ is the number of $p$-sized aggregates of morphology $m$, $c_i$ is the total number of aggregates (including all morphologies) of size $i$ (i.e.,~${c_i = \sum_{m \in \mathcal{M}} n_{im}}$),  and $v_\text{ref}$ is a reference volume.
The Kronecker delta $\delta_{nn'}$ enforces the constraint that only states with $n'=n$ total surfactant molecules are considered, where ${n' = \sum_{m\in \mathcal{M}} \sum_{p \in P_m} pn_{pm}}$ is the total number of surfactants in the microstate considered.

The partition function above allows for all possible combinations of aggregate morphology and size.
Both the case that all surfactants are completely dispersed ($c_1 = n$) as well as the case that all surfactants assemble into a single macroscopic cluster of size $n$ ($c_n = 1$) are considered, although these states are extraordinarily rare. 
To arrive at the standard free energy of classical aggregation theory, we approximate the partition function using the combination of $c_i$ that maximally contribute to the sum, which we denote as $c_i^{\dagger}$.
We will also later enforce the constraint on the total number of surfactant molecules during a free energy minimization, allowing us to drop the factor of $\delta_{nn'}$ in Eq.~\eqref{eq: partition function}. 
The unconstrained partition function is thus approximated as ${Z \approx \prod_{m\in \mathcal{M}}\prod_{p \in P_m} \left(Q_{pm} V/v_\text{ref}\right)^{n_{pm}}/n_{pm}!}$. 
The resulting free energy density $\beta f = -\ln Z/V$ then takes the form:
\begin{equation}
\label{eq:free energy density}
    \beta f v_0 = \sum_{m \in \mathcal{M}} \sum_{p \in P_m} \dfrac{\phi_{pm}}{p} \left[\ln \phi_{pm} - 1 + \beta F_{pm} \right],
\end{equation}
where $\phi_{pm}$ is the volume fraction occupied by a $p$-sized aggregate of morphology $m$ and $\beta F_{pm} = -\ln{Q_{pm}}$ is the \textit{free energy of formation} of that aggregate.
For simplicity, we have taken $v_{\rm ref} = pv_0$ and note that other natural choices (e.g.,~$v_{\rm ref} = v_0$) will result in minor quantitative differences.

We can determine the most probable volume fraction of each allowable aggregate size for each morphology through a constrained minimization of our free energy density. 
The constraint, which results from the fixed total number of surfactants and system volume, is simply ${\sum_{m\in \mathcal{M}} \sum_{p \in P_m} \phi_{pm} = \Phi}$ (where $\Phi \equiv nv_0/V$ is the overall surfactant volume fraction) and is enforced through a Lagrange multiplier. 
The resulting equations for determining these optimum volume fractions take the form:
\begin{equation}
\label{eq:coexist}
    \left. \dfrac{\partial f}{\partial\phi_{pm}} \right|_{\{\phi_{pm}^{\dagger}\}} = \mu^{\rm coexist} ,
\end{equation}
which holds for all ${m \in \mathcal{M}}$ and $ { p \in P_m }$.
Here, ${\{\phi_{pm}^{\dagger}\}}$ represents the set of optimum solutions for each $\phi_{pm}$.
We can interpret $\partial f/\partial \phi_{pm}$ as the aggregate chemical potential and the Lagrange multiplier, $\mu^{\rm coexist}$, as the coexistence chemical potential that must also be self-consistently determined. 
We have precisely the same number of unknowns (each $\phi_{pm}^{\dagger}$ and $\mu^{\rm coexist}$) as equations (including the constraint) and can thus determine the optimal concentration of each aggregate morphology and size.

It is now convenient to take the limit of continuous cluster sizes ${\mathcal{P}_m = \{ p \in \mathbb{R} \mid p_m^{\rm min} \leq p \leq p_m^{\rm max} \}}$ with ${F_{pm} \rightarrow F_m(p)}$ and ${\phi_{pm} \rightarrow \phi_m(p)}$.
In this limit, the free energy density is:
\begin{equation}
\label{eq:free_energy_cont}
    \beta f v_0 = \sum_{m \in \mathcal{M}} \int_{\mathcal{P}_m} dp \dfrac{\phi_{m}(p)}{p} \left[\ln \phi_{m}(p) - 1 + \beta F_{m}(p) \right],
\end{equation}
and the coexistence criteria take the form:
\begin{equation}
\label{eq:coexist2}
    \left.\dfrac{\delta f}{\delta \phi_m}\right|_{\{\phi_m^{\dagger} (p)\}}  = \mu^{\rm coexist},
\end{equation}
which holds for all ${m \in \mathcal{M}}$ and ${p \in P_m}$ and must be solved along with the constraint ${\Phi = \sum_{m \in \mathcal{M}} \int_{\mathcal{P}_m}dp\phi_m(p)}$.
The set ${\{\phi_m^{\dagger} (p)\}}$ represents the set of optimum solutions for each function $\phi_m (p)$.
We can substitute the precise form of the chemical potential ${\delta f/\delta \phi_m}$ into the criteria to obtain:
\begin{equation}
\label{eq:phimexplicit}
\phi_m^{\dagger}(p) = \exp \left( \beta \left [v_0 p\mu^{\rm coexist} - F_m(p)\right] \right), 
\end{equation}
for all ${m \in \mathcal{M}}$ and ${ p \in P_m }$.
Importantly, the morphology distributions remain coupled to one another implicitly through the coexistence chemical potential. 
We can eliminate the coexistence chemical potential and express Eq.~\eqref{eq:phimexplicit} in the familiar ``law of mass action'' form:
\begin{equation}
\label{eq:massaction}
\frac{\phi_m^{\dagger}(p)}{(\phi_{\rm mon}^{\dagger})^{p}}
= \exp\left[ -\beta (F_m(p) - p F_{\rm mon}) \right],
\end{equation}
for all ${m \in \mathcal{M}}$ and ${p \in P_m }$. 
In Eq.~\eqref{eq:massaction} we have introduced the monomer volume fraction, which, from Eq.~\eqref{eq:phimexplicit}, takes the form ${\phi_{\rm mon}^{\dagger} = \exp \left( \beta \left [v_0 \mu^{\rm coexist} - F_{\rm mon}\right] \right)}$ where $F_{\rm mon}$ is the monomer formation energy.

For many conditions, the size distribution for each morphology is expected to be relatively sharp about an optimum size~\cite{Wang1988SizePacking}. 
This most probable size can be probed through further examination of the solutions resulting from Eq.~\eqref{eq:coexist2}.
From Eq.~\eqref{eq:phimexplicit}, these are the sizes that maximize the argument of the exponential, ${v_0 p\mu^{\rm coexist} - F_m(p) }$, for each morphology.
As the coexistence chemical must be self-consistently determined from the full solution, these optimum sizes are not solely determined from the free energy of formation. 
However, minima in the free energy of formation that are ``deep''---as measured in units of ${v_0p\mu^{\rm coexist}}$---will be closely related to the most probable size. 

Determining the optimum size without first determining the full solutions to Eq.~\eqref{eq:coexist2} or Eq.~\eqref{eq:phimexplicit} can allow us to develop a theory with far fewer degrees of freedom. 
We can find these optimum sizes by treating them as an adjustable variable in a mean-field limit, which assumes the volume fraction of each morphology has the form ${\phi_m(p) = \phi_m \delta (p-p_m)}$.
Each morphology is now simply characterized by two variables: $\phi_m$ and $p_m$.
The continuous free energy in Eq.~\eqref{eq:free_energy_cont} reduces to:
\begin{equation}
\label{eq:free_energy_mf}
    \beta f v_0 = \sum_{m \in \mathcal{M}} \dfrac{\phi_{m}}{p_m} \left[\ln \phi_{m} - 1 + \beta F_{m}(p_m) \right],
\end{equation}
where we now only have a single volume fraction ($\phi_m$) and aggregate size ($p_m$) that must be determined for each morphology.
For convenience, we define a vector, $\mathbf{x}$, that contains all of the degrees of freedom---each aggregate size and volume fraction.
Optimizing the free energy with respect to these degrees of freedom results in the following criteria: 
\begin{subequations}
\label{eq:coexist3}
    \begin{align}
        \left. \frac{\partial f}{\partial \phi_m}\right|_{\mathbf{x}^{\dagger}}  &= \mu^{\rm coexist} ,\\
        \left. \frac{\partial f}{\partial p_m}\right|_{\mathbf{x}^{\dagger}}  &= 0,
    \end{align}
\end{subequations}
which hold for all ${m \in \mathcal{M}}$ and must be solved along with the constraint ${\sum_{m\in \mathcal{M}}  \phi_{m} = \Phi}$.
Here, $\mathbf{x}^{\dagger}$ represents the degrees of freedom evaluated at their optima.
The stability of these solutions can be determined through the  eigenvalues of the Hessian of the free energy, as detailed in the SI.

It is instructive to consider the strong aggregation limit in which a single morphology of a single size dominates the system. 
In this limit $\phi_m = \Phi$ and thus the only adjustable variable in the free energy [Eq.~\eqref{eq:free_energy_mf}] is the size of the large aggregate, $p_m$. 
The variable part of the free energy is then simply ${fv_0/\Phi  = F_m(p_m)/p_m}$.
It then follows that the optimum size must satisfy the following criteria:
\begin{subequations} 
\label{eq:stability}
\begin{align}
    \left. \dfrac{\partial (F_m/p)}{\partial p}\right|_{p_m^{\dagger}} &= 0 ,\\    
    \left. \dfrac{\partial^2 (F_m/p)}{\partial p^2}\right|_{p_m^{\dagger}} &> 0,
    \end{align}
\end{subequations}
where $p_m^{\dagger}$ is the optimum size and we take $p$ (rather than $p_m$) to be the argument of $F_m$ for convenience. 
These conditions state that the \textit{per molecule} free energy of formation must have a clear minimum.
As we will later see, many per-molecule formation energies, while convex, do not display a global minima at finite aggregate sizes and only asymptotically approach a limiting minimum value. 
For these aggregates, the system can arbitrarily reduce its free energy with increasing aggregate size. 
Physically, this will lead to the formation of \textit{macroscopic} structures that will only be limited in size by the total amount of surfactants and, in practice, kinetic limitations. 
We can thus use the condition in Eq.~\eqref{eq:stability} to understand how tunable molecular parameters can shape the free energy of formation in ways that result in a transition from mesoscale aggregates to macroscopic structures.
As previously alluded to, understanding of this mesoscale-macroscale transition is inherently not possible in even coarse-grained molecular dynamics simulations due to limitations on the accessible length and time scales. 
We are now ready to turn our focus to constructing a theory for the free energy of formation for the morphologies under consideration, which can reveal the nature of this mesoscale-macroscale transition.

\subsection{Free Energies of Formation}
\label{sec:FreeEnergy}

To construct the free energy of formation for each of the morphologies in the set $\mathcal{M}$, we will first adopt a simple physical picture of an inverse polymer surfactant.
While such a coarse-grained description may not provide quantitative accuracy, it will allow us to capture the general trends of the morphological phase diagram of these systems as a function of molecular parameters.

We take the hydrophobic head group to be rigid and approximately spherical with a radius $r_h$.
Neighboring pairs of head groups are taken to interact with an effective attraction strength that has an energetic scale of magnitude $\epsilon$.
Each head group is attached to a linear polymer chain with $N$ statistical segments of (Kuhn) length $b$.
The chains are taken to be in good solvent conditions, which is captured by the excluded volume parameter, $v$~\cite{Rubinstein2014PolymerPhysics}.
An isolated chain will thus have a characteristic size of $r_t = v^{1/5} b^{2/5} N^{3/5}$~\cite{Rubinstein2014PolymerPhysics}.

In the mean field limit, we take the perspective that the free energy of formation for every morphology is the sum of two contributions: one from the internal core formed by the hydrophobic head groups and the other from the corona of hydrophilic polymer chains:
\begin{equation}
    \label{eq:mean_field_F}
    F_m(p) = F_m^\text{core}(p) + F_m^\text{corona}(p).
\end{equation}
Additionally, the free energy of an isolated molecule is taken to be ${F_{\rm mon} = 0}$ and thus the above core and corona contributions will be implicitly in reference to this free energy.
In the following subsections, we outline specific forms of the core and corona contributions to the free energy of formation for spherical, cylindrical, and planar aggregate morphologies before introducing the key dimensionless groups that emerge from these expressions.
We will then discuss how the resulting free energies of formation affect the morphological phase diagram and the transition from mesoscopic to macroscopic structures for inverse surfactant molecules.

\subsubsection{Core Free Energy} 
\label{sec:core}

We will determine the core free energy of spheres, fibers, and cylinders by describing the number of pairwise interactions between attractive rigid groups as a function of the core geometry.
It will later be useful to identify the geometric dimensions of the core of each morphology and their dependence on $p$. 
The geometry of a spherical core is solely characterized by its radius, $R_s$, while a fiber is characterized by the length of its long dimension, $\ell$, and the diameter of its short dimension, $d$.
Finally, the planar structures are assumed to be circular with a radius $R$ and a finite thickness of $\delta$. 

The core free energy depends on the number of distinct pairwise interactions between neighboring head groups in the aggregate core.
We assume the head groups adopt a dense packing arrangement within the core to maximize the number of attractive interactions.
That molecules must adopt arrangements that keep the hydrophilic polymer tails outside the core places natural constraints on the number of molecules in an aggregate.
For example, the maximum number of particles in a spherical core is set by the Newton number, which states that up to $12$ non-overlapping spheres of equal size can pack into a larger sphere in $3$ dimensions, with none of the smaller spheres being buried inside the core~\cite{Conway1999SphereGroups, Pfender2004Kissing2}.
While the precise value of this upper bound depends on the head-group geometry, for any finite ratio $r_h/r_t$ there necessarily exists a maximum sphere size beyond which spherical packing becomes energetically unfavorable or geometrically impossible, since adding further surfactants would induce substantial head--tail overlap or force a departure from spherical symmetry.
In the limit of  $r_h/r_t\rightarrow \infty$, the tail group becomes inconsequential and three-dimensional aggregation of unbounded extend (including phase separation) is permitted.
In this work, we constrain ourselves to finite $r_h/r_t$.
We define the set of possible sizes for each morphology as ${P_\text{sphere} = \{p \in \mathbb{R} \mid 2 \leq p \le 12\}}$, ${P_\text{fiber} = \{p \in \mathbb{R} \mid p > 12\}}$, and ${P_\text{planar} = \{p \in \mathbb{R} \mid p > 12\}}$, where we have continued to approximate $p$ as a continuous variable.

We note that aggregates such as dimers, trimers, etc. are grouped under the umbrella of our ``spherical'' morphology for convenience.
Moreover, by limiting the maximum sphere size to $12$, we have implicitly assumed that the polymer energetics are suitably strong to prevent the transition from a finite-size sphere to macroscopic spherical aggregates (i.e.,~phase separation).
Our perspective thus applies to systems in which the head and tail strongly contrast in solvent affinity such that the tail must be exposed to solvent.  
Rather than limiting the maximum sphere size, one could allow spheres of arbitrary size but include a steeply repulsive free energy of formation for larger spheres that reflects these energetic considerations. 

To estimate the core free energy we now describe the number of distinct interactions between neighboring head groups for each morphology.
For the spherical morphology, all head groups are assumed to be in close contact, resulting in $p (p-1)/2$ distinct interacting pairs and a core free energy of $-\epsilon p (p-1)/2$.
We can estimate the radius of the sphere by assuming that the packing fraction of the spherical heads takes the form ${\eta_{3D} = p (r_h/R_s)^3}$.
In detail, the packing fraction of a finite number of close-packed spheres will depend on the precise packing arrangement and the number of spheres. 
Here, we neglect these details (which we have found to be quantitatively negligible) and simply assume $\eta_{3D}$ to be a constant that is independent of $p$.
This allows us to express the radius of a spherical core as ${R_s(p) = r_h (p/\eta_{3D})^{1/3}}$ where we emphasize that this radius is a function of the aggregate size, $p$. 

We take the perspective that fibers and planar aggregates are formed by the one- and two-dimensional aggregation of monodisperse units of size $p_s$, respectively.
This perspective is informed by our coarse-grained molecular dynamics simulations in which the cores are found to have a thickness similar to that of spherical micelles (see the representative snapshots in Fig.~\ref{fig:simulation phase diagrams}). 
Accordingly, within these units, each head group interacts with its neighbors with a total energy similar to that of a spherical core.  
We will distinguish between units at the ends of a structure and those in the interior, with the latter anticipated to have a deeper energy.
A fiber should recover the spherical energy as $p \rightarrow p_s$ and thus the energy should approach $-\mathcal{E} = -\epsilon p_s (p_s-1)/2$.  
In this limit, the ``fiber'' can be thought to consist of two spherical endcaps (thus forming a sphere) and we thus take $-\mathcal{E}$ to represent the energy of the ends of the fiber. 
As the fiber grows we expect a linear increase in the core energy proportional to number of interior units with $-\mathcal{E}_f(p/p_s-1)$.
We have introduced $\mathcal{E}_f$, the magnitude of the core energy associated with each interior fiber unit.
We anticipate that the head groups are likely able to adopt more favorable packing arrangements in the fiber's interior, with $\mathcal{E}_f \gtrsim \mathcal{E}$.
The fibers considered here will have a diameter of ${d = 2R_s(p=p_s)}$ and length ${\ell(p) = (p/p_s)d}$ where we emphasize that only the length is $p$ dependent.

We consider planar structures, much like fibers, to be formed of units of size $p_s$ but now spanning two dimensions. 
Edge effects are more pronounced for these planar structures, with the number of edge units proportional to $p^{1/2}$ while those in the interior scale linearly with $p$. 
To ensure that we again recover the spherical energy as $p\rightarrow p_s$, we estimate the edge contribution to the core energy as $-\mathcal{E}(p/p_s)^{1/2}$ while that arising from the interior as $-\mathcal{E}_p(p/p_s-(p/p_s)^{1/2})$.
Here, $\mathcal{E}_p$ is the magnitude of the energy of an interior unit of size $p_s$ within the planar structure. 
We can define the number of interior and edge units as ${I(p) \equiv p/p_s - (p/p_s)^{1/2}}$ and ${E(p) = (p/p_s)^{1/2}}$, respectively.
The total planar core free energy is then simply ${-\mathcal{E}_pI(p) - \mathcal{E}E(p)}$. 
The thickness of the short-dimension follows from the size of the individual spheres units which we take to be ${\delta = 2R_s(p=p_s)}$ and the radius of the circular planar structure takes the form ${R= \delta (1 + (p/p_s-1)^{1/2})/2\eta_{2D}}$ where $\eta_{2D}$ is the two-dimensional packing fraction of the projected area of the units packed in the planar structure.
The precise values of $\eta_{2D}$ will depend on the particular packing arrangement and $p$ for finite-sized aggregates but we will again take it to be a constant.
This form for $R$ and $\delta$ allow us to progressively depart from the spherical limit ($p=p_s$ and $R=\delta/2$) with increasing $p$. 

We summarize the core contribution to Eq.~\eqref{eq:mean_field_F} for each morphology below:
\begin{equation}
    \label{eq:core}
    F_m^{\rm core} (p) = 
    \begin{cases}
    -\epsilon p (p-1)/2, & m = {\rm sphere}, \forall p \in P_{\rm sphere}\\\\
    -\mathcal{E}_f(p/p_s-1) - \mathcal{E}, & m = {\rm fiber}, \forall p \in P_{\rm fiber}
    \\\\
    -\mathcal{E}_pI(p) - \mathcal{E}E(p), & m = {\rm planar}, \forall p \in P_{\rm planar}.
    \end{cases}
\end{equation}
For the remainder of this Article, we will set ${p_s= 12}$ corresponding to the minimum in the free energy per molecule of the spherical morphology.
This choice of $p_s$ results in ${\mathcal{E} = 66 \epsilon}$.
We set ${\eta_{3D} = 0.4}$ informed by proofs of the densest packing arrangements of equally-sized spheres confined to a spherical volume~\cite{Schutte1951AufPlatz}.
For planar structures, we set ${\eta_{2D} = \pi/(2\sqrt{3})}$, consistent with a 2-dimensional hexagonal packing arrangement of monodisperse disks. 

Before proceeding to discuss the corona free energy it is worth emphasizing the importance of $\mathcal{E}_f$ and $\mathcal{E}_p$.
If head groups in the interior of planar and fiber structures pack in precisely the same way as they do in spheres then, within our model, we would have $\mathcal{E}_p = \mathcal{E}_f = \mathcal{E}$.
In this scenario, there would be no driving force for head groups to form fiber or planar cores as the same core free energy can be achieved on a per molecule basis by forming a sphere of size $p_s$. 
These spherical states will ultimately be more favorable due the increased translational entropy of smaller aggregates.
It is thus necessary for $\mathcal{E}_p$ and $\mathcal{E}_f$ to be greater than $\mathcal{E}$ to drive the formation of planar and fiber aggregates.
Physically, this deeper core free energy in fibers and planar structures likely reflects the better packing arrangements of the head groups within these structures that may not be accessible in spherical cores.
This arrangements could be closer-packed or even crystalline arrangements and will depend on the details of the head group. 
For simplicity, we set $\mathcal{E}_p = \mathcal{E}_f = \alpha \mathcal{E}$ where we have introduced a phenomenological \textit{packing efficiency parameter}, $\alpha$. 
When $\alpha > 1$, the core of fibers and planar aggregates has a deeper free energy than those of spheres.
Moreover, because of our choice of setting the energy scale of the fiber ends and planar edge to $\mathcal{E}$, $\alpha > 1$ also penalizes these end effects relative to the interior.
To go beyond a phenomenological parameter, one would need to specify the precise packing arrangements of each morphology for a particular head group chemistry and use this to inform the construction of $\mathcal{E}_p$ and $\mathcal{E}_f$. 

\begin{figure*}
	\centering
	\includegraphics[width=1\textwidth]{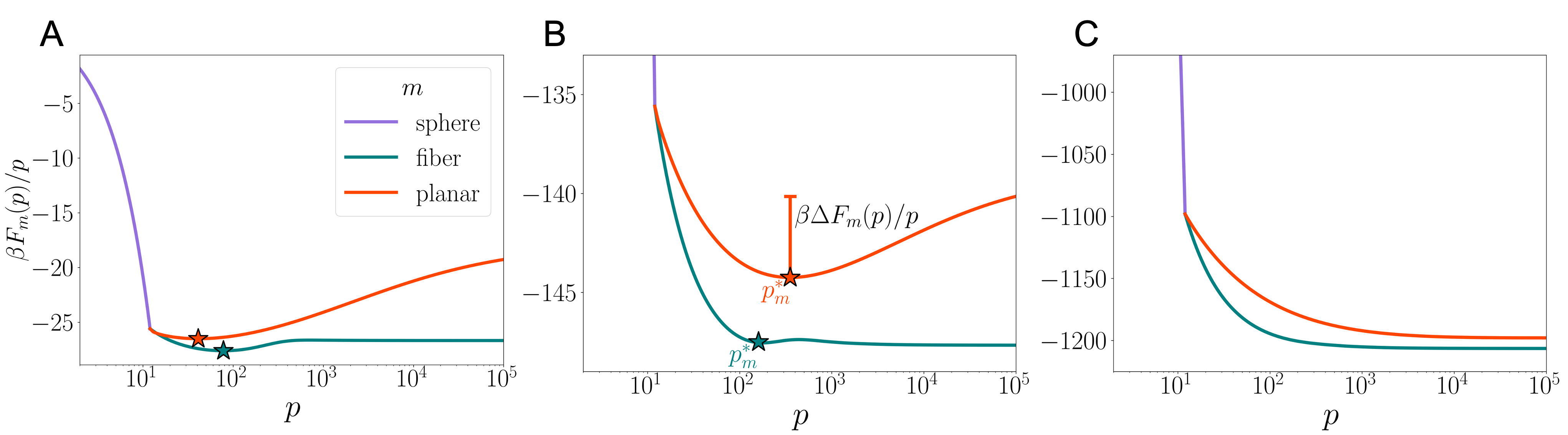}
	\caption{Per molecule free energies of formation for three representative values of $\beta \epsilon$ of (A) $5.0$, (B) $25.0$, and (C) $200.0$ with stars indicating the per-molecule minima for each morphology.
    All curves are shown for $\alpha = 1.1$ and $r_h/r_t = 0.06$. Stars denote the location of local minima.}
	\label{fig:free energies}
\end{figure*}

\subsubsection{Corona Free Energy} 
\label{sec:corona}

We take the perspective that the hydrophilic polymer chains surrounding the core can be treated as a polymer brush in good solvent conditions so that we may describe the corona free energy using concepts from polymer brush theory~\cite{Alexander1977AdsorptionDescription, DeGennes1976ConformationSolvents, deGennes1980ConformationsInterface, Daoud1982StarDependence, Rubinstein2014PolymerPhysics}.
We note that while the experimental system in Sections~\ref{sec:SANS} explored short tails (a $1140$ Da PEG chain features approximately $10$ Kuhn segments) the polymer scaling theory we use to find $F_m^\text{corona}$ for each morphology is better suited to describing systems with longer polymer chains.
We again emphasize that, while our models for the free energy of formation should only be taken as approximate, we are after qualitative trends for amphiphiles with inverted structures.

We now recapitulate the scaling theories for the brush free energy on flat~\cite{Alexander1977AdsorptionDescription, deGennes1980ConformationsInterface, DeGennes1976ConformationSolvents}, cylindrical~\cite{Pryamtisyn2009ModelingNanoparticlesc, Zhulina2005DiblockSolution}, and spherical~\cite{Daoud1982StarDependence, Witten1986ColloidPolymers, Witten1986MacrocrystalSolutions, Bug1987TheoryAggregates, Wang1988SizePacking, Pryamtisyn2009ModelingNanoparticlesc, Zhulina2005DiblockSolution} surfaces, which we will use to describe the free energy of the corona surrounding each morphology.
We omit the full derivation of the brush free energies here and refer the reader to the SI and the original works for further details.
Here, we recall that the free energy of chains grafted to a flat surface with grafting density $\sigma$ takes the form ${F^{\rm brush} =k_B T p \sigma^{5/6} v^{1/3} b^{2/3} N}$, where the planar brush has a height of ${H = \sigma^{1/3} v^{1/3} b^{2/3} N}$~\cite{Rubinstein2014PolymerPhysics}.
The chains stretch to reduce the overlap between neighboring chains and the interaction energy of the brush.
On a spherical surface with radius $R_0$, the brush free energy takes the following form:
\begin{equation}
    \label{eq:sphere_brush}
    \beta F^{\rm brush}_{\rm sph}(\sigma, R_0) = \dfrac{3}{5} p \sigma^{1/2} R_0 \ln\left( \dfrac{5 H}{3 R_0} + 1 \right),
\end{equation}
while on a cylindrical surface of radius $R_0$ we have:
\begin{equation}
    \label{eq:cylinder_brush}
    \beta F_{\rm cyl}^{\rm brush} (\sigma, R_0) = 2 p \sigma^{1/2} R_0 \left[\left(\dfrac{4 H}{3 R_0} + 1 \right)^{3/8} - 1\right],
\end{equation}
where we emphasize the dependencies of each free energy on $\sigma$ and $R_0$ for later convenience.
In both of the above expressions, $H$ remains the height of the brush on a flat surface and $H/R_0$ appears as a natural dimensionless group. 
In the low curvature limit ($H/R_0 \rightarrow 0$), both expressions recover the planar brush free energy (see SI).
Brushes in this limit have been referred to a ``crew-cut'' brushes~\cite{Zhulina2005DiblockSolution}, while coronas of micelles with larger curvatures are referred to as ``star-shaped'' brushes~\cite{Daoud1982StarDependence, Witten1986ColloidPolymers}. 
Physically, the curved surfaces reduce the degree of overlap between neighboring chains with increasing distance from the surface.
The reduced overlap between neighboring chains reduces the degree of chain stretching and overall free energy of the brush. 

We can use these brush free energy expressions to construct the corona free energy for each morphology. 
The grafting density for each morphology is simply the number of chains, $p$, divided by the surface area of the core. 
The relevant core surface area for a sphere, fiber, and planar structure are proportional to $R_s^2$, $d\ell$, and $R^2$, respectively. 
In the limit of small fibers and planar structures ($p\rightarrow p_s$) the morphologies are approximately spherical and, accordingly, the sphere brush free energy is appropriate with $R_0=d/2$ for fibers and $R_0 = R$ for planar aggregates.
As planar structures grow radially, the spherical brush free energy will naturally transition to the planar form as $H/R \rightarrow 0$ (see SI).
In contrast, growing fibers should transition from the spherical brush free energy to the cylindrical free energy as $d/\ell \rightarrow 0$. 
We thus model the sphere and planar corona free energies using Eq.~\eqref{eq:sphere_brush} while the fiber corona interpolates between Eq.~\eqref{eq:sphere_brush} and Eq.~\eqref{eq:cylinder_brush}.  

The corona free energies can be summarized as:
\begin{widetext}
    \begin{equation}
        \label{eq:chain entropy}
        F_m^{\rm corona} (p) = 
        \begin{cases}
            F_{\rm sph}^{\rm brush} (\sigma = p/4\pi R_s^2, R_0 = R_s)& m = {\rm sphere}, \forall p \in P_{\rm sphere},\\\\
            F_{\rm cyl}^{\rm brush}(\sigma = p/\pi d \ell, R_0 = d/2 ) w(p) + F_{\rm sph}^{\rm brush}(\sigma = p/\pi d^2, R_0 = d/2)[1 - w(p)] &  m = {\rm fiber}, \forall p \in P_{\rm fiber},\\\\
            F_{\rm sph}^{\rm brush} (\sigma = p/4\pi R^2, R_0 = R) & m = {\rm planar}, \forall p \in P_{\rm planar}.
        \end{cases}
    \end{equation}
\end{widetext}
While the combined area of the two larger sides of our circular planar structures is $2\pi R^2$, it is convenient to use an area of $4\pi R^2$ in defining the planar grafting density so that we precisely recover the spherical corona free energy as $p\rightarrow p_s$. 
We have introduced a switching function, ${w(p) = \left(1 + \tanh[(p - a)/\Delta]\right)/2}$, that allows us to interpolate between the spherical and cylindrical brush free energies at a transition size $a$.
Throughout this Article, we take ${a = 120}$ corresponding to fibers comprised of ten linearly connected spheres of size $p_s=12$.
We choose $\Delta = 200$ for a smooth transition between the spherical and cylindrical brush free energies.

\subsubsection{Free Energies of Formation: Summary}

It is important to emphasize that, for every morphology, the core contribution to the aggregate free energy $F_m (p)$ is strictly negative and thus will \textit{always} drive aggregation while the corona contributions are strictly positive.
Having established the functional forms of each contribution, 
we may identify the dimensionless groups that allow us to understand how varying the molecular parameters affects the morphological phase diagram.
Our molecular model now contains seven total parameters (the thermal energy in addition to the six molecular parameters: $r_h, \epsilon, N, b, v,$ and $\alpha$). 
Our aim is to identify the dimensionless groups common to each morphology, and thus morphology-specific parameters that we hold fixed will be excluded from this analysis (e.g.,~$\eta_{2D}$ and $\eta_{3D}$). 

We can first appreciate that the core and corona free energy for each morphology scales linearly with $\epsilon$ and $k_BT$, respectively.
Moreover, the core free energy depends on a single dimensionless parameter, $\alpha$, while the corona free energy depends on $H/R_0$.
Straightforward substitutions allow us to identify that ${H/R_0 \propto (r_t/r_h)^{5/3}}$, where again ${r_t = v^{1/5} b^{2/5} N^{3/5}}$ is the size of the isolated chain (the tail group) in a good solvent and $r_h$ is the size of the head group.
The corona free energy for each morphology (see SI) thus depends on a single dimensionless molecular parameter, $r_h/r_t$.
We thus have that the dimensionless formation energy takes the form ${\beta F_m = \epsilon \overline{F}_m^{\rm core}(\alpha) + \overline{F}_m^{\rm corona}(r_h/r_t)}$ where $\overline{F}_m^{\rm core}$ and $\overline{F}_m^{\rm corona}$ are dimensionless morphology dependent functions. 
The free energy of formation for each aggregate is thus solely controlled by $\beta\epsilon$, $\alpha$, and $r_h/r_t$. 

While the interpretation of $\beta\epsilon$ is clear and $\alpha$ was discussed when it was introduced, the physical importance of $r_h/r_t$ merits additional discussion. 
The area per head group on the surface of a micelle is proportional to $r_h^2$ while the projected area of the chain using its isolated size is proportional to $r_t^2$.
We can in fact identify that the grafting density takes the form $\sigma \propto 1/r_h^2$ while the \textit{overlap grafting density}---the grafting density at which neighboring chains will begin to overlap---is $\sigma^*\propto 1/r_t^2$.
When $(r_h/r_t)^2 \equiv (\sigma^*/\sigma)< 1$, the chains in the corona will stretch to avoid overlapping.
Chains on micelles with curved surfaces will experience less overlap (resulting in less unfavorable chain stretching) and thus decreasing $(r_h/r_t)^2$ will naturally promote the formation of fibers and spherical aggregates. 
In contrast, increasing $(r_h/r_t)^2$ reduces chain overlap by reducing the grafting density, and consequently reduces the bias towards structures with curved surfaces. 
This quantity is similar to Israelachvili's packing parameter, which also is a ratio of head group and tail group areas and can serve as a useful heuristic for predicting morphologies~\cite{Israelachvili1975AMembranes, Israelachvili1975TheoryBilayers, Israelachvili2011IntermolecularForces}.

To summarize, while we have seven total parameters, the free energy of formation of each morphology depends on three dimensionless quantities: ${\beta \epsilon}$, $\alpha$, and ${r_h/r_t \equiv r_h/(v^{1/5} b^{2/5} N^{3/5})}$.
We are now prepared to understand the impact of these dimensionless groups on the morphological phase diagram of inverse surfactants. 

Figure \ref{fig:free energies} displays the free energy of formation per molecule for each aggregate morphology as a function of the head group attraction, $\beta \epsilon $.
In the SI, we display the explicit dependencies of the formation energies on the two other molecular parameters, $\alpha$ and $r_h/r_t$, as well as the distinct core and corona contributions.
The spherical free energy, by construction, does not extend beyond $p=12$, which also coincides with the lowest sphere free energy. 
In contrast, the curves for the fiber and planar free energies begin at ${p=p_s=12}$, coinciding with the spherical free energy, and extend to arbitrarily large values of $p$.
For fibers and planar aggregates, the free energy \textit{per molecule} asymptotes to a finite value as $p \rightarrow \infty$.

\subsection{Theoretical Phase Diagrams}
\label{sec:PhaseDiagrams}

With the free energies of formation and classical aggregation theories, we are prepared to construct our phase diagrams.
The tunable dimensionless quantities are the overall volume fraction, $\Phi$, and the three molecular parameters, $\beta\epsilon$, $\alpha$ and $r_h/r_t$. 
We will first use the criteria found in Eq.~\eqref{eq:stability} to determine the regions of the phase diagram where we expect \textit{macroscopic} fibers and planar structures.
We will then construct the phase diagram for the finite-size region by minimizing our mean-field free energy [Eq.~\eqref{eq:free_energy_mf}] subject to constraint of constant $\Phi$ to obtain the most probable values of each morphology's volume fraction and size.

\subsubsection{Mesoscale-Macroscale Transition}

We can determine the regions in the parameter-space where we expect \textit{macroscopic aggregates} to form by determining when finite aggregate sizes no longer minimize the per molecule formation energy [see Eq.~\eqref{eq:stability}].
As previously mentioned, our model only allows for planar and fiber structures to undergo a macroscopic transition as a sphere with $p>12$ cannot exist in a close-packed state without a high degree of unfavorable head-tail interactions.  
At low attraction, the fiber and planar morphologies each have a distinct and finite aggregate size that minimizes the per molecule formation energy, as shown in Figs.~\ref{fig:free energies}A .  
We denote these minimizing sizes as $p_m^*$.
Aggregate sizes near these finite $p_m^*$ will indeed be stable solutions, in accordance with Eq.~\eqref{eq:stability}.
Moreover, these aggregate sizes represent the \textit{global minimum} in the per molecule formation energy and thus finite-sized aggregation is expected to be the globally stable solution. 

With increasing $\beta \epsilon$, we see a shift towards higher $p_m^*$ and a diminishing difference between the free energy minimum at $p_m^*$ and the asymptotic free energy as $p\rightarrow \infty$.
In Fig.~\ref{fig:free energies}B, we see that while both structures continue to feature a local minimum, the minimum is no longer a global minimum in the per molecule formation energy in the case of fibers. 
At these conditions, increasing the aggregate size beyond $p_{\rm fib}^*$ we encounter a small barrier (local maximum) in the free energy and subsequently find that further increasing $p$ results in a free energy that asymptotically approaches a value that is lower than that at $p_{\rm fib}^*$.
Here, finite-sized fibers are \textit{metastable} in comparison to \textit{macroscopic} fibers.
Eventually, a large enough head group attraction or head group size (see SI for details on how $r_h/r_t$ affects the formation energy) entirely eliminates the minimum at finite values of $p$ for both fibers and planar aggregates [see Fig.~\ref{fig:free energies}C].
In the absence of minima at finite sizes, aggregate sizes are only limited by the effects of translational entropy and the amount of material in the system (and, in practice, kinetics). 
It is at these conditions where we expect to observe \textit{macroscopic} aggregates; for lower attractions, where $p_m^*$ is finite, we expect finite-sized or \textit{mesoscale} aggregate sizes.

To understand why a global minimum at finite values of $p$ can be eliminated, we now examine the dependencies of the core and corona contributions to the formation energy. 
It is important to appreciate that $F_m^{\text{core}}/p$ monotonically decreases with $p$ for both fibers and planar structures as, for $\alpha > 1$, the relative importance of penalizing edge effects is diminished [see Eq.~\eqref{eq:core} and SI figures]. 
The local minima in the per molecule formation energy arises when the corona effects are appreciable enough to compete with the core contribution. 
Physically, as a fiber or planar aggregate grow, the transition from spherical to extended shapes results in additional polymer overlap which increases the corona free energy. 
This pushes aggregates to an optimal size that balances both contributions to the core and corona contributions to the free energy. 
Increasing $\beta\epsilon$ naturally diminishes the relative importance of the corona contribution---which is necessary for forming stable finite-sized aggregates---and eventually results in the loss of a minimizing finite aggregate size. 
Similarly, increasing $\alpha$ acts to increasing the magnitude of the core contribution and results in a similar trends as increasing $\beta\epsilon$.
We can also appreciate that increasing $r_h/r_t$ reduces the corona contribution to the free energy by increasing the spacing between tail groups on the micelle (reducing chain overlap and consequently chain stretching).
The transition from mesoscale to macroscopic aggregates can thus be induced by increasing $\beta \epsilon$, $\alpha$, and/or $r_h/r_t$ which all act to reduce the relative contribution of the corona free energy relative to that of the core.

The conditions for determining the precise parameters at which the mesoscale-macroscale transition occurs can be found beginning from the stability criteria of Eq.~\eqref{eq:stability}.
A stable minimum will have a vanishing derivative of $F_m/p$ with aggregate size and a positive curvature. 
We observe that for both fibers and planar aggregates, local minima can no longer be found for large head group attraction (as well as large $r_h/r_t$ and $\alpha$ as detailed in the SI). 
As we discussed earlier, $p^*_m$ can represent a metastable local minimum with macroscopic aggregates representing the globally stable state.
We can identify the location at which the globally stable state shifts towards macroscopic aggregates by defining the ``well depth'' of the free energy as the difference between the free energy per molecule at $p_m^*$ and that for an infinitely large aggregate as:
\begin{equation}
\label{eq: cutoff}
   \Delta F_m(p)/p = \bigg( \lim_{p \rightarrow \infty} F_m(p)/p \bigg) - F_m(p_m^*)/p_m^*.
\end{equation} 
When this energy difference is positive, finite-sized aggregates minimize the free energy per molecule.
We refer to the parameter space in which ${\Delta F_m(p)/p > 0}$ as the ``finite-size'' or ``mesoscopic'' regime.
Assembly under these conditions is also referred to as ``self-limited growth''~\cite{Hagan2021EquilibriumAssembly}.
As the energy difference decreases, the driving force to form finite-size aggregates of size $p_m^*$ in comparison to aggregates of macroscopic extent diminishes. 
Eventually, when the well depth is less than zero, aggregates can arbitrarily lower their per-molecule formation energy towards the asymptotic value with increasing size.
We thus define the \textit{mesoscale-macroscale transition} to occur when $\Delta F_m(p)/p = 0$.

\begin{figure}
	\centering
	\includegraphics[width=.48\textwidth]{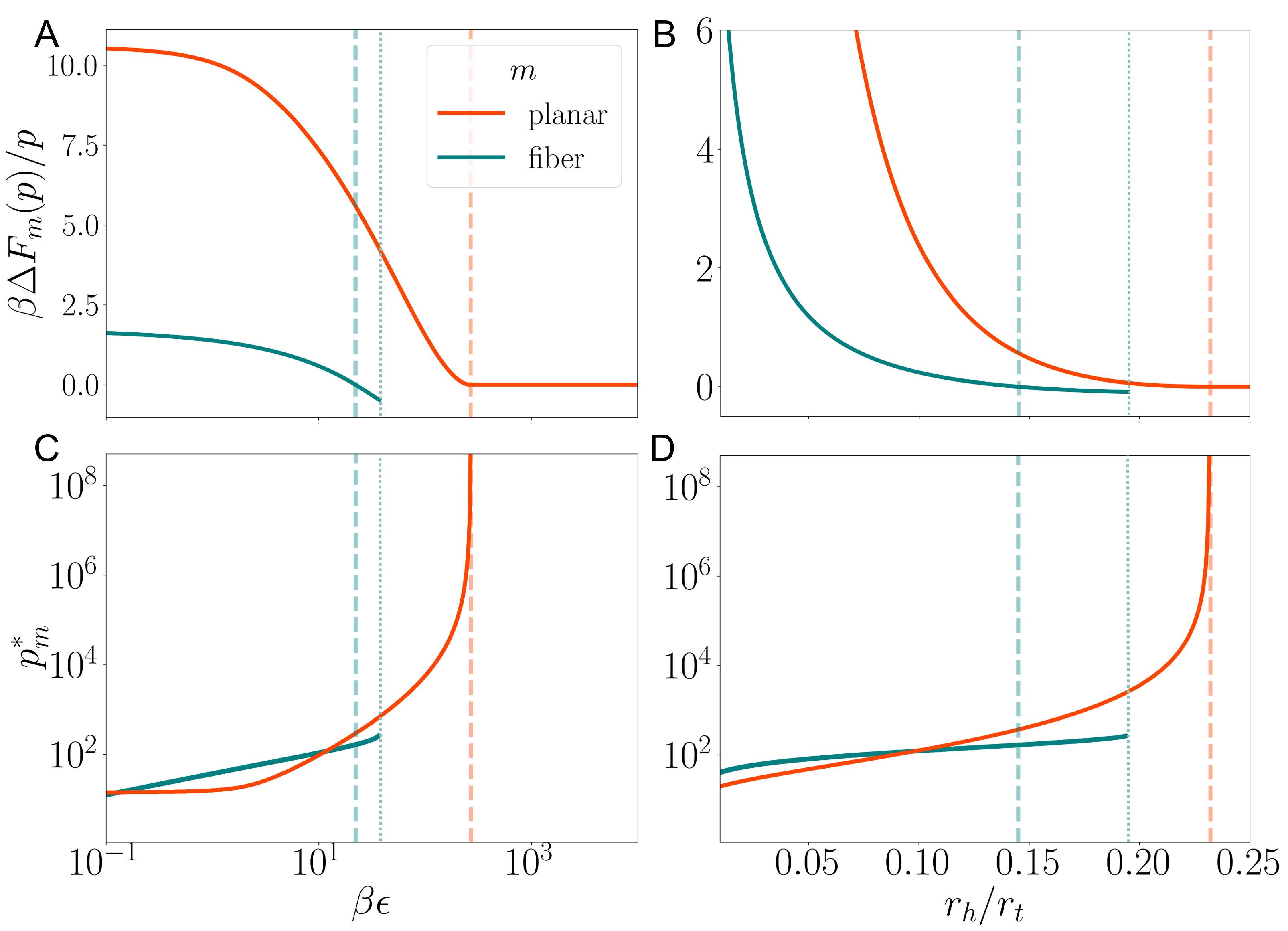}
    \caption{Dependence of the fiber and planar well depth on (A) $\beta\epsilon$ (for ${r_h/r_t = 0.06}$) and (B) $r_h/r_t$ (for ${\beta \epsilon = 7.0}$). Panels (C) and (D) show the minimizing aggregate sizes corresponding to panels (A) and (B), respectively.
    All plots are made using $\alpha = 1.1$.
    The dashed vertical orange and green lines represent the location of the mesoscopic-macroscopic transition for planar and fiber structures, respectively.
    The space between the dotted and dashed green lines marks the region for which the well depth is negative and the finite-sized fibers are metastable.
    At conditions in which local minima are absent, Eq.~\eqref{eq: cutoff} is ill defined and we report no value for it.
    }
	
	\label{fig: meso_macro}
\end{figure}

Figure~\ref{fig: meso_macro} displays the well depth, ${\Delta F_m (p)/p}$, and the aggregate size that minimizes the free energy of formation per molecule, $p_m^*$, as a function of $\beta \epsilon$ and $r_h/r_t$ for the fiber and planar morphologies.
As the dependence on $\alpha$ is similar to that of $\beta \epsilon$, the $\alpha$ dependence is shown in the SI.
Increasing both $\beta \epsilon$ and $r_h/r_t$ is found to decrease the well depth (Figs.~\ref{fig: meso_macro}A and B) and increase $p_m^*$ (Figs.~\ref{fig: meso_macro}C and D) for both morphologies.
The point at which $\Delta F_m (p)/p$ reaches $0$ marks the onset of the macroscopic regime, beyond which aggregate growth becomes unbounded as increasing size continues to decrease the free energy per molecule.
For fibers, we can see that a finite local minimum remains present after the transition as previously discussed.
This occurs over a narrow region in which $\Delta F_m (p)/p < 0$ in Fig.~\ref{fig: meso_macro}A and B, indicating that finite-sized fibers are a metastable local minimum in the free energy per molecule.
The lifetime of these metastable finite-sized aggregates will likely be controlled by the shallow free energy barrier dividing the local minima from larger aggregates [see Fig.~\ref{fig:free energies}B].
The origins of this barrier are rooted in the transition from the spherical to the larger cylindrical corona brush free energy with increasing fiber length. 
Whether this is a robust feature of fiber assembly or specific to our particular model for the fiber corona free energy is an open question. 
For planar structures, we see that the mesoscale-macroscale transition coincides with the complete elimination of a local minimum and thus finite-sized planar aggregates are not even metastable beyond the transition point. 

The mesoscale-macroscale transition consistently occurs at lower head group attraction strengths and sizes for fibers in comparison to planar aggregates.
We attribute this to the difference in the form of the corona free energy between the two morphologies.
While the core contributions are similar, the corona contribution will always provide a great free energy penalty for planar structures in comparison with finite-curvature fibers.

\begin{figure}
	\centering
	\includegraphics[width=.48\textwidth]{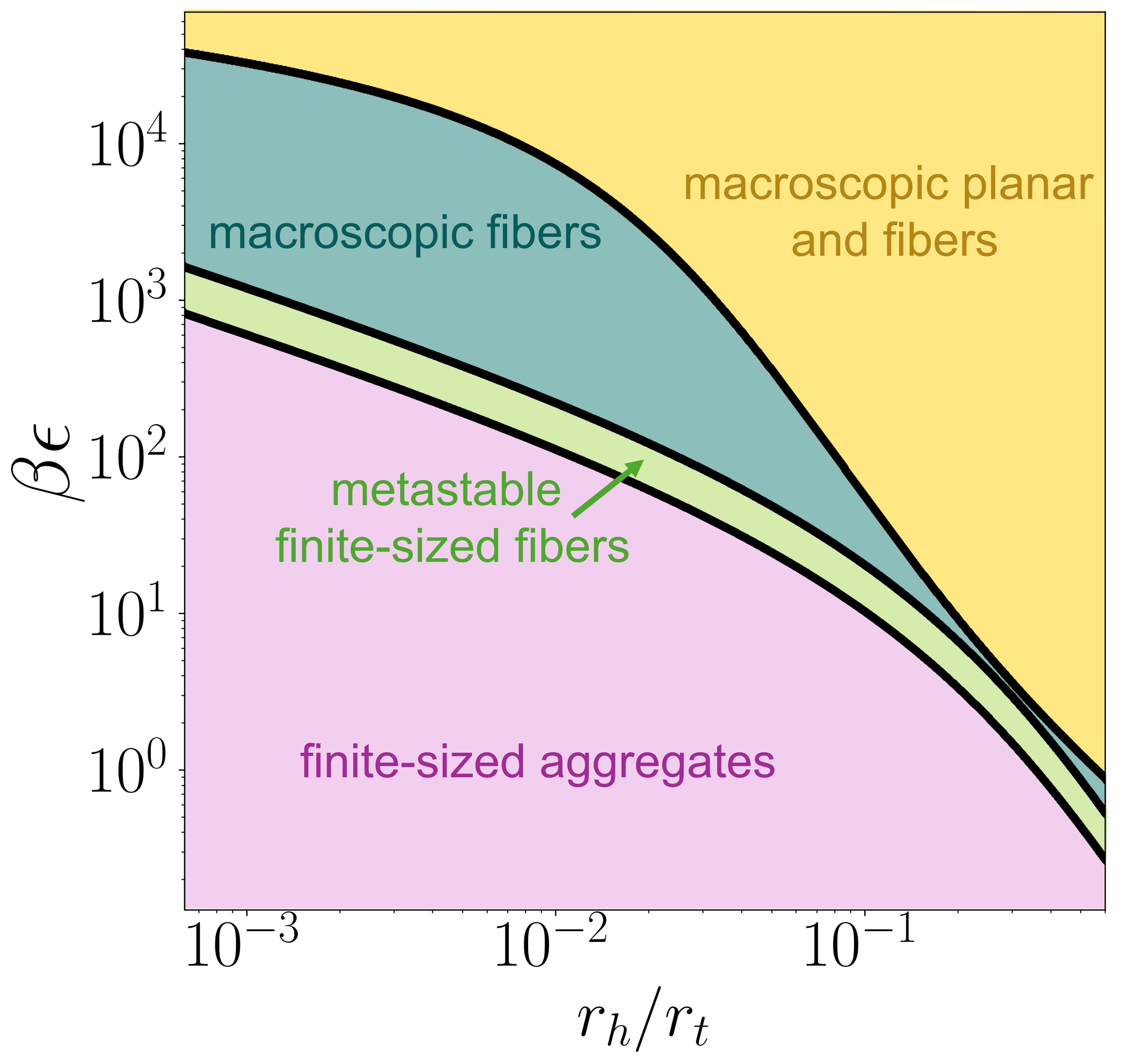}
	\caption{Morphological phase diagram describing the transition from mesoscopic to macroscopic structures at fixed $\alpha = 1.1$. 
    The globally stable state in the \textit{metastable finite-sized} region is dominated by macroscopic fibers, as shown in the SI.}
	\label{fig:theory_MM}
\end{figure}

We can now describe the complete dependence of the mesoscale to macroscale transition on $\beta \epsilon$ and $r_h/r_t$.
Figure~\ref{fig:theory_MM} shows that small variations in $r_h/r_t$ dramatically affect the stability of macroscopic structures.
This is qualitatively consistent with both the experiments and coarse-grained simulations of inverse surfactants discussed in Sec.~\ref{sec:ModelSystem}, which revealed that varying the size of the head group while fixing the polymer chain length affected dramatic transitions in the aggregate morphologies.
In the SI, we show that increasing $\alpha$ also acts to promote the formation of large aggregates and the onset of macroscopic aggregation, much like $\beta \epsilon$.
For both planar and fiber morphologies, the transition from mesoscopic to macroscopic aggregates occurs at lower $\beta \epsilon$ with increasing $r_h/r_t$, as shown in Fig.~\ref{fig:theory_MM}.
For \textit{all} values of $r_h/r_t$, the macroscopic fiber transition occurs at lower $\beta \epsilon$ in comparison to that of planar aggregates, in agreement with our observation in Fig.~\ref{fig: meso_macro}.
Indeed, only in the limit of large $r_h/r_t$, where the curvature effects responsible for this different are absent, does the mesoscale-macroscale transition of both structures coincide.
As the size ratio of the head and tail groups approaches one, the transition line occurs at head group attraction strengths less than the thermal energy.
It is important to emphasize that, at these conditions, the translational entropy effects, which do not enter our definition of these transitions, would prevent the observation of macroscale aggregates upon optimizing the complete system free energy.
One may thus continue to see mesoscale aggregates beyond the curves displayed in Fig.~\ref{fig: meso_macro}, particularly at low values of $\beta\epsilon$ and $\Phi$.

\subsubsection{Finite-size Region}

\begin{figure*}
	\centering
	\includegraphics[width=.95\textwidth]{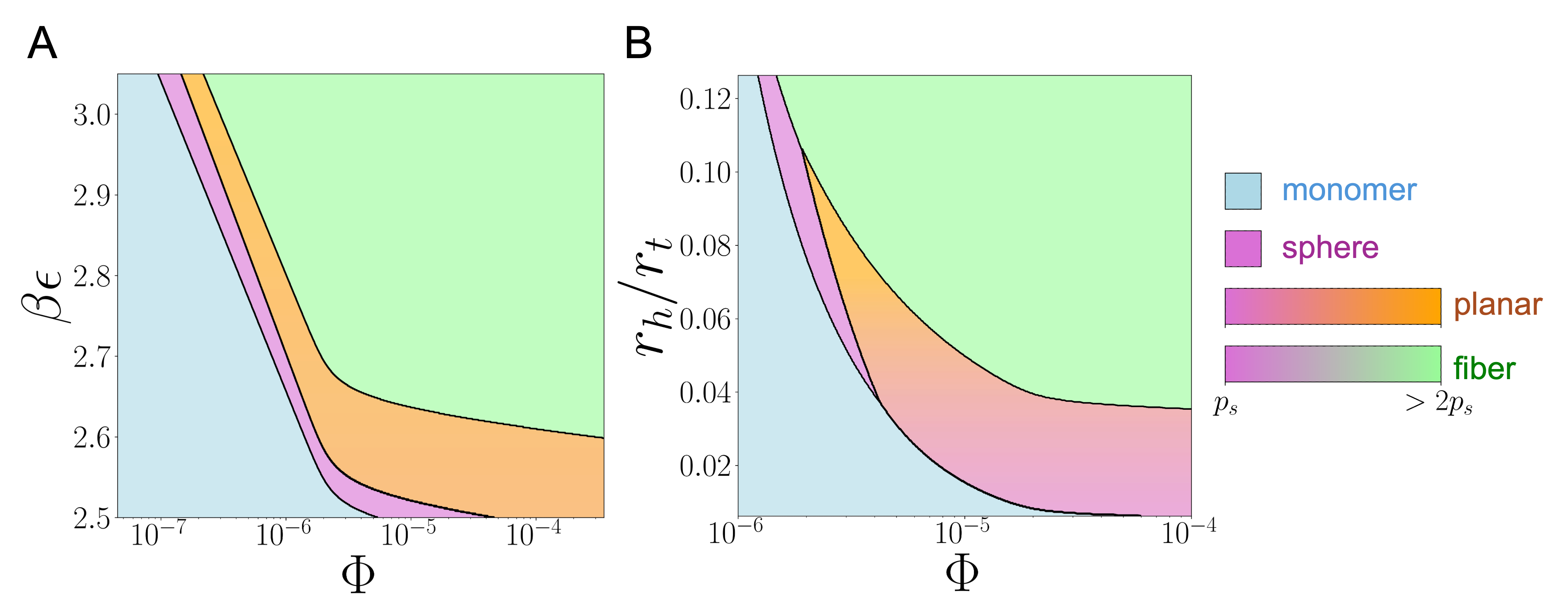}
	\caption{Morphological phase diagrams showing the dominant morphology as a function of $\Phi$ and (A) $\beta \epsilon$ at fixed $r_h/r_t = 0.06$ and (B) $r_h/r_t$ at fixed $\beta \epsilon = 2.5$.
    The molecular packing efficiency $\alpha = 1.1$ . The effects of varying $\alpha$ are provided in the SI, along with the volume fractions of each morphology for the phase diagrams provided above.
    The black lines represent conditions at which morphologies have equal volume fractions.
    While the monomer and sphere aggregation numbers are fixed (at $1$ and $p_s$, respectively) the fiber and planar sizes vary. 
    We use a color gradient to emphasize the departure of fiber and planar aggregation numbers from the spherical limit of $p_s$ to values in excess of $2p_s$.
    The complete fiber and planar size dependencies are presented in Fig.~\ref{fig:pstar}.} 
	\label{fig:theory_finite}
\end{figure*}

We now consider the dependencies of the morphology concentrations in the finite-size region of the phase diagram (i.e.,~in the lower left region of Fig.~\ref{fig:theory_MM} where macroscopic aggregates are not predicted) on $\beta\epsilon$, $r_h/r_t$, and $\Phi$. 
The conditions for finding the optimal concentrations $\{\phi_m^{\dagger}\}$ (four volume fractions, including that of the monomers) and sizes $\{p_m^{\dagger}\}$ (three total sizes) for our free energy [Eq.~\eqref{eq:free_energy_mf}] were provided in Eq.~\eqref{eq:coexist3}.
In practice, we can go beyond the mean-field limit and consider aggregates of a range of sizes for each morphology. 
We would simply need to equate the chemical potential of each aggregate considered. 
We nevertheless consider only the optimum aggregate size and further simplify by approximating the optimal aggregate size as the size that minimizes the per-molecule formation free energy, i.e.,~${p_m^\dagger \approx p_m^*}$.
Further numerical details are provided in the SI.

\begin{figure*}
    \centering
    \includegraphics[width=0.75\textwidth]{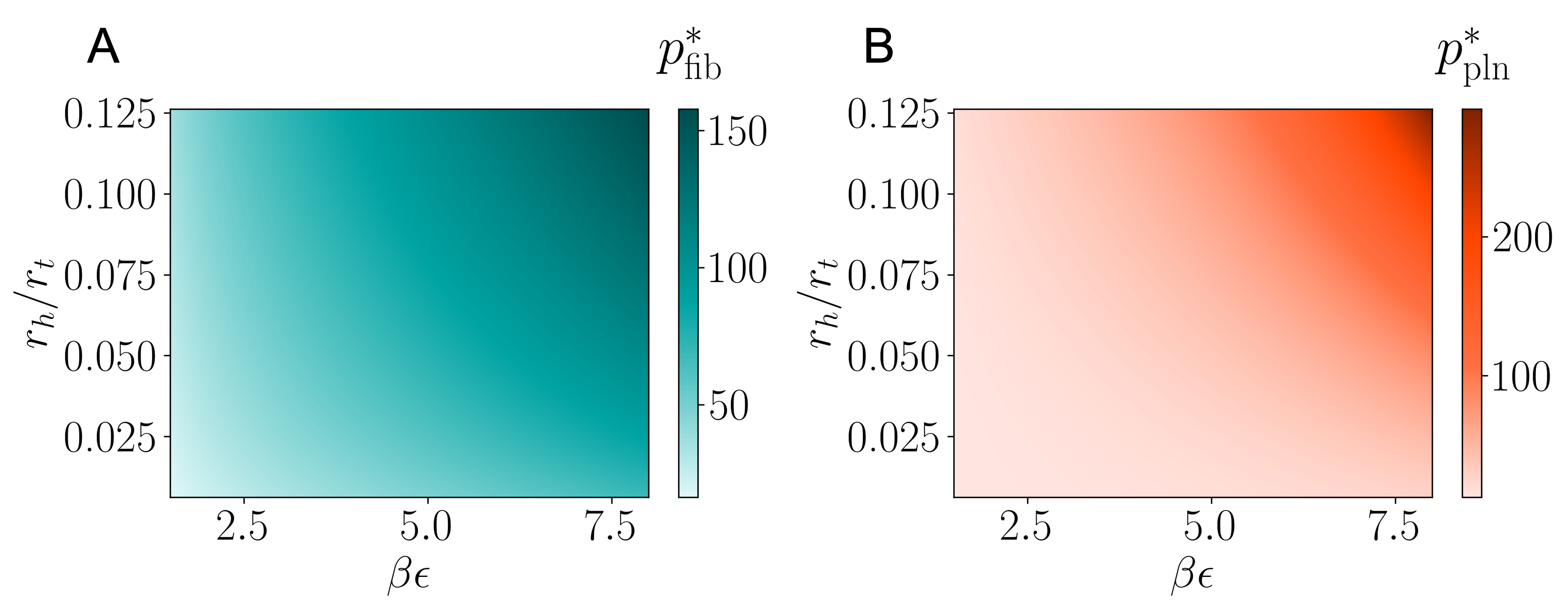}
    \caption{Effects of $\beta \epsilon$ and $r_h/r_t$ ($\alpha = 1.1$) on the size minimizing the free energy of formation per molecule for A) fiber and B) planar structures.}
    \label{fig:pstar}
\end{figure*}

Figure~\ref{fig:theory_finite} displays the morphological phase diagram as a function of $\Phi$, $\beta \epsilon$, and $r_h/r_t$.
In the SI, we report the precise composition of each morphology as a function of $\Phi$, $\beta \epsilon$, and $r_h/r_t$ as well as $\alpha$.
Here, we use the morphology with the largest volume fraction to label the regions shown in Fig.~\ref{fig:theory_finite}.
We emphasize that, particularly near the morphological boundaries, multiple morphologies can coexist with appreciable concentrations, as shown in the SI.
The boundaries shown represent conditions under which morphologies exist at equal volume fraction and only at conditions far from any boundaries does a single morphology truly dominate. 
Importantly, the $\beta \epsilon$ and $r_h/r_t$ dependence of the \textit{sizes} of the fiber and planar structures that minimize the respective per-molecule formation energies are shown in Figs.~\ref{fig:pstar}A and \ref{fig:pstar}B, respectively.
One can appreciate that in the limit of small $\beta \epsilon$ and $r_h/r_t$ there is little distinction between fibers and planar aggregates with spheres as their optimal sizes approach $p_s$. 
We thus color fibers and aggregates in Fig.~\ref{fig:theory_finite} as spheres when their aggregation number approaches $p_s$.

At fixed $r_h/r_t$, (Fig.~\ref{fig:theory_finite}A) monomers comprise the largest volume fraction at the lowest concentrations and head group attractions shown. 
This is consistent with the expectation that translational entropy favors a disaggregated state at weak attraction.
As $\Phi$ and $\beta \epsilon$ increase, the dominant morphology transitions to spherical structures.
This transition requires higher head group attraction with decreasing concentration. 
In this regime, spherical micelles provide a balance between planar aggregates and fibers, which may have a low formation energy but also have low translational entropy, and monomers.
This spherical micelle regime is relatively narrow and, with increasing attraction and concentration, gives way to a regime dominated by planar aggregates that is also relatively narrow, particularly at low concentrations. 
Upon modestly increasing concentration and head group attraction, we observe that fibers are the dominant morphology for the broadest range of conditions.

Figure~\ref{fig:theory_finite}B shows that the morphological states are also sensitive to the shape of the molecule.
While we broadly expect to find similar trends with $r_h/r_t$ as we did with $\beta \epsilon$, and we do indeed find that increasing $r_h/r_t$ promotes aggregation, there are some interesting differences. 
At low $r_h/r_t$ and $\Phi$, we again find that monomers dominate the system as expected.
Intriguingly, with increasing $r_h/r_t$, we observe a direct transition from a monomer-dominated regime to a planar aggregate regime.
We emphasize that at the monomer-planar boundary, spherical micelles are present at appreciable concentrations (see SI) but we only report the dominant structure in the regions labeled in both panels of Fig.~\ref{fig:theory_finite}.
Moreover, for small $r_h/r_t$, these ``planar'' structures have an aggregation number that is approximately $p_s$ and therefore, in effect, their morphology and formation energy is nearly identical to that of a spherical micelle.
Planar structures that are nearly spherical in shape have a slightly lower core formation energy compared to ``true spheres'' for $\alpha \gtrapprox1$ which is why they appear to be the preferred morphology.
It is only with increasing  $r_h/r_t$ (and $\beta \epsilon$) that both planar and fiber morphologies begin to appreciably deviate from the spherical limit and represent genuinely distinct morphologies.
At the largest concentrations and head group sizes, we again find the phase diagram to be dominated by fibers.

We now have a clear picture of the thermodynamic ground state for an inverse surfactant molecule that assembles into spheres, fibers, and planar aggregates, which we can compare to our simulation results.
Just as in simulation, we broadly find that aggregation is induced with increasing concentration ($\Phi$), head group attraction ($\beta \epsilon$), and the head-to-tail size ratio ($r_h/r_t$). 
More specifically, we find spherical and planar aggregates occupy a relatively narrow region between the monomer regime located at small $\Phi$, $\beta \epsilon$, and/or $r_h/r_t$ and a fiber regime located at large $\Phi$, $\beta \epsilon$, and/or $r_h/r_t$. 
The eventual dominance of fibers predicted by our theory is in agreement with our simulations and the experimental reports of fibers (including macroscopic fibers).
We note that this finding likely justifies, for the surfactants considered here, our earlier assumption that the packing arrangements of the head groups in both fibers and planar aggregates result in similar energies (i.e.,~$\mathcal{E}_p \approx \mathcal{E}_f$).
For other head groups, planar structures may offer more optimum packing arrangements which would drive the formation of these structures.

Our theory predicts that increasing the relative head-group size can promote the formation of planar aggregates which have aggregation numbers appreciably larger than spheres [see Fig.~\ref{fig:theory_finite}B].
The formation of planar structures that compete with spherical micelles just outside the monomer regime with increasing head group size was indeed found in our simulations (see Fig.~\ref{fig:simulation_phase_diagrams}).
That our relatively simple scaling theory can capture these details is notable and suggests that the salient physical features are captured by our theoretical perspective. 
Moreover, the theory is able to show that the head group size (or more precisely, $r_h/r_t$) also plays a crucial role in the transition from \textit{mesoscopic} to \textit{macroscopic} aggregates, an insight we could not obtain from simulations.

It is worth emphasizing that in our experiments, increasing the DPCA content in our head group results in an increase in $\beta \epsilon$ \textit{and} $r_h/r_t$.
Our theory predicts that increasing both of these parameters, as well as the overall concentration, should act to promote the formation of spherical micelles followed by fibers and planar aggregates.
These trends are qualitatively consistent with our SANS fits, displayed in Fig.~\ref{fig:sans}, and detailed in the SI.
We also note that while our theory, in principle, allows for coexistence between various morphologies, we nevertheless find that a single morphology often dominates the system.
The concentrations of the different morphologies at the same conditions in our morphological phase diagrams [Fig.~\ref{fig:theory_finite}] are shown in the SI.
In contrast, both our simulations and SANS experiments are consistent with broader regions of morphological coexistence.
It may be possible that the increased morphological heterogeneity in both simulation and experiment is the result of kinetic limitations inherent to self-assembly under dilute conditions.
These kinetic limitations are, of course, unaccounted for in our thermodynamic theory and would require a kinetic theory for molecular exchange between micelles~\cite{Lund2006LogarithmicMicelles, Choi2010MechanismLength}.

Finally, we can now also appreciate the impact of the PEG molecular weight on our phase behavior through Fig.~\ref{fig:theory_finite}.
The increased molecular weight examined here (\SI{1140}{\dalton}) compared to earlier works (\SI{750}{\dalton})~\cite{Cheng2019SupramolecularRegenerationc, DeFrates2022TheProdrugc} amounts to approximately a $29\%$ increase in $r_t$ [$(1140/750)^{3/5}$], which will generally reduce the sizes of fibers and planar structures (see Fig.~\ref{fig:pstar}) but is not expected to qualitatively alter the phase behavior outside of a narrow range of $\Phi$ (see Fig.~\ref{fig:theory_finite}).

\section{Summary and Conclusions}
\label{sec:conclusions}

We have outlined an approach to predicting inverse surfactant self-assembly that is informed by results from SANS experiments, computer simulations, and extensions of classical aggregation theory.
The SANS data determined that all of the molecules in our study can make a rich mosaic of morphologies including spherical, cylindrical, and planar structures.
These findings were qualitatively confirmed by our simulations.
Together, the experiments and simulations shown here reveal that aggregate morphology is sensitive to head group size.
Our theory uses concepts from spherical packing theory and polymer scaling to construct the free energy of formation for each morphology as a function of molecular parameters. 
These formation energies, combined with a classical aggregation theory (extended to allow for multiple competing structures), allow for the determination of the morphological phase diagram in both the mesoscale and macroscopic regions of parameter space.

Our theory corroborates our computational and experimental findings that the head group size plays a crucial role in the micelle morphology, with a quantity similar to Israelachvili's packing parameter naturally emerging in the corona free energies.
Importantly, the theory provides insight into the mesoscale to macroscale transition, the description of which naturally eludes simulation and experiment. 
This transition is driven by both the head-group attraction strength relative to the polymer entropy and the size ratio of the head to the tail; increasing either of these molecular parameters promotes the formation of larger structures.

It is our hope that the emerging thermodynamic picture offered by our work can serve as a basis for future investigations aimed to disentangle thermodynamic and kinetic effects.
While our perspective considered spherical, cylindrical, and planar aggregate morphologies, it can naturally be extended to more complex architectures by modifying the free energies of formation based on the considerations we provided.
We hope the insights offered in this work contribute to the growing literature on rational design principles for de novo self-assembled materials.

\acknowledgments
We thank Emma Vargo, Beihang Yu, and Tom Russell for helpful comments on the neutron scattering data.
N.A.S. acknowledges support from the NSF Graduate Research Fellowship under DGE 1752814 and DGE 2146752. 
Partial funding for this project was provided by National Institutes of Health grant R01DE021104 and the UC Berkeley Bakar Spark Fellows Award 2022.
This project used the Savio computational cluster resource provided by the Berkeley Research Computing program. 
Neutron scattering research was conducted under the general user proposal IPTS-32408.1 at the Bio-SANS CG-3 instrument (Center for Structural Molecular Biology), a DOE Office of Science, Office of Biological and Environmental Research resource (FWPERKP291), used resources at the High Flux Isotope Reactor, a DOE Office of Science, Scientific User Facility operated by the Oak Ridge National Laboratory.
Synthesis and characterization of deuterated polyethylene glycol was supported as part of a user project at the Center for Nanophase Materials Sciences (CNMS), which is a US Department of Energy, Office of Science User Facility at Oak Ridge National Laboratory.

\end{document}


\maketitle

\pagenumbering{gobble}

\setcounter{secnumdepth}{2}
\setcounter{tocdepth}{2}
{
	\hypersetup{
		linkcolor=Black,
		citecolor=Black
	}
	\vspace{-10pt}
	\tableofcontents
}
\newpage 

\pagenumbering{arabic}

\section{Experimental Methods}

\subsection{Synthesis of Deuterated Polyethylene Glycol}

\subsubsection{Materials and Methods}
All reagents were used as received from the suppliers without further purification unless otherwise noted. 
Purification of methanol (Sigma-Aldrich), tetrahydrofuran (Sigma-Aldrich) and ethylene oxide-\textit{d$_4$}  (Cambridge Isotope Laboratories) has been reported elsewhere \cite{Hadjichristidis2000AnionicTechniques}.
Naphthalene (Sigma-Aldrich) was sublimed twice under vacuum and collected in a glass ampule. 
Isopropyl alcohol-{\textit{d$_8$}} was purified with \ce{CaH2} for \SI{24}{\hour} and was distilled under vacuum in glass ampules.

\subsubsection{Synthesis of deuterated isopropoxy-poly(ethylene glycol) (PEG)}
The synthesis of low molecular deuterated PEGs was conducted \textit{via} anionic polymerization methods and high vacuum techniques and the introduction of all reagents was done through break seals in custom made glass reactors. 
Deuterated isopropoxy potassium (\ce{IPO$^-$K+-\textit{d$_7$}}) was used as the initiator for the synthesis of the deuterated PEG and the reaction was left under vigorous stirring at \SI{50}{\celsius} for \SI{2}{\day}.  
The synthesis of the initiator included the activation of the hydroxy group of isopropyl alcohol with potassium naphthalene which was obtained from the reaction of naphthalene with potassium mirror. 
The average molecular weight of the polymer was found to be \SI{1140}{\dalton} with a dispersity $\textit{\DJ} (M_w / M_n) = 1.03$ using MALDI-TOF and end-group analysis from \ce{$^{13}$C}-NMR. \ce{$^{13}$C{$^1$\{H\}}} NMR \ce{(CDCl3)}: $\delta 71.8$  (pent, $J_{CD}$ = \SI{21.4}{\hertz}, \ce{-\textit{C}D2CD2OCD(CD3)2}), $71.2$ (trip, $J_{CD}$ = \SI{21.2}{\hertz} Hz, \ce{-\textit{C}D(CD3)2}), $69.8$ (pent, $J_{CD}$ = \SI{21.3}{\hertz}, \ce{-O\textit{C}D2CD2-}), $66.6$ (pent, $J_{CD}$ = \SI{21.5}{\hertz}, \ce{-\textit{C}D2OCD(CD3)2}), $61.0$ (pent, $J_{CD}$ = \SI{21.1}{\hertz},  \ce{-\textit{C}D2OH}), $21.1$ (sept, $J_{CD}$ =  \SI{19.1}{\hertz}), \ce{-CD(\textit{C}D3)2}).

\subsubsection{Synthesis of deuterated isopropoxy-PEG-\ce{CO2H}}
The procedure followed was based upon that described by Sharma and co-workers \cite{Kumari2014SynthesisBehavior} for the oxidation of Mw $1000$ poly(ethylene glycol) to poly[ethylene glycol bis(carboxmethyl) ether] by \ce{KMnO4}, with several modifications.
To a \SI{500}{\milli\litre} round bottom flask containing a Teflon-coated stirbar was added the perdeuterated isopropoxy-PEG above (\SI{16.8}{\gram}, ca \SI{14}{\milli\mol}) and \SI{140}{\milli\litre} mL DI water.
After the PEG had dissolved, the solution was cooled in an ice-water bath, and NaOH pellets (\SI{2.24}{\gram}, \SI{56}{\milli\mol}) added with stirring.  \ce{KMnO4} (\SI{12.85}{\gram}, \SI{84}{\milli\mol}) was divided into 12 portions of ca. \SIrange{107}{108}{\milli\gram}, and each portion added at \SI{20}{\minute} intervals over a \SI{4}{\hour} period to the stirred solution at \SI{0}{\celsius}. 
The solution color turns from green to brown with sediment formation during this time. 
The ice-water bath was then removed, and the reaction mixture stoppered and allowed to warm to ambient temperature and stir for \SI{20}{\hour} in this manner. 
The brown suspension was filtered through Celite $545$, and the Celite pad washed with DI water (3 $\times$ \SI{100}{\milli\liter}). 
The washings were combined with the initial filtrate and the volume reduced to \SI{210}{\milli\litre} by rotary evaporation at \SI{55}{\celsius}.
As a small amount of suspended brown fines remained, the solution was filtered again through Celite and washed with DI water (2 $\times$ \SI{100}{\milli\liter}).
The pH of the filtrate was reduced to about pH $2$ by the addition of \SI{70}{\milli\litre} of $1$ N HCl. 
The aqueous solution was extracted once with \SI{250}{\milli\litre} dichloromethane. 
The dichloromethane solution was separated, dried through a column of anhydrous granular sodium sulfate, and the dichloromethane removed by rotary evaporation to afford ca. 
\SI{10}{\gram} of colorless viscous oil. 
Further drying under high vacuum afforded \SI{9.83}{\gram} ($58\%$) viscous oil which solidified to a waxy solid over time. 
\ce{$^{13}$C{$^1$\{H\}}} NMR \ce{(CDCl3)}: $\delta$ $172.0$ (s, \ce{\textit{C}=O}), $71.2$ (trip, $J_{CD}$ = \SI{21.6}{\hertz}, \ce{-\textit{C}D(CD3)2}), $70.3$ (pent, $J_{CD}$ = \SI{21.3}{\hertz}, \ce{-\textit{C}D2CD2OCD(CD3)2}) overlaps with $69.7$ (pent, $J_{CD}$ = \SI{21.1}{\hertz}, \ce{-O\textit{C}D2CD2-}), $68.3$ (pent, $J_{CD}$ = \SI{21.7}{\hertz}, \ce{-\textit{C}D2CO2H}), $66.5$ (pent, $J_{CD}$ = \SI{21.3}{\hertz},  \ce{-\textit{C}D2OCD(CD3)2}), $21.1$ (sept, $J_{CD}$ = \SI{19.1}{\hertz}), \ce{-CD(\textit{C}D3)2}).\\

\subsubsection{Nuclear Magnetic Resonance (NMR)}
NMR spectra were obtained at the Center for Nanophase Materials Sciences on a Bruker Avance NEO NMR console coupled to a \SI{11.74}{\tesla}  actively shielded magnet (Magnex Scientific/Varian) operating at \SI{499.717}{\mega\hertz} for proton. 
All spectra were acquired at \SI{298}{\kelvin} in \ce{CDCl3} (\SI{7.27}{\ppm} $^1$H reference and \SI{77.23}{\ppm} \ce{$^{13}C$} reference). 
Carbon NMR spectra were obtained using inverse-gated decoupling with pw = $90$ and a recycle delay of \SI{30}{\second}.

\subsubsection{MALDI-TOF}
Matrix-assisted laser desorption ionization time-of-flight (MALDI-TOF) mass spectra were acquired on a Bruker Autoflex Speed in positive ion reflection mode with a laser ($\lambda =$ \SI{337}{\nano\meter}). 
The sample solutions were prepared at \SI{2}{\milli\gram\per\milli\litre}, and the matrix trans-2 [3-(4-\textit{tert}-Butylphenyl)-2-methyl-2-propenylidene] malononitrile (DCTB) at \SI{2}{\milli\gram\per\milli\litre}.
The solution was mixed in a $1:1$ (matrix:sample) ratio and repeated vortex mixing to ensure dissolution. 
Aliquots of the solutions (\SI{1}{\micro\litre}) was deposited on the ground stainless steel target for analysis. 
The spectra were obtained by summing at $2000$ shots into the sum buffer prior to analysis.
The mass spectra were analyzed in Bruker Flex Analysis software.

\subsection{Synthesis of Amphiphilic Molecules}
\subsubsection{Materials}
All chemicals were obtained from suppliers and used without further purification unless otherwise noted. 
Chemicals used were 8-Aminoquinoline (TCI). Diethyl ethoxymethylenemalonate ($99$\%+, Acros Organics).
Diphenyl ether (ReagentPlus, $\geq99$\%, Sigma-Aldrich). 
Activated carbon (powder/USP, Fisher Chemical). 
Sodium hydroxide (NaOH; BioXtra, $\geq 98$\%, Sigma-Aldrich).
Hydrochloric acid (\ce{HCl}; $36.5-38.0$ \%, BioReagent, for molecular biology solutions, Sigma-Aldrich).
1,1’-carbonyldiimidazole (CDI; Oakwood Chemical). 
Poly(ethylene glycol) monomethyl ether (PEG-OH; Sigma Aldrich, $202495$; Mn=\SI{750}{\gram\per\mol}). 
(2,2,6,6-Tetramethylpiperidin-1-yl)oxyl (TEMPO; $98$\%, Sigma-Aldrich). 
Sodium bromide (NaBr; $ \geq 99.0$\%, Sigma Aldrich). 
Sodium hypochlorite solution (\ce{NaClO}; available chlorine $10-15$\%, Sigma Aldrich). 
2-Amino-2-(hydroxymethyl)propane-1,3-diol (TRIS; BioUltra, for molecular biology, $\geq 99.8$\% (T), Sigma-Aldrich). 
2-Amino-1,3-propanediol ($\geq 98$\%, Sigma-Aldrich). 
Ethanolamine ($\geq98$\%, Sigma-Aldrich). 
N,N-Diisopropylethylamine (DIPEA; ReagentPlus, $\geq 99$\%, Sigma-Aldrich).
N,N,N'N'-Tetramethyl-O-(\ce{1H}-benzotriazol-1-yl)uronium hexafluorophosphate (\ce{HBTU}; $\geq 99.0$ \%, Sigma-Aldrich). 
Sodium chloride (\ce{NaCl}; BioXtra, $\geq 99.5$ \% (AT), Sigma-Aldrich). 
Sodium sulfate (\ce{Na2SO4}; anhydrous for analysis EMSURE ACS, ISO, Reag. Ph Eur). 
Sodium hydride (\ce{NaH}; $60$\% dispersion in mineral oil, Sigma-Aldrich). 
Triethylamine (\ce{TEA}; $\geq 99.5$ \%, Sigma-Aldrich). 
Solvents used were Hexanes (ACS, VWR Chemicals BDH). 
Dichloromethane (DCM; ACS, AR, Macron Fine Chemicals). 
Diethyl ether (anhydrous, certified ACS, Fisher Chemical). 
N-N-Dimethylformamide (DMF; $99.8$\%, extra dry, anhydrous, SC, Thermo Scientific Chemicals).
Methanol (\ce{MeOH}; $\geq 99.8$\% ACS for production, VWR Chemicals BDH).
Ethanol (\ce{EtOH}; $200$ proof, USP, Koptec). 
Unless otherwise noted, water used in this study was ultrapure Milli-Q water at \SI{<0.05}{\micro\siemens\per\centi\meter}.

\subsubsection{Synthesis of DPCA and imidazole (Im)-DPCA}
DPCA and Im-DPCA synthesis procedures were adapted from prior work \cite{Cheng2019SupramolecularRegeneration, DeFrates2022TheProdrugc}. 
8-Aminoquinoline (\SI{12.5}{\gram}, \SI{87.7}{\milli\mol}) and diethyl ethoxymethylenemalonate (\SI{19.1}{\milli\litre}, \SI{95.37}{\milli\mol}) heated to \SI{100}{\celsius} under reflux with stirring for \SI{2}{\hour}. 
Warmed diphenyl ether (\SI{150}{\milli\litre}) was added to the reaction mixture and refluxed at \SI{250}{\celsius} for \SI{6}{\hour} and cooled to room temperature. 
The reaction mixture was washed with hexanes (\SI{400}{\milli\litre} twice) and DCM (\SI{400}{\milli\litre}) and vacuum filtered. 
This crude product was purified with activated carbon in DCM (\SI{1}{\gram} activated carbon in \SI{500}{\milli\litre} DCM per \SI{1}{\gram} product). 
Activated carbon was removed using vacuum filtration followed by three rounds of gravity filtration. 
Solvent was reduced under vacuum, and product was collected via vacuum filtration and washing with additional DCM and diethyl ether.
The resulting product (\SI{10}{\gram}, \SI{37.27}{\milli\mol}) was combined with \ce{NaOH} (\SI{11.19}{\gram}, \SI{279.68}{\milli\mol}) and water (\SI{500}{\milli\litre}) and refluxed at \SI{110}{\celsius} for \SI{2}{\hour}. 
Additional \ce{NaOH} (\SI{1}{\gram}) and water (\SI{100}{\milli\litre}) were added and refluxed for an additional \SI{2}{\hour} to dissolve remaining solids. 
The mixture was cooled to room temperature and vacuum filtered twice. 
Crushed ice and \ce{HCl} were added to reach pH $2$, and the solid product was washed with water, isolated by centrifugation, and dried under vacuum, yielding DPCA.

\begin{figure}
    \centering
    \includegraphics[width=1\linewidth]{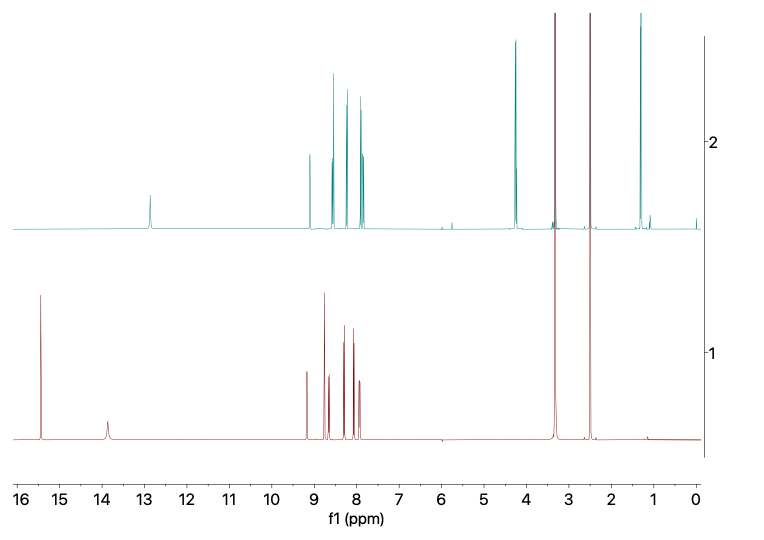}
    \caption{$^1$H NMR spectra of DPCA (bottom) and Im-DPCA (top)}
    \label{fig:NMR_DPCA}
\end{figure}

For coupling, DPCA (\SI{5}{\gram}, \SI{20.82}{\milli\mol}) was activated with \ce{CDI} (\SI{10.13}{\gram}, \SI{62.45}{\milli\mol}) in DMF (\SI{300}{\milli\litre}) at \SI{110}{\celsius} for \SI{3}{\hour}. 
Product was separated by centrifugation, washed with DMF twice (\SI{280}{\milli\litre}, \SI{200}{\milli\litre}) and with diethyl ether (\SI{320}{\milli\litre}), and dried under vacuum, yielding Im-DPCA.
\ce{1H} NMR (\SI{500}{\mega\hertz}, \ce{DMSO-d6}); 
H NMR spectra for the products is provided in Fig.~\ref{fig:NMR_DPCA}.
DPCA: $\delta 15.43$ (s, \ce{1H}), $13.84$ (s, \ce{1H}), $9.17$ (dd, $J=4.5$, \SI{1.6}{\hertz}, \ce{1H}), $8.75$ (s, \ce{1H}), $8.65$ (dd, $J=8.3$, \SI{1.6}{\hertz}, \ce{1H}), $8.28$ (d, $8.8$ Hz,\ce{1H}), $8.06$ (d, $J=$ \SI{8.8}{\hertz}, \ce{1H}), $7.93$ (dd, $J=8.3$, \SI{4.3}{\hertz}, \ce{1H}); Im-DPCA: $13.19$ (s, \ce{1H}), $9.15$ (dd, $J=4.3$, \SI{1.6}{\hertz}, \ce{1H}), $8.62$ (dd, $J=8.4$, \SI{1.7}{\hertz}, \ce{1H}), $8.45$ (d, $J=$\SI{2.7}{\hertz}, \ce{1H}), $8.27-8.21$ (m, \ce{2H}), $7.97$ (d, $J=$\SI{8.8}{\hertz}, \ce{1H}), $7.89$ (dd, $J=8.3,$\SI{4.3}{\hertz}, \ce{1H}), $7.68$ (t, $J=$\SI{1.5}{\hertz}, \ce{1H}), $7.07$ (dd, $J=1.7$, \SI{0.9}{\hertz}, \ce{1H}).

\subsubsection{Synthesis of PEG-COOH, \ce{PEG-(OH)x}, and \ce{d-PEG-(OH)x}}
\begin{figure}
    \centering
    \includegraphics[width=1\linewidth]{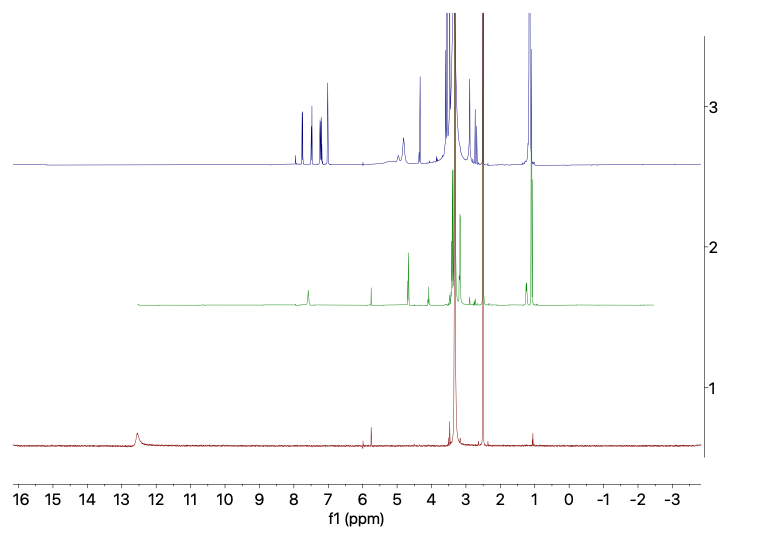}
    \caption{$^1$ {H NMR spectra of (bottom to top) \ce{d-PEG-COOH}, \ce{d-PEG-(OH)_1}, and \ce{d-PEG-(DPCA)_1}.}
    For \ce{$^1$H} NMR and MALDI-TOF MS of non-deuterated PEG-\ce{(OH)x}, see DeFrates et al.\cite{DeFrates2022TheProdrugc}}
    \label{fig:NMR_linker}
\end{figure}

Synthesis of \ce{PEG-(OH)1}, \ce{PEG-(OH)2}, \ce{PEG-(OH)3}, \ce{d-PEG-(OH)1}, \ce{d-PEG-(OH)2,} and \ce{d-PEG-(OH)3} was adapted from previous work \cite{Cheng2019SupramolecularRegeneration, DeFrates2022TheProdrugc}.
PEG-OH was oxidized to PEG-COOH for further coupling.
PEG-OH (\SI{15}{\gram}, \SI{20}{\milli\mol}) was dissolved in water (\SI{300}{\milli\litre}) with TEMPO (\SI{1}{\gram}, \SI{6}{\milli\mol}) and \ce{NaBr} (\SI{2}{\gram}, \SI{20}{\milli\mol}) for \SI{2}{\hour} with stirring at RT. 
Oxidation was initiated by adding \ce{NaClO} (\SI{80}{\milli\litre}) with stirring, and pH was adjusted to $11$ throughout the reaction with \ce{NaOH} ($30$ wt\% solution). 
After \SI{1}{\hour}, the reaction was quenched with \SI{20}{\milli\litre} \ce{EtOH}, and pH was adjusted to $2$ with \ce{HCl} ($10$ vol\% solution). 
\ce{NaCl} (\SI{100}{\gram}) was added to reduce polymer solubility, and product was extracted four times with DCM (\SI{200}{\milli\litre}, \SI{100}{\milli\litre} thrice). 
The organic phase was dried with \ce{Na2SO4}, vacuum filtered, concentrated under vacuum, and precipitated in diethyl ether (\SI{200}{\milli\litre}) for \SI{1}{\hour}. 
Following centrifugation and decanting, a final DCM dissolution and ether precipitation were conducted if needed, resulting in PEG-COOH.
PEG-COOH or deuterated (d)-PEG-COOH (\SI{2}{\milli\mol}) were combined with the appropriate linker: TRIS (\SI{4}{\milli\mol}), 2-Amino-1,3-propanediol (\SI{4}{\milli\mol}), or ethanolamine (\SI{4}{\milli\mol}) in dry DMF (\SI{100}{\milli\litre}) and dissolved thoroughly.
DIPEA (\SI{4}{\milli\mol}) and HBTU (\SI{3}{\milli\mol}) were added and stirred at \SI{37}{\celsius} overnight under nitrogen. 
The reaction mixture was dropped into cold diethyl ether (\SI{40}{\milli\litre}), stored at \SI{-20}{\celsius} for \SI{1}{\hour}, and decanted. 
Crude product was redissolved in DCM (\SI{100}{\milli\litre}) and extracted thrice with \ce{HCl} ($5$ vol\%) and washed with brine. 
The organic phase was dried with \ce{Na2SO4}, vacuum filtered, concentrated under vacuum, and precipitated in diethyl ether (\SI{100}{\milli\litre}) overnight at \SI{-20}{\celsius}, and decanted. 
Finally, the product was dissolved in \ce{MeOH} (\SI{10}{\milli\litre}), syringe filtered (hydrophobic \ce{PTFE}, \SI{0.2}{\micro\meter}), precipitated into diethyl ether (\SI{100}{\milli\litre}), decanted, and dried, yielding \ce{PEG-(OH)x} or \ce{d-PEG-(OH)x}. 
The \ce{1H} NMR spectrum for \ce{d-PEG-COOH} and a representative spectrum for the deuterated polymer with the appropriate linker, \ce{d-PEG-(OH)_1}, is shown in Fig.~\ref{fig:NMR_linker}.
$^1$\ce{H} NMR (\SI{500}{\mega\hertz}, \ce{DMSO-d6}); \ce{d-PEG-(OH)x}: $\delta$ (approx. from \ce{PEG-(OH)1}) $7.6$ (\ce{1H}, linker amide bond), $4.67$ (\ce{2H}, \ce{O-CH2-CO}), 3.4 (\ce{2H}, linker \ce{CH2-CH2-OH}), 3.15 (\ce{2H}, linker \ce{NH-CH2-CH2}).

\subsubsection{DPCA Coupling for \ce{PEG-(DPCA)x} and \ce{d-PEG-(DPCA)x}}
\begin{figure}
    \centering
    \includegraphics[width=1\linewidth]{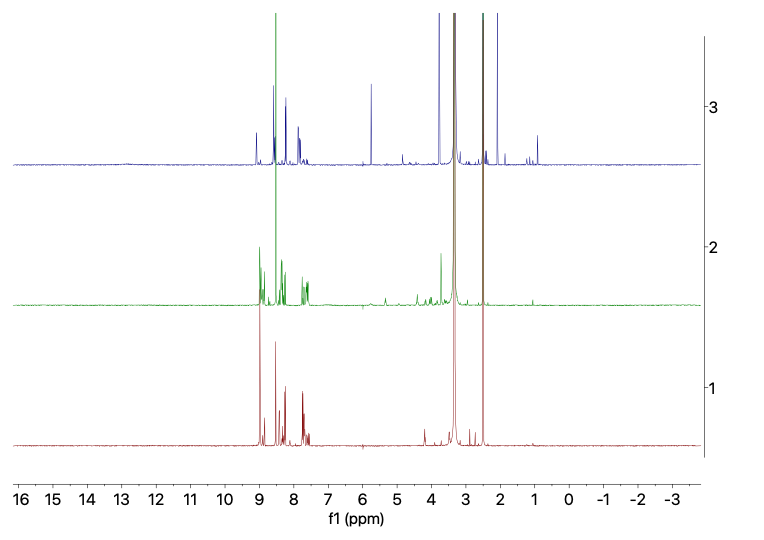}
    \caption{$^1$H NMR spectra of (bottom to top) d-PEG-(DPCA)$_1$, d-PEG-(DPCA)$_2$, and d-PEG-(DPCA)$_3$.}
    \label{fig:NMR_dPEG-DPCA}
\end{figure}

DPCA coupling procedures were adapted from prior work\cite{Cheng2019SupramolecularRegeneration, DeFrates2022TheProdrugc}. 
\ce{PEG-(OH)x} or \ce{d-PEG-(OH)x} (\SI{0.1}{\milli\mol}) was dissolved in \ce{DMF} (\SI{3}{\milli\litre}) at \SI{50}{\celsius} and dried under vacuum for \SI{15}{\minute}.
\ce{NaH} ($60$\% in oil, $2$ molar eq. to \ce{-OH}) was added while stirring at \SI{50}{\celsius} for \SI{10}{\minute}. 
Im-DPCA ($1.1$ molar eq. to \ce{-OH}) was added and reacted at \SI{50}{\celsius} for \SI{45}{\minute}.
Solutions with higher degrees of DPCA coupling became viscous, and solution color darkened slightly.
The reaction mixture was precipitated in to cold diethyl ether ($\geq10:1 $ether to crude mixture by volume) and kept at \SI{-20}{\celsius} for \SI{1}{\hour}. 
\ce{PEG-(DPCA)x} or \ce{d-PEG-(DPCA)x} were separated by centrifugation.
When further purification (removal of uncoupled DPCA) was necessary, crude product (\SI{80}{\milli\gram}) was dissolved in DCM (\SI{30}{\milli\litre}) followed by the addition of water (\SI{12}{\milli\litre}) and TEA (\SI{50}{\micro\litre}) and stirred for \SI{30}{\minute}. 
The DCM phase containing the product was then separated and dried under vacuum, yielding \ce{PEG-(DPCA)x} or \ce{d-PEG-(DPCA)x}.
The \ce{1H} NMR spectra of each of the products is shown in Fig.~\ref{fig:NMR_dPEG-DPCA}.
$^1$\ce{H} NMR (\SI{500}{\mega\hertz}, \ce{DMSO-d6}); $\delta$ (approx. from PEG-(DPCA)1) $7.5-9.1$ (\ce{6H}, aromatic protons), $8.1$ (\ce{1H}, linker amide bond), $4.2$ (\ce{2H}, \ce{O-CH2-CO}), $3.5$ (\ce{2H}, linker \ce{CH2-CH2-OH}) $3.15$ (\ce{2H}, linker \ce{NH-CH2-CH2}).

MALDI-TOF MS [m/z] max. mass peak; \ce{d-PEG-(DPCA)1}: \SI{1330}{\gram\per\mol}; \ce{d-PEG-(DPCA)2}: \SI{1485}{\gram\per\mol}; \ce{d-PEG-(DPCA)3}: \SI{1594}{\gram\per\mol}; 

\subsubsection{Nuclear magnetic resonance (NMR)}
All reaction products and starting materials were analyzed using solution state $^1$\ce{H} NMR in \ce{DMSO-\textit{d$_6$}}.
Chemical shifts are given in ppm relative to the solvent residual peak (\ce{DMSO-\textit{d$_6$}}: \SI{2.50}{\ppm}). 
Spectra were collected on a Bruker \SI{500}{\mega\hertz} NMR spectrometer at the Pines Magnetic Resonance Center’s Core NMR Facility in the College of Chemistry at UC Berkeley.

\subsubsection{Matrix-assisted laser desorption/ionization time-of-flight mass spectrometry (MALDI-TOF MS)}
\begin{figure}
    \centering
    \includegraphics[width=1\linewidth]{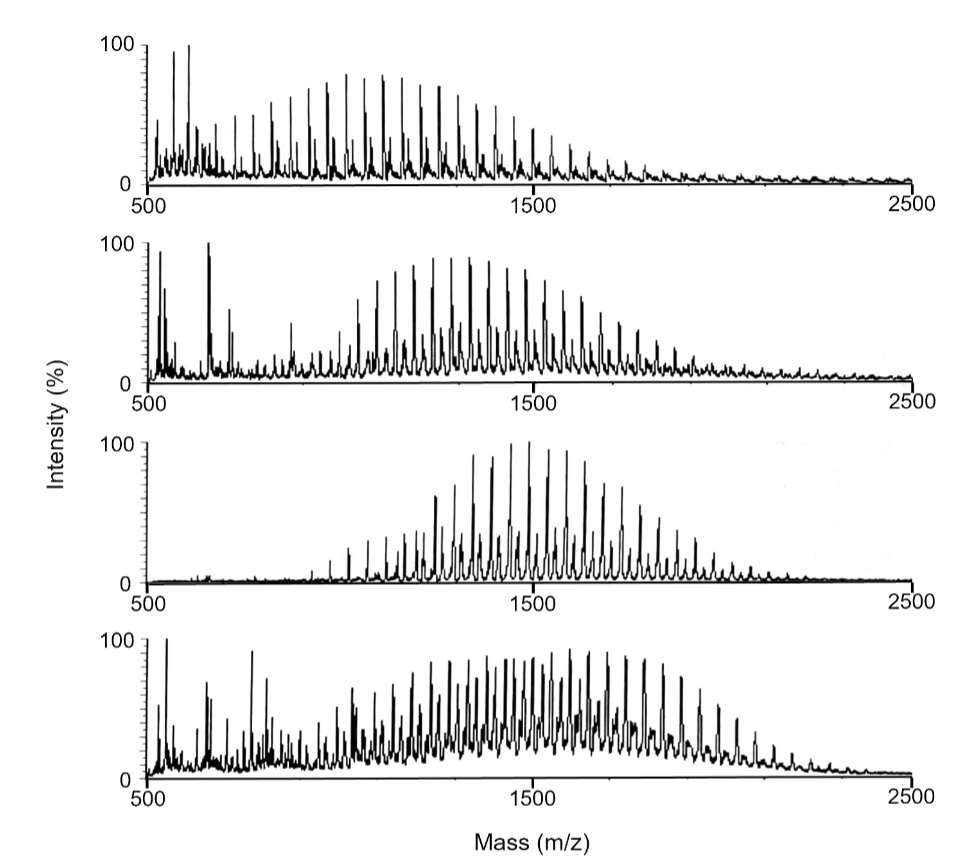}
    \caption{MALDI-TOF MS spectra of (top to bottom) d-PEG-COOH, d-PEG-(DPCA)$_1$, d-PEG-(DPCA)$_2$, and d-PEG-(DPCA)$_3$.}
    \label{fig:chapter3_S4}
\end{figure}

MALDI-TOF mass distribution information was obtained for reaction products and starting materials using an Applied Biosystems Voyager DE Pro in the QB3/Chemistry Mass Spectrometry Facility at UC Berkeley. 
MALDI-TOF MS spectra are shown in Fig.~\ref{fig:chapter3_S4}.
Samples were prepared in water/acetonitrile mixtures sandwiched between layers of $\alpha$-cyano-4-hydroxycinnamic acid (CHCA) matrix from saturated solution in \ce{EtOH}.

\subsection{SANS Measurements}
Small-angle neutron scattering (SANS) measurements were carried out on the Bio-SANS (CG-3) instrument at the High Flux Isotope Reactor (HFIR) at Oak Ridge National Laboratory (ORNL, Oak Ridge, TN, USA).
A neutron wavelength of \SI{6}{\angstrom} with a wavelength spread of $13.2\%$ ($\Delta \lambda/\lambda$) was used. 
Dilute samples were exposed to the beam for 3600 or 5400 seconds, depending on the contrast; concentrated samples were exposed for 600 seconds. 
The detection system consisted of three detector arrays: main, mid-range, and wing. 
The main detector is planar and movable parallel to the beam path. 
It was located \SI{15.5}{\meter} from the sample position and offset by 350 mm from the beam center. 
The mid-range and wing detector arrays are curved and located at fixed distances from the sample of \SI{4}{\meter} and \SI{1.13}{\meter}, respectively, which are also their respective radii of curvature. They rotate about the sample position and were rotated to $1^\circ$ and $5.5^\circ$ from the direct beam, on opposite sides, to optimize $q$-overlap and provide access to a scattering vector range of \SI{0.003}{\per \angstrom} $< q < $ \SI{0.85}{\per\angstrom} in a single exposure.
Scattered neutrons were recorded on the position-sensitive 2D detector arrays, which were azimuthally averaged to produce 1D intensity profiles, $I(q)$.

Three polymer surfactants were studied, each consisting of a $\sim$\SI{1140}{\dalton} deuterated PEG backbone (dPEG) conjugated to $1, 2,$ or $3$ hydrophobic groups, respectively.
To isolate scattering from the hydrophobic cores of assembled structures, we performed contrast matching by preparing samples in a $99\%$ D$_2$O $/ 1\%$ H$_2$O solvent mixture, which suppresses scattering from the deuterated PEG corona.
This ratio was determined by calculating the neutron scattering length density for each part of the molecule and comparing to the scattering length density for various solvent mixtures.

Samples were loaded into \SI{1}{\milli\meter} path length Hellma cylindrical (banjo) cells and analyzed at concentrations at which macroscopic aggregates were expected, as determined from coarse-grained molecular dynamics simulations, the theoretical results presented in the main text, and the experimental critical micelle concentrations estimated in Ref.~\cite{DeFrates2022TheProdrugc}.

Data processing was conducted using ORNL facility developed software, drtSANS~\cite{Heller2022}, and the resulting 1D intensity profiles were placed on an absolute scale.
Background subtraction was performed in Igor Pro (Wavemetrics, Inc.)~\cite{Kline2006ReductionPro}.
The structural features were inferred by analyzing power-law regimes and crossover points in the scattering curves.

\subsubsection{Physical Picture and Fit Procedure}

For any concentration of the three molecules examined, it is possible for aggregates of different sizes and different morphologies to coexist. 
We take the perspective that aggregates of globular (or spherical), elongated (or cylindrical/fiber-like), and planar shapes may also coexist at a given concentration and that changing the concentration or the amount of DPCA per molecule simply alters the relative population and preferred sizes of the morphologies.
We anticipate that the radius of the spheres ($R_s$), the cross-sectional diameter of the fibers ($d$), and the thickness of the planar sheets ($\delta$) will be within our resolution, but that the longer dimensions of the fibers ($\ell$) and sheets ($R$) likely exceed our size resolution of  approximately \SI{200}{\nano\meter}.
Our previous work has indeed revealed that fiber lengths can exceed several micrometers~\cite{DeFrates2022TheProdrugc}.
We also note that while the DPCA of isolated molecules (i.e.,~``monomers'' as described in the main text) is within the spatial resolution of the measured SANS data, the intensity of these unaggregated molecules is likely too low to resolve (especially PD1, which has only a single DPCA group and thus lower contrast in comparison to PD2 and PD3).
Our fits therefore only assume the presence of spheres, fibers, and planar aggregates, neglecting monomers.

To consider multiple morphology populations, we fit our intensity $I(q)$ by summing the modified Guinier functions associated with each morphology that are appropriate for dilute systems:
\begin{subequations}
\label{eq:guinier}
\begin{align}
\label{eq:guinier_I}
I(q) &= \sum_{m} w_m I_{m,0}\, q^{-\alpha_m} 
\exp\!\left[-\frac{q^{2} R_{m,g}^{2}}{(3 - \alpha_m)}\right] 
+ I_{\mathrm{bkg}},\\[8pt]
\label{eq:guinier_alpha}
\alpha_m &= 
\begin{cases}
0, & m = \mathrm{sphere},\\[4pt]
1, & m = \mathrm{fiber},\\[4pt]
2, & m = \mathrm{planar},
\end{cases}\\[10pt]
\label{eq:guinier_Rg}
R_{m,g} &= 
\begin{cases}
R_s\sqrt{3/5}, & m = \mathrm{sphere},\\[4pt]
d\sqrt{1/8}, & m = \mathrm{fiber},\\[4pt]
\delta\sqrt{1/12}, & m = \mathrm{planar}.
\end{cases}
\end{align}
\end{subequations}
where $I_{m,0}$ is the intensity scalar of the Guinier (spheres) or modified Guinier (fibers and planar), 
$\alpha_m$ is the power-law exponent that defines the shape of the aggregate, and $R_{m,g}$ represents the characteristic radius of gyration of the $m$th morphology (which we can then directly connect to the geometric parameters of interest [see Eq~\eqref{eq:guinier_Rg}]), and $I_{\mathrm{bkg}}$ is the background intensity in the flat high-$q$ region. 
We have also introduced an indicator function, $w_m$, which takes a value of one if the $m$th morphology is considered in the fit and a value of zero otherwise. 
The use of these indicator functions allows us to assume that different morphological combinations may dominate a given system (e.g.,~for a given set of conditions, spheres may dominate the system with very little intensity due to fibers or planar aggregates).
While, in principle, changes in the relative populations of the morphologies should be captured in a changing $I_{m,0}$, use of the indicator function practically reduces the number of fitting parameters while providing physical insight into the minimal number of morphologies that need to be considered to fit the data.
We emphasize that while our (modified) Guinier functions describe globular and elongated aggregates, we refer to these shapes as sphere and fiber for simplicity.  

Below, we discuss the quantitative aspects of our fit that were omitted from our main text discussion. 
A summary of all our fit parameters is provided in Table~\ref{tab:sans_fit_summary}.
The data and resulting fits are all shown in the main text with one additional sample shown in Fig.~\ref{fig:PD2 2 mg/ml}.
In the following discussion we recall that the zero-angle intensity scalar takes the form $I_{m,0} = \Delta \rho^2 \phi_m V_m$ where $\Delta \rho$ is the contrast, $\phi_m$ is volume fraction occupied by the aggregates of type $m$, and $V_m$ is the volume of the aggregate.  
Comparing the intensity scalar between systems with identical contrast thus allows us to make statements on relative product of the volume fractions and volumes.
As we can determine the entire volume of spheres (assuming they are smaller than our beamline resolution of \SI{200}{\nano\meter} as we anticipate), we can further identify the relative volume fractions. 
Finally to contextualize the below size estimates, the length of an individual DPCA molecule (i.e.,~the head group of PD1) is estimated to be approximately \SI{1}{\nano\meter}.

\begin{table}[htbp]
\centering
\caption{Fit parameters for PD1–PD3 at indicated concentrations. Intensities are absolute (\si{cm^{-1}}); sizes are in \si{nm}. Block shading indicates the set of morphologies present.}
\label{tab:sans_fit_summary}
\small
\setlength{\tabcolsep}{6pt}

\begin{tabular}{llcccl}
\toprule
System & Conc. (mg/mL) & $R_s$ (nm) & $d$ (nm) & $\delta$ (nm) & Intensity (\si{cm^{-1}}) \\
\midrule

\rowcolor{blue!6}
 &  &  &  &  & sph: $(12.4\pm3.6)\times10^{-4}$ \\[-2pt]
\rowcolor{blue!6}
PD1 & 3 & $1.55\pm0.36$ &  &  & fib: \\[-2pt]
\rowcolor{blue!6}
 &  &  &  &  & pln: \\[-2pt]
\rowcolor{blue!6}
 &  &  &  &  & bkg: $(5.6\pm0.5)\times10^{-4}$ \\[4pt]

\rowcolor{cyan!10}
 &  &  &  &  & sph: $(14.7\pm3.7)\times10^{-4}$ \\[-2pt]
\rowcolor{cyan!10}
PD1 & 7 & $1.47\pm0.27$ &  & $27.7\pm2.1$ & fib: \\[-2pt]
\rowcolor{cyan!10}
 &  &  &  &  & pln: $(8.7\pm0.6)\times10^{-6}$ \\[-2pt]
\rowcolor{cyan!10}
 &  &  &  &  & bkg: $(5.6\pm0.5)\times10^{-4}$ \\[4pt]

\rowcolor{olive!12}
 &  &  &  &  & sph: \\[-2pt]
\rowcolor{olive!12}
PD1 & 10 &  & $2.4\pm0.8$ & $40.5\pm3.1$ & fib: $(2.6\pm0.5)\times10^{-4}$ \\[-2pt]
\rowcolor{olive!12}
 &  &  &  &  & pln: $(27.3\pm2.3)\times10^{-6}$ \\[-2pt]
\rowcolor{olive!12}
 &  &  &  &  & bkg: $(2.7\pm1.0)\times10^{-4}$ \\
\midrule

\rowcolor{orange!10}
 &  &  &  &  & sph: \\[-2pt]
\rowcolor{orange!10}
PD2 & 1.5 &  &  & $55.4\pm3.5$ & fib: \\[-2pt]
\rowcolor{orange!10}
 &  &  &  &  & pln: $(14.2\pm1.4)\times10^{-6}$ \\[-2pt]
\rowcolor{orange!10}
 &  &  &  &  & bkg: $(1.6\pm1.0)\times10^{-4}$ \\[4pt]

\rowcolor{blue!6}
 &  &  &  &  & sph: $(5.9\pm3.2)\times10^{-4}$ \\[-2pt]
\rowcolor{blue!6}
PD2 & 2 & $1.3\pm0.6$ &  &  & fib: \\[-2pt]
\rowcolor{blue!6}
 &  &  &  &  & pln: \\[-2pt]
\rowcolor{blue!6}
 &  &  &  &  & bkg: $(2.6\pm0.7)\times10^{-4}$ \\[4pt]

\rowcolor{cyan!10}
 &  &  &  &  & sph: $(4.3\pm2.9)\times10^{-4}$ \\[-2pt]
\rowcolor{cyan!10}
PD2 & 3.5 & $1.3\pm0.6$ &  & $39.5\pm4.2$ & fib: \\[-2pt]
\rowcolor{cyan!10}
 &  &  &  &  & pln: $(6.0\pm0.8)\times10^{-6}$ \\[-2pt]
\rowcolor{cyan!10}
 &  &  &  &  & bkg: $(5.2\pm0.5)\times10^{-4}$ \\[4pt]

\rowcolor{violet!10}
 &  &  &  &  & sph: $(14.1\pm2.3)\times10^{-4}$ \\[-2pt]
\rowcolor{violet!10}
PD2 & 7 & $0.67\pm0.11$ & $20.0\pm5.4$ & $46.8\pm5.5$ & fib: $(2.8\pm1.4)\times10^{-4}$ \\[-2pt]
\rowcolor{violet!10}
 &  &  &  &  & pln: $(14.6\pm1.2)\times10^{-6}$ \\[-2pt]
\rowcolor{violet!10}
 &  &  &  &  & bkg: $(5.4\pm0.6)\times10^{-4}$ \\
\midrule

\rowcolor{cyan!10}
 &  &  &  &  & \makecell[l]{sph: $(28.1\pm3.4)\times10^{-4}$\\sph: $(15.9\pm1.1)\times10^{-2}$} \\[-2pt]
\rowcolor{cyan!10}
PD3 & 3 & \makecell[l]{$1.35\pm0.12$\\$12.75\pm0.35$} &  & \makecell[l]{$43.0\pm1.0$\\$129.9\pm1.4$} & fib: \\[-2pt]
\rowcolor{cyan!10}
 &  &  &  &  & \makecell[l]{pln: $(102\pm3)\times10^{-6}$\\pln: $(905\pm2.2)\times10^{-6}$} \\[-2pt]
\rowcolor{cyan!10}
 &  &  &  &  & bkg: $(4.2\pm0.4)\times10^{-4}$ \\[4pt]

\rowcolor{violet!10}
 &  &  &  &  & sph: $(47.0\pm9.8)\times10^{-4}$ \\[-2pt]
\rowcolor{violet!10}
PD3 & 7.5 & $2.09\pm0.31$ & $19.5\pm0.8$ & $39.5\pm1.7$ & fib: $(33.9\pm2.9)\times10^{-4}$ \\[-2pt]
\rowcolor{violet!10}
 &  &  &  &  & pln: $(63.8\pm1.3)\times10^{-6}$ \\[-2pt]
\rowcolor{violet!10}
 &  &  &  &  & bkg: $(3.3\pm1.4)\times10^{-4}$ \\[4pt]

\rowcolor{violet!10}
 &  &  &  &  & sph: $(46.7\pm13.8)\times10^{-4}$ \\[-2pt]
\rowcolor{violet!10}
PD3 & 10 & $2.08\pm0.38$ & $15.0\pm0.9$ & $29.1\pm1.7$ & fib: $(23.0\pm3.6)\times10^{-4}$ \\[-2pt]
\rowcolor{violet!10}
 &  &  &  &  & pln: $(72.8\pm1.3)\times10^{-6}$ \\[-2pt]
\rowcolor{violet!10}
 &  &  &  &  & bkg: $(4.8\pm1.3)\times10^{-4}$ \\

\bottomrule
\end{tabular}
\end{table}

\subsubsection{dPD1 SANS analysis}

For the \SI{3}{\milli\gram\per\milli\liter} PD1 sample, 
the data can be well-described (recognizing the large uncertainty in this low intensity data) solely considering spherical aggregates (i.e.,~we set the indicator function for fibers and planar aggregates to zero). 
In contrast, the \SI{7}{\milli\gram\per\milli\liter} and \SI{10}{\milli\gram\per\milli\liter} samples were fit to a combination of spherical (high-$q$) and planar (low-$q$) structures. 
For the \SI{7}{\milli\gram\per\milli\liter} sample, the size and concentration of the spherical aggregates remain unchanged from the \SI{3}{\milli\gram\per\milli\liter} sample; however, an additional population of planar aggregates is detected at \SI{7}{\milli\gram\per\milli\liter}. 
The thickness of the planar structure at \SI{7}{\milli\gram\per\milli\liter} is $\delta = \SI{27.7 \pm 2.1}{\nano\meter}$.
Increasing the concentration from \SI{3}{\milli\gram\per\milli\liter} to \SI{7}{\milli\gram\per\milli\liter} resulted in a $20\%$ increase in spherical intensity, and because the volume decreased by a factor of $\sim 0.9 = (11.4^3/12.0^3)$, the volume fraction of the spherical aggregates consequently increased by $40\%$. 
At \SI{10}{\milli\gram\per\milli\liter}, the data are well described using a combination of fibers and planar aggregates.
The trend of increasing spherical aggregate concentration followed by transition to fiber-like structures with increasing overall concentration is consistent with our theoretical expectations. 

\subsubsection{dPD2 SANS analysis}

\begin{figure}
    \centering
    \includegraphics[width=0.5\linewidth]{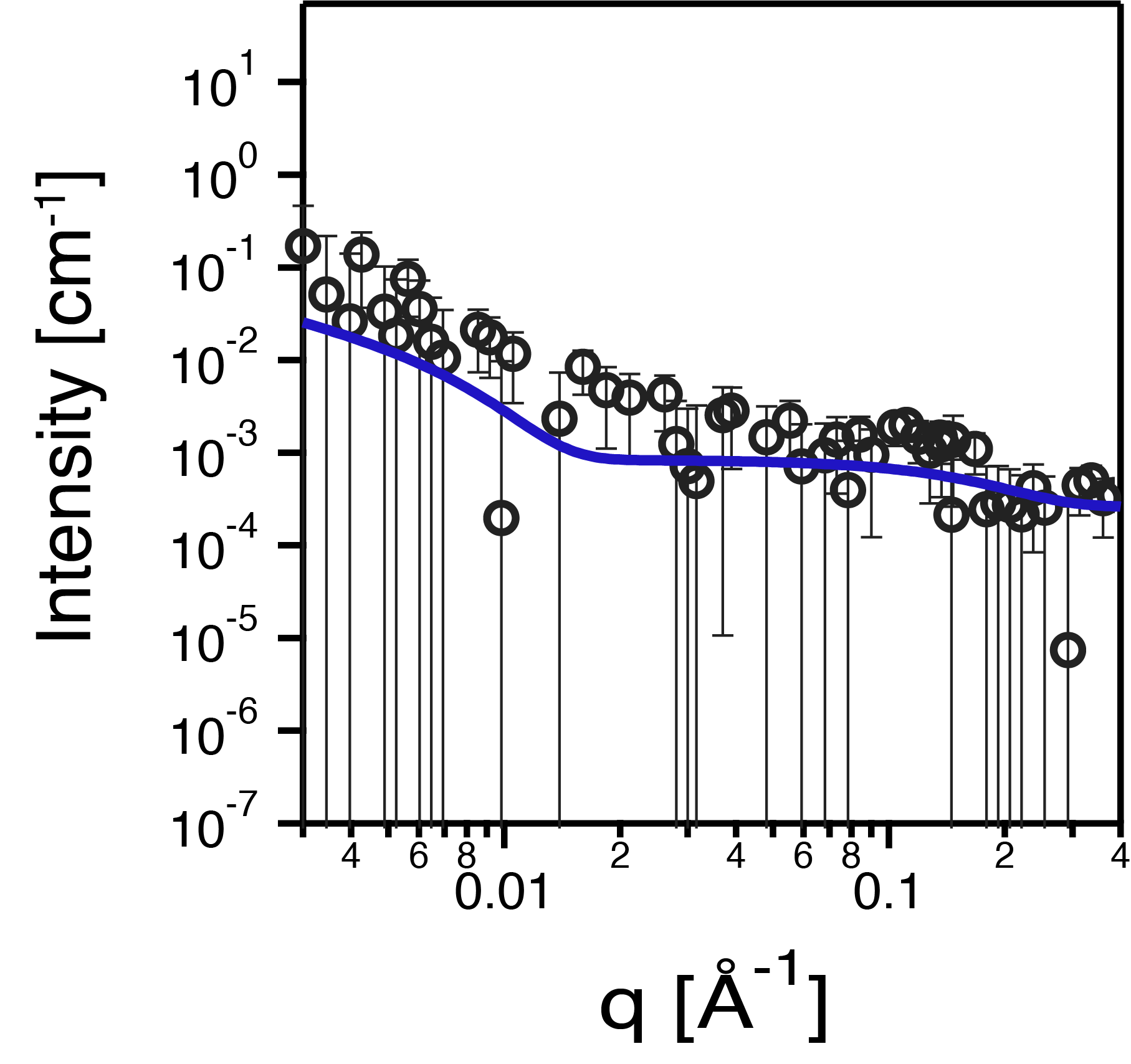}
    \caption{SANS profile of PD2 sample at 2 mg/ml. Solid line shows the fit curve.}
    \label{fig:PD2 2 mg/ml}
\end{figure}

The analyses of PD2 samples reveal the emergence of elongated aggregates co-existing with spherical and planar aggregates at higher concentrations. 
The \SI{1.5}{\milli\gram\per\milli\liter} sample was fit using a combination of spheres and planar aggregates. 
In contrast, the \SI{2}{\milli\gram\per\milli\liter} sample was best fit with a combination of spheres and elongated aggregates, though other combinations are reasonable given the high degree of uncertainty for this sample. 
The data and corresponding fit are shown in Fig.~\ref{fig:PD2 2 mg/ml}. 
In the \SI{3.5}{\milli\gram\per\milli\liter} sample, both spherical and planar aggregates were detected, and the concentration of spherical aggregates decreased by a factor of $\sim 1.2$ relative to the \SI{2}{\milli\gram\per\milli\liter} sample. 
The \SI{7}{\milli\gram\per\milli\liter} sample fit consisted of all the aggregate shapes considered: spherical, elongated, and planar. 
Unlike the PD1 series, where the spherical aggregate volume fraction decreased with increasing concentration, a significant increase in the spherical aggregate volume fraction was observed for the PD2 series with an increase by a factor of $\sim 11 = [(14.1/4.3)/(6.3/9.5)^3]$ when increasing concentration from \SI{3.5}{\milli\gram\per\milli\liter} to \SI{7}{\milli\gram\per\milli\liter}.

\subsubsection{dPD3 SANS analysis}

Interestingly, the \SI{3}{\milli\gram\per\milli\liter} sample was fit best by assuming a \textit{bimodal} size population of \textit{both} spheres and planar aggregates.
The smaller spherical aggregates were fit to a radius of \SI{1.35}{\nano\meter} while the larger spherical aggregates have a radius of \SI{12.75}{\nano\meter}.
We note that the smaller aggregates are similar in size to those of the PD1 system, which were speculated to consist of just a few molecules (perhaps dimers or trimers).
As PD3 has three times the number of DPCA groups per molecule in comparison to PD1, the smaller spherical ``aggregates'' observed here may correspond to the DPCA content of isolated PD3 molecules. 
The larger of two planar aggregate sizes was fit to a thickness of \SI{130}{\nano\meter}, which appears to be larger than our theoretical picture would suggest is possible as we expect the thickness should be within an order of magnitude of the head group size.
This may reflect the presence of impurities---such as DPCA that is not conjugated with PEG and is thus free to aggregate without bound. 
For both the \SI{7.5}{\milli\gram\per\milli\liter} and \SI{10}{\milli\gram\per\milli\liter} samples, the data are well described by a fit assuming coexistence of all three morphologies.
The spherical sizes are slightly larger than the smaller sphere sizes observed at \SI{3}{\milli\gram\per\milli\liter}, and could possibly reflect PD3 aggregating into dimers.

We observe non-monotonic trends in the scattering intensity as a function of concentration for the systems with $2$ and $3$ hydrophobic groups.
In both cases, the lowest concentrations (\SI{1.5}{\milli\gram\per\milli\litre} for PD2 groups and \SI{3}{\milli\gram\per\milli\litre} for PD3) demonstrate the highest scattering intensities.
This initially seems counterintuitive as one would expect that increasing concentration should lead to larger aggregates with stronger scattering signals.
However, this behavior can be explained by considering the limits of the length scales in SANS measurements.
The smallest scattering vector for this beamline is $q = 0.003~\mathrm{\AA}^{-1}$.
Aggregates larger than \SI{200}{\nano\meter} scatter at $q$ values lower than the minimum accessible $q$ and therefore elude detection.
As the concentration increases, the systems with $2$ and $3$ hydrophobic groups likely form aggregates that exceed this detection limit.
While we do not expect aggregates with dimension(s) that exceed an order of magnitude of the head group dimensions, impurities, such as DPCA that is not conjugated to PEG, may result in the formation of larger aggregates that elude detection.
As a result, the measured scattering intensity in the low $q$ is diminished because the aggregates whose scattering intensity is detected are in fact not the largest (and most intense signal) in the sample.

\subsection{Small Angle X-Ray Scattering}

\begin{figure}
    \centering
    \includegraphics[width=0.8\linewidth]{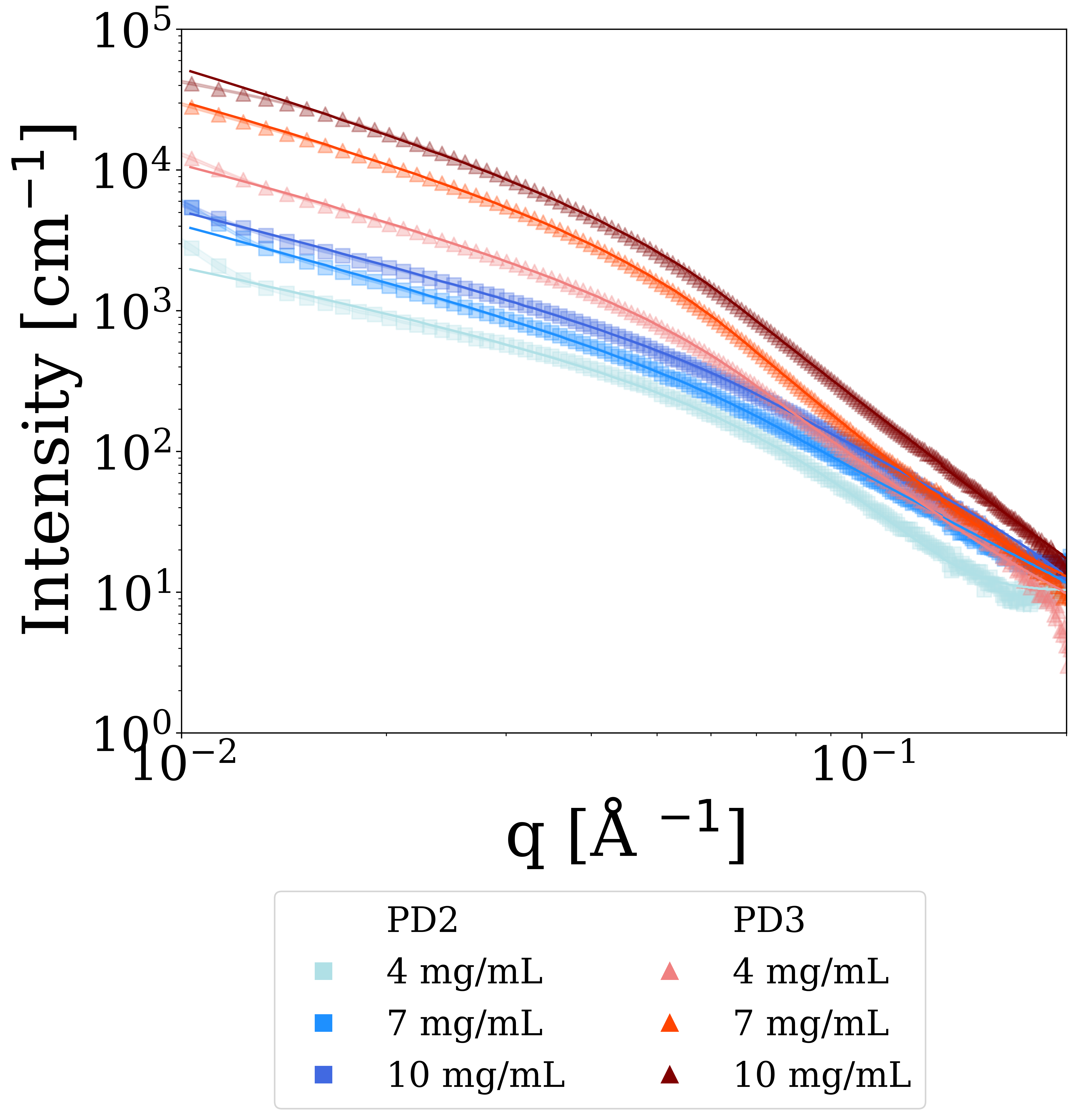}
    \caption{SAXS scattering data and fit functions for PD2 and PD3 samples. Solid lines are Guinier-Porod fits.}
    \label{fig:SAXS}
\end{figure}
SAXS was carried out at beamline 7.3.3 at the Advanced Light Source (Lawrence Berkeley National Lab). 
Samples were measured in 1.5 mm thin-walled quartz capillary tubes and subject to single $60$ second exposures. 
2D scattering results were azimuthally averaged to produce one-dimensional SAXS profiles. 
Superimposable profiles were averaged and then subtracted from the background data. 
The 1D profiles in Fig.~\ref{fig:SAXS} were fit to a Guinier-Porod model of the form:
\begin{equation}
I(q) =
\begin{cases}
G\, q^{-(3-\alpha)} \exp\left( -\dfrac{q^2 R_g^2}{3-\alpha} \right), & q \le q_1, \\[6pt]
\dfrac{G}{q_1^{\,(3-\alpha-m)}} \exp\left( -\dfrac{q_1^2 R_g^2}{3-\alpha} \right) q^{-m}, & q > q_1,
\end{cases}
\quad
q_1 = \frac{1}{R_g} \sqrt{\frac{\big(m - (3-\alpha)\big)(3-\alpha)}{2}} .
\end{equation}
where $G$ is the contrast, $m$ is the Porod (power law) exponent, $\alpha$ is the dimension variable, and $R_g$ is the radius of gyration of the generalized form factor. 
$q_1$ is defined by the end of the low-$q$ power law regime. 
Here, we take a different approach in comparison to our SANS analysis. 
Rather than assuming different shape combinations we treat $\alpha$ as a fitting parameter and will interpret trends in how its value changes with varying conditions.

The results of the SAXS measurements qualitatively support the findings from our SANS data.
The Porod exponent is $\sim -3$ for surfactant assemblies with 2 hydrophobic groups and $\sim -4$   for those with 3 hydrophobic groups.  
As a function of concentration for each, the dimension variable increases slightly. 
For those with 2 hydrophobic groups, at 4 and \SI{10}{\milli\gram\per\milli\litre}, $\alpha = 1.0$ and $1.2$, respectively. 
For those with 3 hydrophobic groups, at 4 and \SI{10}{\milli\gram\per\milli\litre}, $\alpha = 1.3$ and $1.5$, respectively, reflecting an increase in the planar character of aggregates.

\section{Morphological Phase Diagrams from Coarse-grained Molecular Dynamics Simulations}
This section outlines the procedure for constructing the morphological phase diagrams from our computer simulations.
To determine the dominant morphological state, we first must identify the distinct self-assembled clusters. 
We can then characterize the shape of each cluster and, ultimately, which shape(s) dominate the system state as a function of $\Phi$ and $\epsilon/k_BT$.
The volume fraction, $\Phi$, is defined as the sum of the volume of each bead in the simulation (both head and tail beads) divided by the system volume. 
The volume of each bead is defined using a diameter of $2^{1/6}\sigma_{\rm LJ}$ where is the Lennard-Jones diameter.
After identifying the morphology of each cluster, we find the volume fraction of each morphology, $\phi_m$, and report the dominant morphological state for each $\Phi$ and $\epsilon/k_B T$.

We identified clusters by grouping molecules with neighboring  attractive beads (beads within a distance ${1.25\sigma_{\text{LJ}}}$) into the same cluster.
The shape of the clusters can then be categorized as spherical (${\lambda_x \approx \lambda_y \approx \lambda_z}$), cylindrical (${\lambda_x \approx \lambda_y \ll \lambda_z}$), or planar (${\lambda_x \ll \lambda_y \approx \lambda_z}$). 
We previously reported that the mean cluster size and the cluster density from these simulations exhibited disparate scaling and was qualitatively distinct, with coarse-grained PD1 and PD2 displaying sublinear growth in the mean cluster size and linear growth in the cluster density while PD3 exhibited linear growth in the cluster size and sublinear growth in the cluster density~\cite{DeFrates2022TheProdrugc}. 
This was taken as evidence of a morphological difference between aggregates made from PD1 and PD2 and those made from PD3, which we can now more precisely quantify.

To determine the shape of each cluster, we compute the radius of gyration tensor, $\mathbf{S}$, using the positions of each bead. 
We can diagonalize this tensor with:
\begin{equation}
    \mathbf{S} = \left[\begin{matrix} \lambda_{xx} & 0 & 0\\
    0 & \lambda_{yy} & 0\\
    0 & 0 & \lambda_{zz}
    \end{matrix} \right],
\end{equation}
where we adopt the convention ${\lambda_{zz} > \lambda_{yy} > \lambda_{xx}}$. 
Note that the scalar radius of gyration is defined as ${R_g^2 = \mathrm{tr}(\mathbf{S})/3}$.
We can now compute normalized asphericity $\hat b$, acylindricity $\hat c$, and the relative anisotropy $\kappa^2$\cite{Rudnick1986TheWalks}. 
\begin{subequations}
\begin{align}
     b & = \lambda_{zz} - \dfrac{1}{2}(\lambda_{xx} + \lambda_{yy}),\\
     c & = \lambda_{yy} - \lambda_{xx},\\
     \kappa^2 &= \dfrac{3}{2}\dfrac{\lambda_{xx}^2 + \lambda_{yy}^2 + \lambda_{zz}^2}{(\lambda_{xx} + \lambda_{yy}+ \lambda_{zz})^2} - \dfrac{1}{2},
\end{align}
\end{subequations}
where we normalize $b$ and $c$ by $\lambda_{zz}$ and $\lambda_{yy}$, respectively.
The asphericity $\hat b$ is a measure of how much a structure deviates from spherical symmetry so that perfectly spherical structures have ${\hat b = 0}$ and infinitely long cylinders have ${\hat b = 1}$. 
Conversely, the acylindricity measures the extent to which a cluster deviates from cylindrical symmetry.
While $\hat b$ and $\hat c$ allow us to quantify the degree of asphericity and acylindricity of a structure, the relative anisotropy allows us to identify planar aggregates---large planar structures will have ${\lambda_{xx}\ll \lambda_{yy} \approx \lambda_{zz}}$ and ${\kappa^2 = 1/4}$.

\begin{figure}
    \centering
    \includegraphics[width=1\linewidth]{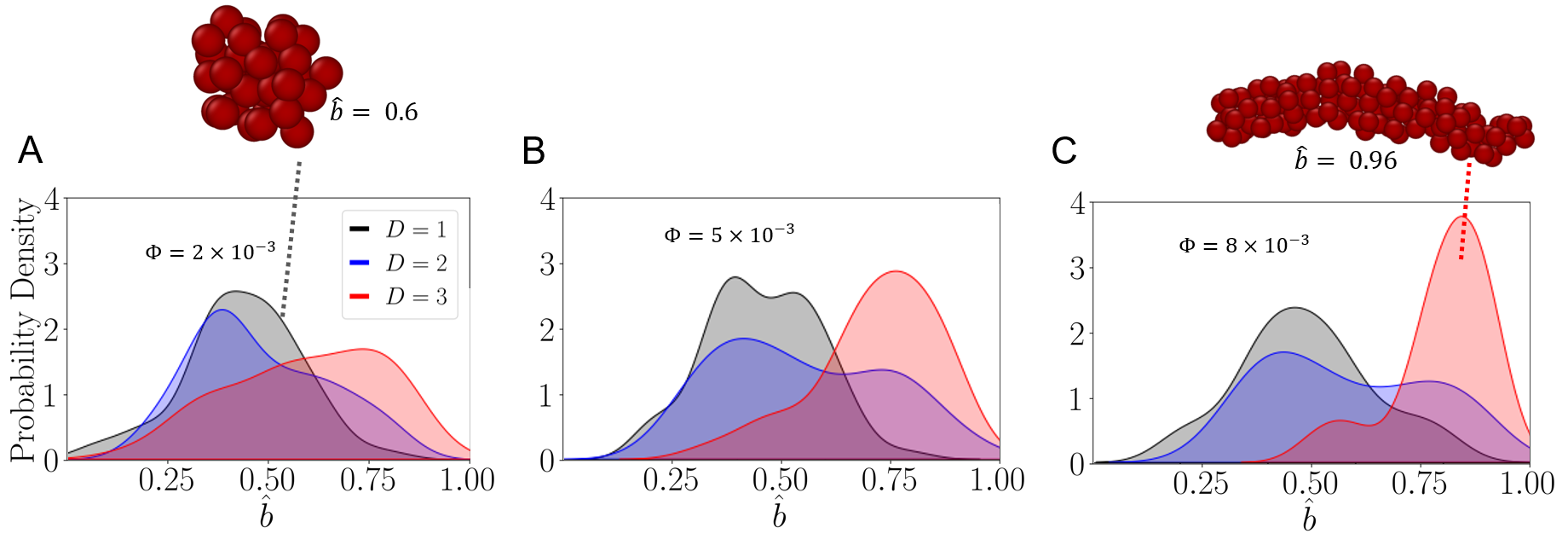}
    \caption{Distributions of $\hat{b}$ for polymer surfactants with $1$, $2$, and $3$ hydrophobic groups and snapshots of aggregate cores for structures with $\hat{b} = 0.6$ and $\hat{b} = 0.96$.
    Polymer is removed from snapshots for clarity, but is present in simulations.}
    \label{fig:asphericity}
\end{figure}

Figure~\ref{fig:asphericity} shows the normalized asphericity distributions for each of the three coarse-grained inverse surfactants in our simulations at three different volume fractions. 
Clusters formed by molecules featuring a single solvophobic bead are largely spherical, with their $\hat b$ peaking near $0.5$ even at densities an order of magnitude higher than the experimentally observed CMC (this corresponds to ${\Phi = 0.008}$ in our simulations). 
Those formed by molecules with $2$ solvophobic beads are slightly less spherical and their probability distributions are bimodal, indicating two populations of aggregate morphologies. 
The probability distribution of clusters formed by molecules with $3$ solvophobic beads displays a markedly different trend, peaking near $0.9$ at even the lowest densities. 

We define spherical structures as those for which $\hat b  < 0.5$, fibers as those for which $\hat b > 0.5$, and planar structures as those for which $0.23 < \kappa^2 < 0.27$.
Molecules not in a cluster are considered monomers.
We then found the volume fraction of each morphology (including monomers).
Next, we compared the four volume fractions ($\phi_\text{mon}, \phi_\text{sph}, \phi_\text{fib}, \phi_\text{pln}$) at each $\Phi$ and $\beta \epsilon$.
In the case that the largest volume fraction was more than $10\%$ larger than the second-largest, a single state was reported.
If the two largest volume fractions were within $10\%$ of one another, both were reported.
The dominant morphologies for each point in parameter space are represented in the simulated phase diagrams in Fig.~3 of the main text.

\section{Theoretical Methods}

\subsection{Stability Criteria of Mean-field Free Energy Solutions}
In this section, we expand on our discussion of the stability criteria to ensure that a given $p_m$ and $\phi_m$ are minima in the free energy density.
Recall that our free energy takes the form:
\begin{equation}
\label{eq:continuous_f}
    \beta f v_0 = \sum_{m \in \mathcal{M}} \dfrac{\phi_m}{p_m} [\ln{\phi_m} - 1 + \beta F_m (p_m)],
\end{equation}
where we have assumed that a single to-be-determined aggregate size for each morphology has a much larger volume fraction that all other sizes.
Extremizing the free energy with respect to these degrees of freedom results in the following criteria: 
\begin{subequations}
\label{eq:coexist3}
    \begin{align}
        \left. \frac{\partial f}{\partial \phi_m}\right|_{\mathbf{x}^{\dagger}}  &= \mu^{\rm coexist} ,\\
        \left. \frac{\partial f}{\partial p_m}\right|_{\mathbf{x}^{\dagger}}  &= 0,
    \end{align}
\end{subequations}
which hold for all ${m \in \mathcal{M}}$ and must be solved along with the constraint ${\sum_{m\in \mathcal{M}}  \phi_{m} = \Phi}$.
Here, $\mathbf{x}^{\dagger}$ represents the degrees of freedom evaluated at their optima.

We can determine the stability of the resulting optimal solutions, $p_m$ and $\phi_m$, by evaluating the components of the Hessian matrix ${\mathbf{H} = \partial^2f/(\partial \mathbf{x} \partial \mathbf{x})}$. 
Stable solutions will be those for which each eigenvalue of the Hessian is positive.
The independence of the morphologies (while the compositions must satisfy the linear constraint, this does not impact the Hessian) results in a relatively sparse Hessian comprised of independent $2 \times 2$ matrices for each morphology, $\mathbf{H}_m$, with elements:
\begin{equation}
\label{eqLmorpho_hessian}
\mathbf{H}_m =  
\begin{bmatrix}
\dfrac{\partial^2 f}{\partial \phi_m \partial \phi_m} & \dfrac{\partial^2 f}{\partial \phi_m \partial p_m} \\ \\
\dfrac{\partial^2 f}{\partial p_m \partial \phi_m} & \dfrac{\partial^2 f}{\partial p_m \partial p_m}
\end{bmatrix}
.
\end{equation}
The stability of solutions can be established by finding the signs of the two eigenvalues for the Hessian of each morphology.
For a solution to be stable, each eigenvalue for every morphology must be positive and thus, for these $2\times2$ Hessians, we require both ${\operatorname{tr}(\mathbf{H}_m)|_{p_m^{\dagger},\phi_m^{\dagger}}>0}$ and ${\det(\mathbf{H}_m)|_{p_m^{\dagger},\phi_m^{\dagger}}>0}$. 
We can explicitly express these requirements with:
\begin{subequations}
\label{eq:hessian_conditions}
\begin{align}
\mathrm{tr}(\mathbf{H}_m) &= 
\dfrac{1}{p_m \phi_m} 
+ \dfrac{2 \phi_m}{p_m^3} \big[\ln{\phi_m} - 1 + \beta F_m (p_m)\big] 
- \dfrac{2\phi_m}{p_m^2} \beta F_m^{\prime} (p_m) 
+ \dfrac{\phi_m}{p_m} \beta F_m^{\prime \prime} (p_m) 
> 0, \label{eq:trace} \\[1mm]
\mathrm{det}(\mathbf{H}_m) &=
\begin{aligned}[t]
& \dfrac{1}{p_m^4} \Big[ 2 (\ln\phi_m - 1 + \beta F_m (p_m)) - 4 p_m \beta F_m^{\prime}(p_m) + 2 p_m^2 \beta F_m^{\prime\prime}(p_m) \Big] \\
& \quad - \Big[ -\dfrac{1}{p_m^2} (\ln \phi_m + \beta F_m (p_m)) + \dfrac{1}{p_m} \beta F_m^{\prime} (p_m) \Big]^2
> 0.
\end{aligned}
\end{align}
\end{subequations}
The above criteria will allow us to determine if optimal solutions  for $\phi_m^{\dagger}$ and $p_m^{\dagger}$ indeed correspond to minima in Eq.~\ref{eq:continuous_f}.
We find that in the limit of large $p$ the leading order term in both the trace and the determinant is the curvature of the per-molecule free energy of formation, which shows that stable solutions to the free energy density require that the free energy of formation per molecule be convex, as emphasized in the main text.

\subsection{Polymer Brush Free Energy Derivation}

We now derive the height and free energy of a polymer brush grafted to planar, cylindrical, and spherical surfaces using scaling theory.
In a good solvent, isolated chains have an end-to-end distance vector with a mean magnitude of ${r_t = N^{3/5} b^{2/5} v^{1/5}}$~\cite{Rubinstein2014PolymerPhysics}.
Upon uniformly grafting $p$ chains to a surface with area $A$, the resulting chain conformations will depend on the grafting density, ${\sigma \equiv p/A}$~\cite{Daoud1982StarDependence, Witten1986ColloidPolymers, Bug1987TheoryAggregates, Wang1988SizePacking, Zhulina2005DiblockSolution}.
When the distance between neighboring chains, $\sigma^{-1/2}$, is less than the isolated chain size of approximately $r_t$, neighboring chains will stretch away from the surface to avoid overlap and reduce the interchain interaction energy. 
The brush height and free energy (relative to the free energy of isolated chains) can be obtained through simple scaling theory.
Following the classical theories~\cite{Alexander1977AdsorptionDescription, DeGennes1976ConformationSolvents, deGennes1980ConformationsInterface}, we envision that each chain will stretch until it no longer overlaps with neighboring chains.
For a planar surface, each chain can thus be thought of consisting of vertically stacked ``correlation blobs'' of size $\xi = \sigma^{-1/2}$. 
Within these blobs, the chain does not ``feel'' neighboring chains and experiences self-avoiding walk statistics with ${\xi = g^{3/5} b^{2/5} v^{1/5}}$ where ${g=\sigma^{-5/6}b^{-2/3}v^{-1/3}}$ is the number of segments per correlation blob.
The number of blobs is simply $N/g$ and the brush height is ${H = \xi N/g = N \sigma^{1/3} b^{2/3} v^{1/3}}$.
The stretching free energy is taken to be the product of the number of correlation blobs and the thermal energy for each of the $p$ chains with ${F^{\rm brush} \approx pk_BT N/g = pk_BTN\sigma^{5/6}b^{2/3}v^{1/3}}$.
With increasing grafting density, the chains adopt increasingly stretched conformations, which is reflected in the strong grafting density dependence of the free energy.

We can straightforwardly determine the brush free energy on a \textit{curved} surface. 
We consider brushes uniformly grafted to cylinders (neglecting the ends) of radius $R_0$ and length $\ell$ and spheres with a radius $R_0$. 
Each system has radial symmetry with radial coordinate $r$.
The concavity of the surfaces results in a reduced average spacing between neighboring chains with increasing $r$.
A fictitious cross-sectional area at $r>R_0$ thus has an apparent reduced grafting density with $\sigma(r)<\sigma(r=R_0)\equiv p/A$ where $A$ is the surface area of a sphere/cylinder with radius $R_0$.
For spheres and cylinders, we have ${\sigma(r) = p/(4\pi r^2)}$ and ${\sigma(r) = p/(2\pi \ell r)}$, respectively.
The blob sizes will thus increase with the distance from the surface as ${\xi(r)=\sigma(r)^{-1/2}}$ and the number of segments per blob will also increase as ${g(r)=\sigma(r)^{-5/6}b^{-2/3}v^{-1/3}}$. 
The height of the brush can now be found by radially integrating the segmental density ${\rho(r)=g(r)/\xi(r)^3 = \sigma(r)^{2/3}b^{-2/3}v^{-1/3}}$ over all space and recognizing that this must return $pN$ with the brush height entering the bounds of the integral.
The free energy can subsequently be founding by defining a spatially varying free energy density that is proportional to the density of correlation blobs, ${f(r) = k_BT/\xi(r)^3 = k_BT\sigma(r)^{3/2}}$, and integrating it over the system volume.
We can see that this approach recovers the expected planar brush results where we no longer have any spatial variations.
The height of the brush can be determined from the constraint on the total number of segments with:
\begin{subequations}
\label{eq:planar_R}
\begin{equation}
    pN = \int_V\rho(\mathbf{r}) \,d\mathbf{r}=A\int_{0}^{H} \rho(z) \, dz 
      = AH\sigma^{2/3}b^{-2/3}v^{-1/3},
\end{equation}
allowing us to identify the planar height as: 
\begin{equation} 
H = N\sigma^{1/3}b^{2/3}v^{1/3},
\end{equation}
in agreement with our earlier expression.
The free energy is then:
\begin{equation}
    F^{\rm brush} = \int_Vf(\mathbf{r}) \,d\mathbf{r}=A\int_{0}^{H} \rho(z) \, dz 
      = k_BTAH\sigma^{3/2} = pk_BTN\sigma^{5/6}b^{2/3}v^{1/3},
\end{equation}
which is again in perfect agreement with our expectations.
\end{subequations}

Using the above approach, we can determine the properties of the cylindrical brush:
\begin{subequations}
\label{eq:cylinder_R}
\begin{equation}
    pN = \int_V\rho(\mathbf{r}) \,d\mathbf{r}=2\pi\ell\int_{R_0}^{R_0 + H_{\rm cyl}} \rho(r)r \, dr,
\end{equation}
where the height of the cylindrical brush follows as: 
\begin{equation} 
\label{eq:cyl_H}
H_{\rm cyl}/R_0 = \left(1 + \frac{H}{R_0}\right)^{3/4} - 1,
\end{equation}
where $H$ is the planar brush height. 
The free energy of the cylindrical brush is then:
\begin{equation}
\beta F_{\rm cyl}^{\rm brush} = 2\pi \ell \int_{R_0}^{R_0 + H_\text{cyl}} \beta f(r) r \, dr = 2 p \sigma^{1/2} R_0 \left[\left( \dfrac{4}{3}\dfrac{H}{R_0} + 1\right)^{3/8} -1 \right],
\end{equation}
where $\sigma$ is the grafting density at the surface [i.e.,~{$~\sigma \equiv \sigma(r=R_0)$}].
\end{subequations}
The spherical brush height and free energy can also be straightforwardly found with:
\begin{subequations}
\label{eq:sphere_R}
\begin{equation}
    pN = \int_V\rho(\mathbf{r}) \,d\mathbf{r}=4\pi\int_{R_0}^{R_0 + H_{\rm sph}} \rho(r)r^2 \, dr,
\end{equation}
which allows us to identify: 
\begin{equation} 
\label{eq:sphere_H}
H_{\rm sph}/R_0 = \left( 1 + \frac{H}{R_0}\right)^{3/5} - 1.
\end{equation}
The free energy of the spherical brush follows as:
\begin{equation}
\beta F_{\rm sph}^{\rm brush} = \dfrac{3}{5} p \sigma^{1/2} R_0 \ln\left( \dfrac{5 H}{3 R_0} + 1 \right).
\end{equation}
\end{subequations}

We can appreciate that both the cylindrical and spherical brush energies take the form of the planar brush energy in the limit of small $H/R_0$ with a fixed grafting density.
Taylor expanding about zero curvature (small $H/R_0$) we find:
\begin{equation}
    \beta F_{\rm cyl}^{\rm brush} = 2 p \sigma^{1/2} R_0 \left[\left( \dfrac{4}{3}\dfrac{H}{R_0} + 1\right)^{3/8} -1 \right] \approx  p \sigma^{1/2} H = pN\sigma^{5/6}b^{2/3}v^{1/3} = \beta F^{\rm brush},
\end{equation}
as well as:
\begin{equation}
\beta F_{\rm sph}^{\rm brush} = \dfrac{3}{5} p \sigma^{1/2} R_0 \ln\left( \dfrac{5 H}{3 R_0} + 1 \right) \approx p\sigma^{1/2}H = \beta F^{\rm brush}.
\end{equation}
One can also readily verify that $H_{\rm cyl} \approx 3H/4$ $H_{\rm sph} = 3H/5$ in these limits using Eqs.~\eqref{eq:cyl_H} and \eqref{eq:sphere_H}.
As stated in the main text, it is therefore appropriate to describe planar micelles using a spherical corona (replacing $R_0$ with the radius of the planar disk, $R$) as we will smoothly transition from the spherical limit (i.e.,~when $R$ is comparable to the thickness of the disk and ``planar'' structures are approximately spherical) to the planar limit with increasing $R$.

Finally, we can clearly identify the dimensionless groups---in terms of the molecular parameters of the surfactants---that the spherical, cylindrical, and planar brush free energies depend on when representing the free energy of a micelle corona. 
Each free energy is linearly proportional to the thermal energy, reflecting the entropic origins of the brush free energy that is of course independent of geometry. 
The spherical and cylindrical free energies take the form ${\beta F\sim \sigma^{1/2}R_0f(H/R_0)}$ where $f(x)$ is a geometry-dependent function. 
We can express the planar brush height as ${H = \sigma^{1/3}r_t^{5/3}}$ and recognize that for spheres we have ${R_0 = R_s}$ and ${\sigma = p/(4\pi R_s^2)}$, for fibers we have ${R_0 = d/2}$ and ${\sigma = p/(\pi d \ell)}$, and for planar aggregates we have ${R_0 = R}$ and ${\sigma = p/(4 \pi R^2)}$.
We can appreciate that each geometric length scale ($R_s, d, \ell, R$) is proportional to the size of the surfactant head group, $r_h$. 
We thus have that $\sigma \propto r_h^{-2}$ and $R_0 \propto r_h$ for each morphology.
One can then immediately verify that $\sigma^{1/2}R_0$ is a dimensionless numerical constant for each morphology and that $H/R_0 \propto (r_t/r_h)^{5/3}$.
Each brush free energy is thus proportional to $k_BT$ and solely controlled by the dimensionless molecular length scale $r_h/r_t$.

\subsection{Formation Energy Dependencies}

\begin{figure}
    \centering
    \includegraphics[width=1\linewidth]{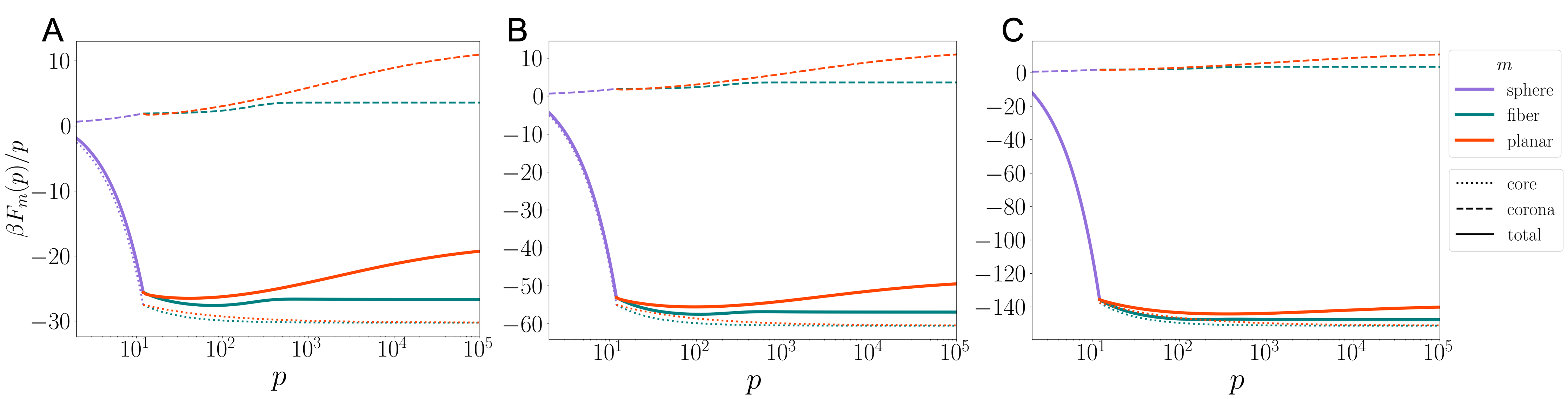}
    \caption{Core and corona contributions to the per molecule free energies of formation shown in Fig.~4 of the main text.
    The panels depict representative values of $\beta \epsilon$ at A) $5.0$, B) $10.0$ and C) $25.0$ at fixed $\alpha = 1.1$ and $r_h/r_t = 0.06$.}
    \label{fig:free_energy_contributions}
\end{figure}

\begin{figure}
    \centering
    \includegraphics[width=1\linewidth]{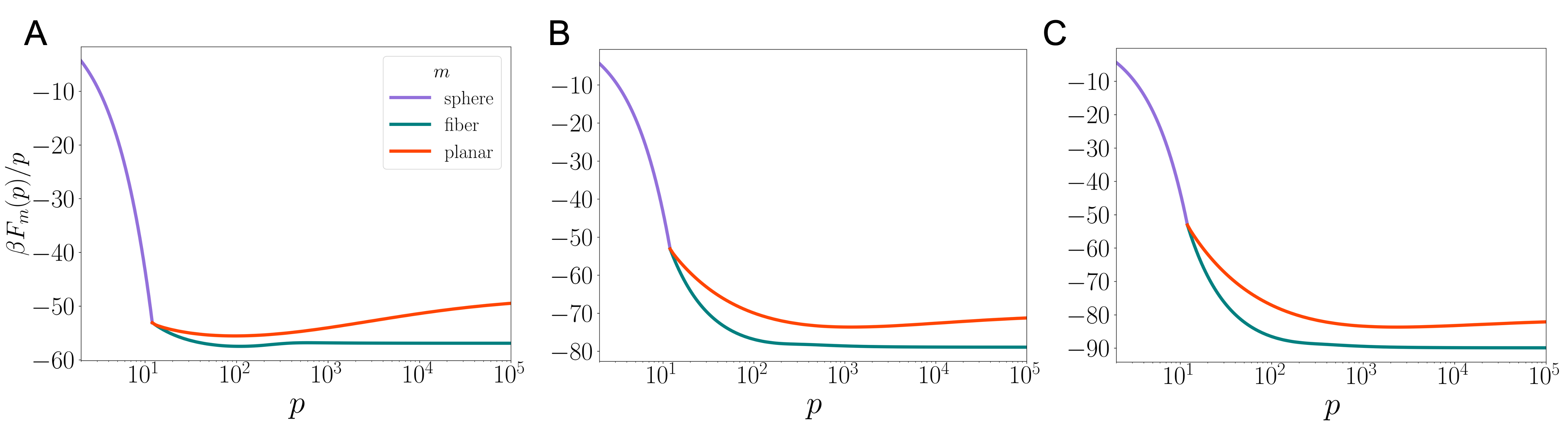}
    \caption{Effects of varying the the molecular packing efficiency parameter $\alpha$ on the free energies of formation.
    The panels depict values of $\alpha$ at A) $1.1$, B) $1.5$ and C) $1.7$ at fixed $\beta \epsilon = 10.0$ and $r_h/r_t = 0.06$.}
    \label{fig:free_energy_varying_alpha}
\end{figure}

\begin{figure}
    \centering
    \includegraphics[width=1\linewidth]{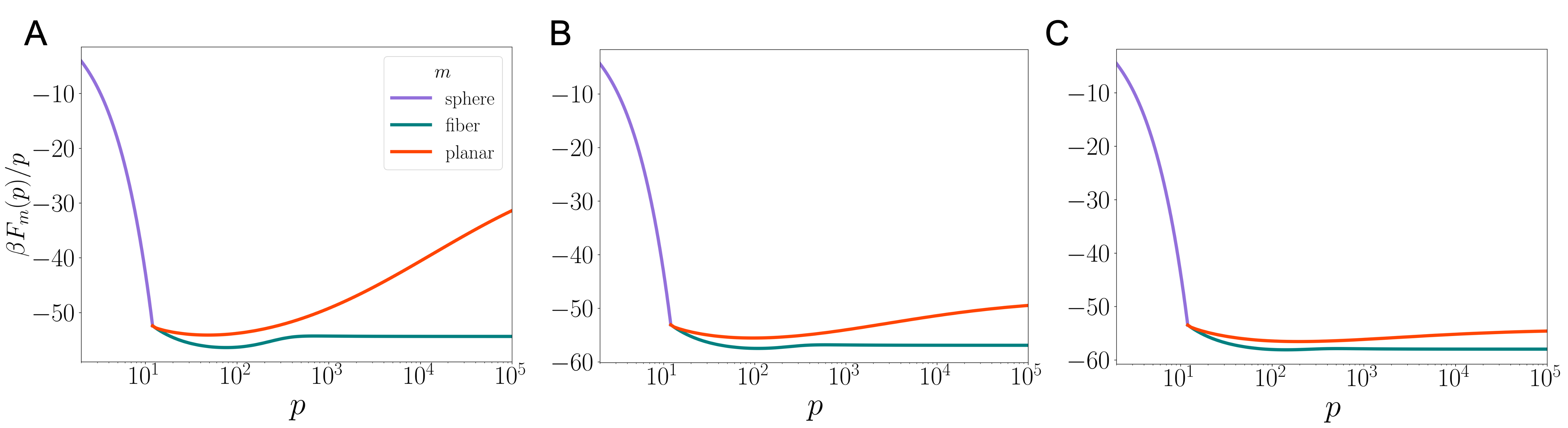}
    \caption{Effects of varying the ratio of head group to tail group size on the free energies of formation.
    The panels depict values of $r_h/r_t$ at  A) $0.03$, B) $0.06$ and C) $0.09$ at fixed $\beta \epsilon = 10.0$ and $\alpha = 1.1$.}
    \label{fig:free_energy_varying_rhrt}
\end{figure}

Figure~\ref{fig:free_energy_contributions} shows the contributions from the core and corona free energies to the total free energy of formation per molecule for three different values of $\beta \epsilon$.
We see that increasing $\beta \epsilon$ reduces the total free energy of formation per molecule by increasing the relative contribution of the core enthalpy.
Increasing $\alpha$ increases the magnitude of the core contribution for planar aggregates and fibers relative to that of spheres and similarity to the $\beta \epsilon$ trend,  decreases $\beta F_m (p)/p$, as depicted by Fig.~\ref{fig:free_energy_varying_alpha}.
Finally, Fig.~\ref{fig:free_energy_varying_rhrt} shows that increasing $r_h/r_t$ also decreases the free energy of formation per molecule.
In contrast to increasing $\beta \epsilon$ and increasing $\alpha$, which reduces $\beta F_m (p)/p$ by increasing the magnitude of the negative core enthalpy, varying $r_h/r_t$ affects the free energy per molecule by reducing the positive contribution from stretching the polymer chains in the corona.
As $r_h/r_t$ increases, the chain grafting density decreases so that chains stretch less, thereby reducing the brush free energy.

\subsection{Phase Diagram Construction }
\subsubsection{Finite-Size Regions}
\begin{figure}
    \centering
    \includegraphics[width=1\linewidth]{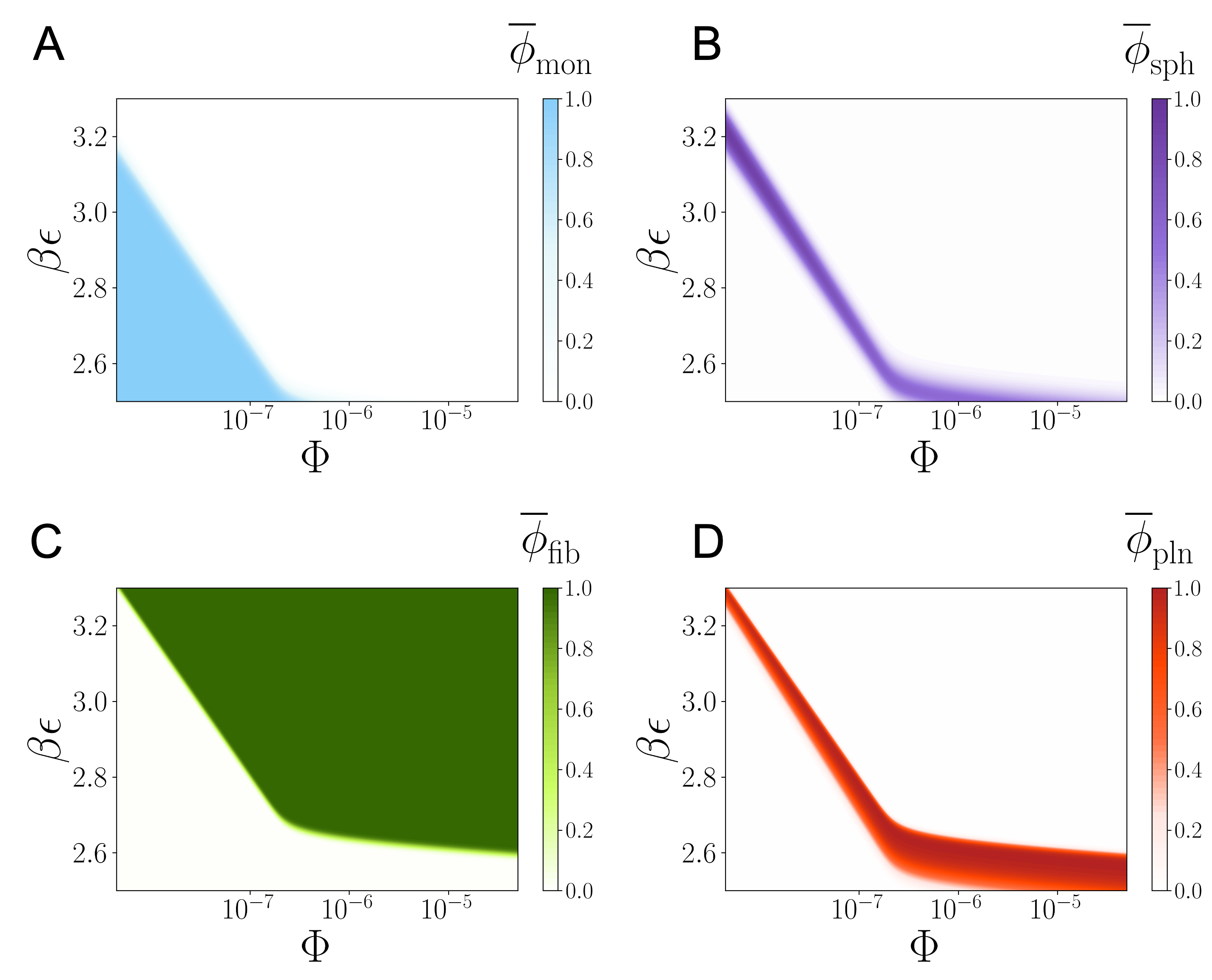}
    \caption{The volume fraction of (A) monomer, (B) sphere, (C) fiber, and (D) planar below the transition to macroscopic aggregates, with varying $\beta \epsilon$ at fixed $r_h/r_t = 0.06$ and $\alpha = 1.1$.
    The normalized volume fractions $\overline{\phi}_m = \phi_m/\Phi$.}
    \label{fig:finite_epsilon}
\end{figure}

\begin{figure}
    \centering
    \includegraphics[width=1\linewidth]{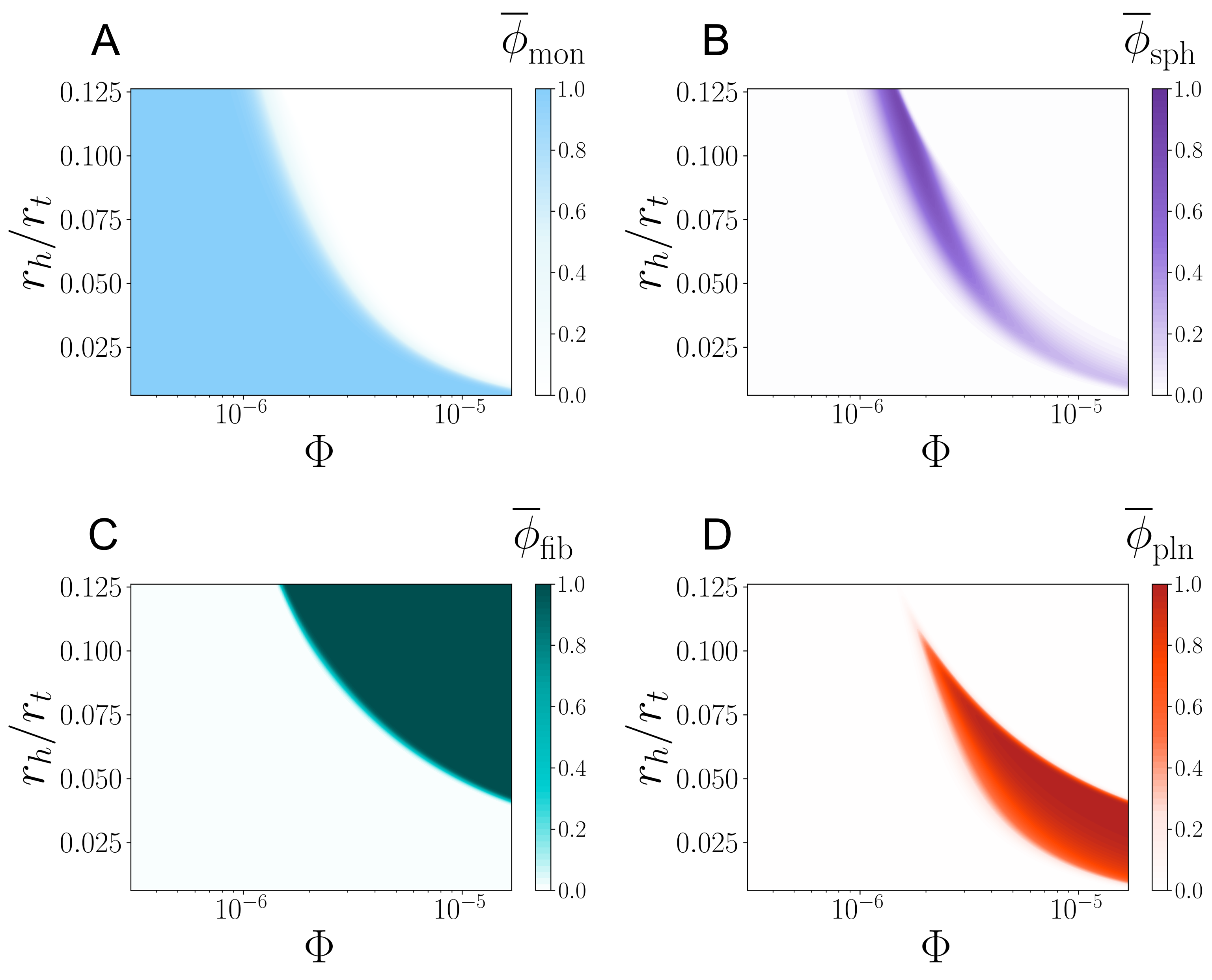}
    \caption{The volume fraction of (A) monomer, (B) sphere, (C) fiber, and (D) planar below the transition to macroscopic aggregates, with varying $r_h/r_t$ at fixed $\beta \epsilon = 2.5$ and $\alpha = 1.1$.
    The normalized volume fractions $\overline{\phi}_m = \phi_m/\Phi$.}
    \label{fig:finite_rh}
\end{figure}

To generate the finite-size phase diagrams shown in the main text we find the $\phi_m$ for each morphology we \textit{select} a value of the coexistence chemical potential and use the equilibrium criterion derived in the main text [see also Eq.~\eqref{eq:coexist3}]:
\begin{equation}
\label{eq:chemical potential}
    \beta v_0 \mu_\text{coexist} = \dfrac{1}{p_m^*}[\ln{\phi_m} + \beta F_m (p_m^*)],
\end{equation}
where the right-hand-side is the chemical potential of an aggregative of morphology $m$ and size $p_m^*$.
While we could use the criterion to obtain the composition of any sized aggregate, we limit our analysis here to the aggregate size that minimizes the formation energy per molecule, $p^*_m$.
Of course, for a monomer we have by definition $p_{\rm mon}^* = 1$ (and $\beta F_m = 0$ such that ${\beta v_0 \mu_\text{coexist} = \ln{\phi_\text{mon}}}$) and for spheres we have $p_{\rm sph}^* = 12$ using our form of the formation energy.
We first compute the free energy of formation for each $\beta \epsilon$ ($r_h/r_t$) and fixed $r_h/r_t$ ($\beta \epsilon)$ and then identify the $p_m^*$ for which $F_m(p)/p$ is minimized.
Figure~8 in the main text displays the value of $p_m^*$ for varying $r_h/r_t$ and $\beta \epsilon$ for the fiber and planar morphologies.
This aggregate size that minimizes $\beta F_m (p)/p$ increases with both increasing $r_h/r_t$ and increasing $\beta \epsilon$.
After identifying $p_m^*$ and $\beta F_m (p_m^*)/p_m^*$ for each point in the parameter space, we sweep through a range of $\mu_\text{coexist}$ and find the $\phi_m$ that satisfy Eq.~\eqref{eq:chemical potential} for each morphology.
The $\phi_m$ are then summed to determine the overall volume fraction, $\Phi$.
We report the morphology with the largest volume fraction in each of the regions shown in our phase diagrams.  
A full breakdown of the $\phi_m$ for each morphology in the $\alpha = 1.1$ case is provided in Figs.~\ref{fig:finite_epsilon} and~\ref{fig:finite_rh}.

\begin{figure}
    \centering
    \includegraphics[width=1\linewidth]{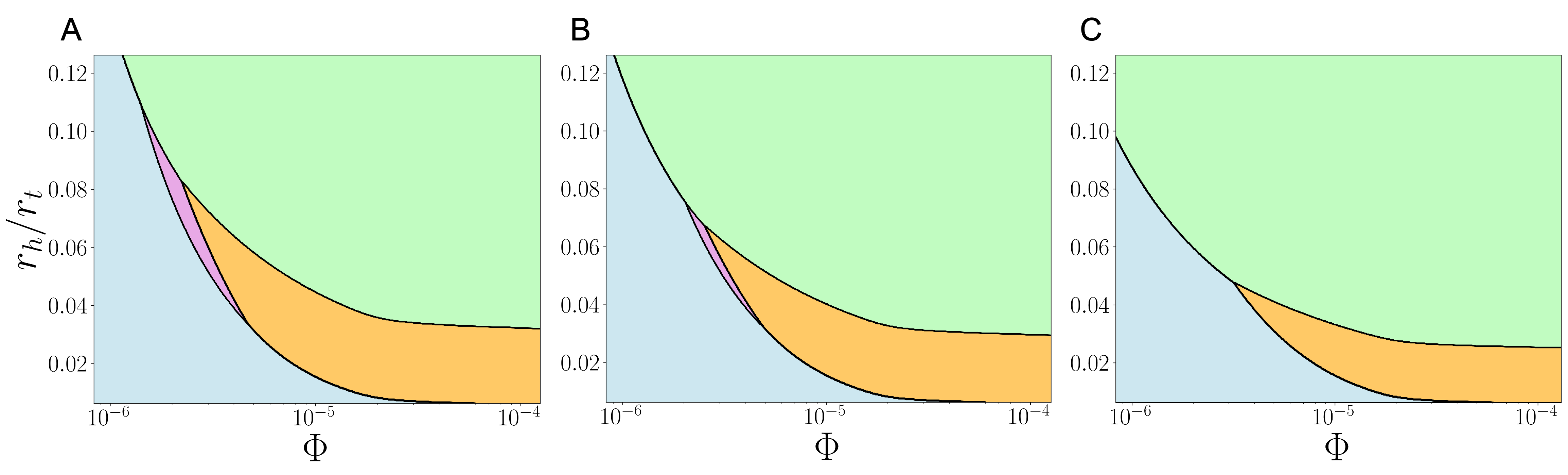}
    \caption{Effects of varying $\alpha$ on the finite-size region of the phase diagram in $\Phi$--$r_h/r_t$ phase space.
    The panels depict values of $\alpha$ at A) $1.12$, B) $1.14$, and C)$1.18$ at fixed $\beta \epsilon = 2.5$.
    In contrast to Fig.~7 in the main text, here we do not color the fiber and planar regions according to the aggregation number.}
    \label{fig:finite_rh_alpha}
\end{figure}

We show in Fig.~\ref{fig:finite_rh_alpha} how varying $\alpha$ affects the finite-size region of the phase diagram.
Increasing $\alpha$ and $r_h/r_t$ pushes $\beta F_m (p)/p$ towards the core contribution, the boundaries between monomers and aggregates shifts to lower and $r_h/r_t$.
We also see the sphere region diminish as the free energy of fibers and planar aggregates is reduced relative to the size of a spherical $12$-mer.

\subsubsection{Macroscopic Regions}
\begin{figure}
    \centering
    \includegraphics[width=1\linewidth]{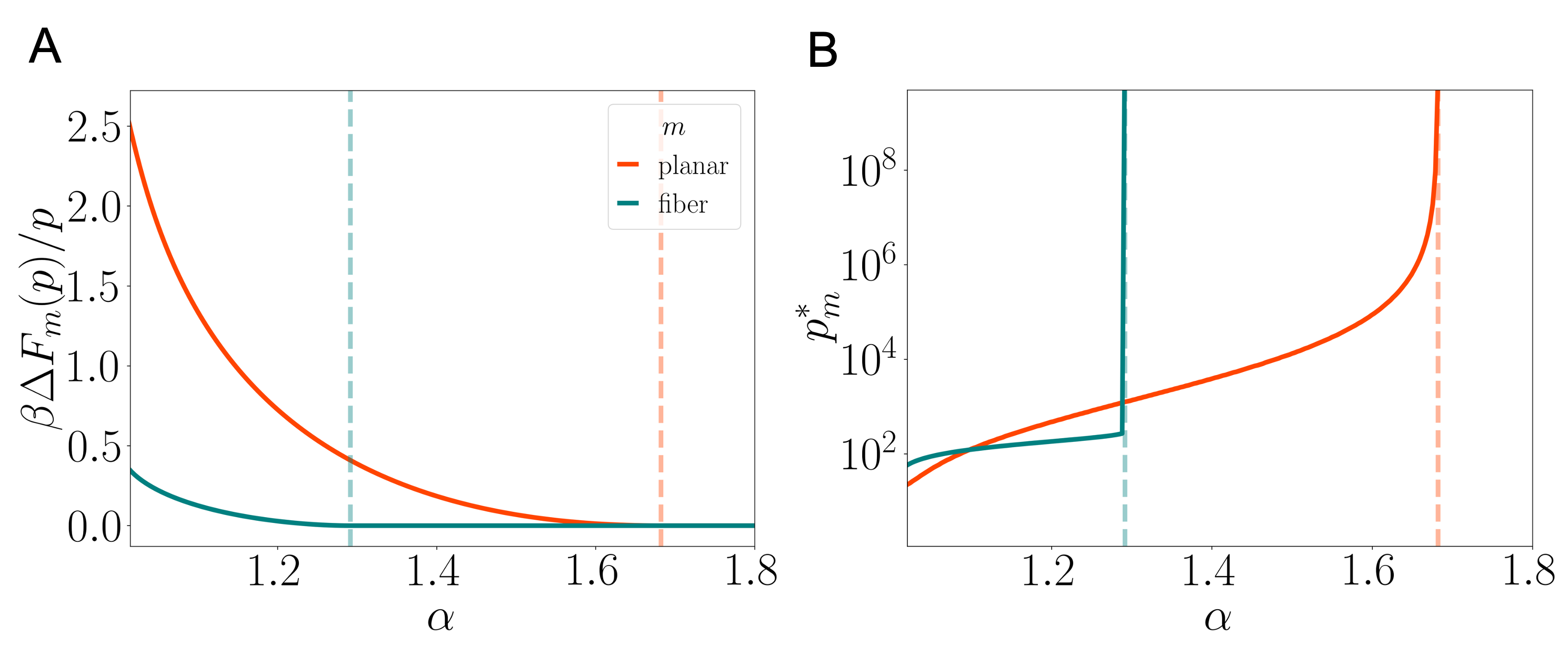}
    \caption{Effects of varying $\alpha$ on the A) well depth and B) $p_m^*$ for planar and fiber morphologies at fixed $r_h/r_t = 0.15$ and $\beta \epsilon = 2.5$. The dashed green and orange lines mark the transition between the finite-sized region and the macroscopic region for fibers and planar structures, respectively.}
    \label{fig:cutoffs_alpha}
\end{figure}

\begin{figure}
    \centering
    \includegraphics[width=1\linewidth]{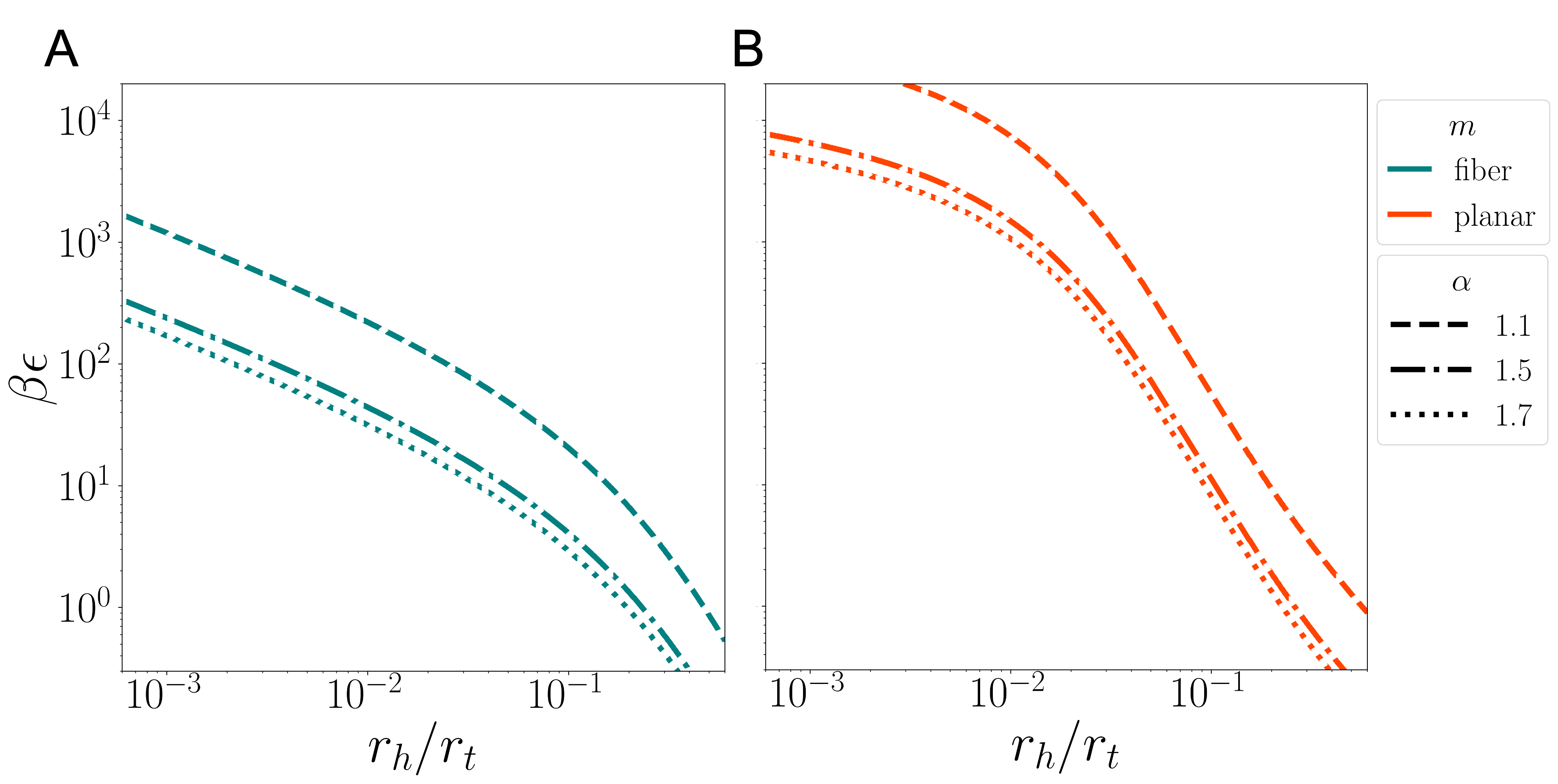}
    \caption{Cutoff between mesoscopic and macroscopic planar structures with varying $\alpha$ for A) fiber and B) planar morphologies.}
    \label{fig:alpha_effects}
\end{figure}

The macroscopic boundaries of fiber and planar aggregates in the main text were made by finding the $\beta \epsilon$ for which the free energy difference between macroscopic and finite-size structures vanished for each $r_h/r_t$.
We show in Fig.~\ref{fig:cutoffs_alpha} that increasing $\alpha$ has similar effects on the well depth and $p_m^*$ for planar and fiber structures as does $\beta \epsilon$.
The difference in the free energy per molecule between finite-sized and macroscopic structures decreases as $\alpha$ increases.
The corresponding minimum size, $p_m^*$, also increases.
Figure~\ref{fig:alpha_effects} shows that varying $\alpha$ only quantitatively effects that phase diagram, with increasing $\alpha$ leading to a lower mesoscale-macroscale transition for each of the structures. 

\clearpage

\addcontentsline{toc}{section}{References}
\bibliographystyle{bibStyle}